\documentclass[onecolumn,amsmath,amssymb,nofootinbib,12pt]{article}
 
 \ifx\pdfoutput\not\undefined
 \pdfoutput=1
\fi
\usepackage{jheppub}
\usepackage{ifpdf}

\usepackage{mdframed}

\usepackage{graphicx,subcaption,comment}
\usepackage{float}
\graphicspath{{images/}{./}}
\usepackage{epsfig}
\usepackage{pdfpages}
\usepackage{bbm}
\usepackage{graphicx,epstopdf}
\usepackage[numbers,sort&compress]{natbib}
\usepackage[makeroom]{cancel}
\usepackage{hyperref}
\hypersetup{colorlinks=true,allcolors=blue}
\usepackage{array}
\usepackage[export]{adjustbox}

\usepackage{amsmath}
\usepackage{amssymb}
\usepackage{amsfonts}
\usepackage{amsthm}
\usepackage{lineno}
\usepackage{dsfont}
\usepackage{bm}
\usepackage{physics}
\usepackage{mathtools}
\usepackage{commath}
\usepackage{color}
\usepackage{xcolor}
\usepackage[normalem]{ulem}

\newcommand{\sy}{\mathsf{y}}

\newcommand{\A}{\textsc{a}}
\newcommand{\B}{\textsc{b}}

\newcommand{\tv}{\tilde{v}}

\newcommand{\ttt}{\tilde{t}}
\newcommand{\tx}{\tilde{x}}
\newcommand{\tz}{\tilde{z}}

\newcommand{\mR}{\mathcal{R}}
\newcommand{\sfX}{\mathsf{X}}
\newcommand{\sfx}{\mathsf{x}}
\newcommand{\sfy}{\mathsf{y}}

\newcommand{\kfA}{K_A^{(f)}}
\newcommand{\kfd}{K_D^{(f)}}
\newcommand{\kfr}{K_R^{(f)}}

\newcommand{\eg}{{e.g.,}\ }
\newcommand{\iee}{{i.e.,}\ }
\newcommand{\mt}[1]{\textrm{\tiny #1}}

\providecommand{\ii}{i} 
\providecommand{\ie}{}
\renewcommand{\ie}{\ii\epsilon}
\newcommand{\cL}{\mathcal{L}}
\newcommand{\cM}{\mathcal{M}}
\newcommand{\cN}{\mathcal{N}}
\newcommand{\GF}[1]{G^{(\mathrm{F})}_\text{#1}}

\providecommand{\dd}{}
\renewcommand{\dd}{\mathrm{d}}
\newcommand{\sA}{\textnormal{\textsc{a}}}
\newcommand{\sB}{\textnormal{\textsc{b}}}
\newcommand{\zA}{z_\sA}
\newcommand{\zB}{z_\sB}

\title{Boundary duals of bulk detectors
}

\author[1,2]{Jacqueline Caminiti,}
\author[1]{Robert C. Myers,}
\author[1,3,4]{and Kelly Wurtz}

\affiliation[1]{Perimeter Institute for Theoretical Physics, Waterloo, Ontario, N2L 2Y5, Canada}
\affiliation[2]{Department of Physics and Astronomy, University of Waterloo,\\ Waterloo, Ontario N2L 3G1, Canada}
\affiliation[3]{Department of Applied Mathematics, University of Waterloo,\\ Waterloo, Ontario, N2L 3G1, Canada}
\affiliation[4]{Institute for Quantum Computing, University of Waterloo, Waterloo, Ontario, N2L 3G1, Canada}

\emailAdd{jcaminiti@perimeterinstitute.ca,
rmyers.perimeter@gmail.com, kwurtz@uwaterloo.ca}

\abstract{What is the boundary dual of an Unruh--DeWitt detector in anti-de Sitter space? We argue that this question is naturally answered by the HKLL reconstruction program. A local detector in the bulk can be associated with a smeared detector in the boundary,
where the choice of smearing depends on the choice of HKLL kernel. 
We use this formalism to analyze entanglement harvesting in the Poincar\'e patch of AdS$_{3}$ for a bulk Unruh--DeWitt detector paired either with a boundary detector or with a second bulk detector. 
We compute the amount of harvesting  in the standard way, namely by evaluating the two-detector density matrix to quadratic order in the coupling $\lambda$.
Additionally, we address an apparent tension between microcausality for spacelike-separated bulk detectors and the fact that the dual HKLL detectors may directly overlap in the boundary theory.
}

\begin{document}
\vspace*{1.5cm}

\maketitle

\section{Introduction}
\label{sec:intro}

\noindent
Unruh--DeWitt detectors provide an operational framework for probing correlations in quantum field theory \cite{Unruh:1976db,DeWitt:1979qg}. In this framework, one couples localized quantum systems (the ``detectors'') to a field and studies the correlations they acquire. Historically, detector models helped elucidate fundamental quantum phenomena such as Hawking radiation \cite{Hawking:1975vcx} and the thermal response of accelerated systems in vacuum \cite{Unruh:1976db}. 
Meanwhile, in relativistic quantum information, detector models are used to study the entanglement structure of quantum fields in a more direct way \cite{Peres:2004cc,Alsing:2012wf}. Typically, this is done using  a protocol known as entanglement harvesting, in which two initially uncorrelated detectors can become entangled through their interaction with a field \cite{Reznik:2002fz,Pozas-Kerstjens:2015gta}. Notably, such entanglement can arise even when the detectors remain spacelike separated throughout their interaction with the field, and therefore the acquired entanglement cannot be attributed to causal signalling between the detectors. 
Instead, it comes from entanglement already present in the state of the quantum field, making the  detectors' behaviour a probe of the field's entanglement structure.

Given that entanglement plays a central role in AdS/CFT, underlying the emergence of bulk geometry from a non-gravitational boundary theory \cite{VanRaamsdonk:2010pw,Rangamani:2016dms}, it is natural to study detector-based protocols in holographic CFTs. 
This perspective is complementary to the more familiar use of subregion entanglement measures in CFT and holography: here, the object of interest is not the entanglement entropy of a spatial region, and no replica trick is required. Instead, at leading order in perturbation theory, the detectors' reduced density matrix is determined using the correlation functions of the local CFT operator to which the detectors couple. Entanglement harvesting in CFTs was developed in \cite{Wurtz:2026ofi}, where pointlike Unruh--DeWitt detectors were coupled to scalar primary operators.
Ref.~\cite{Wurtz:2026ofi} studied how detector correlations depend on conformal data, and how the generalized free field structure of large-$N$ holography can separate genuinely harvested correlations from communication-mediated effects. Related detector-based probes of holographic CFTs, which focused on resource harvesting from a single detector, were considered in \cite{Zhang:2026jll}.
Entanglement harvesting and local Unruh--DeWitt detectors in AdS spacetimes were also studied in, \eg \cite{Henderson:2018lcy,Ng:2018drz,
Pitelli:2021oil,Yang:2025zrl}.
Our perspective is complementary to these earlier studies. Rather than treating protocols for bulk or boundary detectors in isolation, we use HKLL reconstruction to describe local detectors in AdS in terms of spacetime-smeared detectors in the dual boundary theory.

In general, the simplest model of an Unruh--DeWitt detector
is a two-level quantum system coupled to a scalar quantum field; following \cite{Wurtz:2026ofi}, we consider conformal primary scalar fields in holographic CFTs. A convenient idealization is to treat the detector as pointlike, so that the interaction between the qubit and the conformal field takes place along a single worldline in spacetime. However, more generally, finite-size detectors can be treated by introducing additional spatial smearings \cite{Schlicht:2003iy,Pozas-Kerstjens:2015gta,
Martin-Martinez:2020pss,Martin-Martinez:2018gzb}. 
We demonstrate that the boundary dual of such standard, local detectors in the bulk AdS spacetime can be constructed by leveraging the bulk reconstruction program in AdS/CFT, which is also commonly known as HKLL reconstruction (\eg see \cite{Hamilton:2005ju, Hamilton:2006az, Hamilton:2006fh, Harlow:2018fse,Kajuri:2020vxf}).\footnote{This approach was also hinted at in \cite{Zhang:2026jll}.} 
Hence we refer to these detectors in the boundary theory as ``HKLL detectors.'' 
We will find that our HKLL detectors have several unusual features and differ from the standard detector models one would typically use to probe the boundary CFT.

We then apply our general framework of HKLL detectors to two examples of entanglement harvesting in a two-dimensional holographic CFT.
In both cases, we compute the amount of entanglement harvested  using the two-detector density matrix, evaluated to quadratic order in the coupling $\lambda$,
both numerically and using closed-form approximations.
As our first example, we consider harvesting between one pointlike detector and one HKLL detector,
where the latter represents the boundary dual of a local detector in the bulk.
We demonstrate that our results reproduce ordinary CFT harvesting in the limit where the bulk detector is pushed to the boundary.
We also demonstrate that harvesting drops off to zero in the opposite limit, in which the bulk detector is pushed deeper into the bulk.
As our second example, we consider harvesting between two HKLL detectors,
each corresponding to a local detector in the bulk,
and we examine  the sensitivity of harvesting to causal signalling between the two bulk detectors.

A curious feature of HKLL detectors is that they can have overlapping supports in the boundary CFT, even when their bulk duals are spacelike separated. This brings out an apparent tension between microcausality in the bulk and boundary theories.  We address this tension and its resolution in the discussion section and relegate a detailed derivation to an appendix.

The outline of the rest of the paper is as follows:
In section~\ref{sec:setup}, we review the bulk reconstruction program and HKLL kernels in the AdS/CFT correspondence. 
In section~\ref{sec:detector-setup}, we introduce Unruh--DeWitt detectors, smeared Unruh--DeWitt detectors, and HKLL detectors, where the latter serve as the boundary duals of local detectors in the bulk. In section~\ref{sec:harvesting-geometries}, we perform harvesting calculations for two different detector protocols. We conclude in section \ref{sec:discussion} with a summary of our results and an outlook on future directions. 
In appendix \ref{app:causality-overview}, we explain how to recover the bulk two-point function (and hence, bulk microcausality) using HKLL.
In appendix \ref{app:WKB}, we study asymptotic features of Klein-Gordon mode functions in the AdS-Rindler spacetime. 

\section{Bulk reconstruction}
\label{sec:setup}

In this section, we review the ingredients of bulk reconstruction in AdS/CFT that will be needed below. Specifically, we will explain how a local bulk field, and hence a qubit detector coupled to that field (as explored in section~\ref{sec:detector-setup}), can be represented as a smeared operator in the boundary theory. The interested reader may find more extensive discussions of bulk reconstruction in AdS/CFT in
\cite{Hamilton:2005ju,Hamilton:2006az,Hamilton:2006fh,Harlow:2018fse,Kajuri:2020vxf,Dong:2016eik}.

We will work in the Poincar\'e patch of AdS$_{d+1}$, with metric
\begin{equation}\label{eq:metric}
  \dd s^2
\ = \ \frac{\ell^2}{z^2}\,
  \bigl(\dd z^{2} - \dd t^{2}
  + \dd\vec{x}^{\,2}\bigr)\, .
\end{equation}
The conformal boundary of this patch lies at $z=0$ and is $d$-dimensional Minkowski spacetime.
We denote bulk points by $\sfX=(z,t,\vec{x})$ and boundary points by $\sfx=(t,\vec{x})$.

According to the AdS/CFT correspondence \cite{Aharony:1999ti,Ammon:2015wua,Harlow:2018fse}, certain large-$N$ conformal field theories give a nonperturbative definition of quantum gravity in asymptotically AdS spacetime. 
For our present purposes, the most important part of the dictionary is the relation between a bulk scalar field $\hat{\Phi}$ and a (single-trace) conformal primary operator $\hat{\mathcal{O}}$ in the boundary CFT. If $\hat{\Phi}$ has mass $m$, then the conformal dimension $\Delta$ of $\hat{\mathcal{O}}$ is fixed by
\begin{equation}
    \Delta \ = \ \frac{d}{2}+\sqrt{\frac{d^2}{4}+m^2\ell^2}\, .
    \label{eq:confdimen}
\end{equation}
Near the boundary, the two operators are related by the extrapolate dictionary \cite{Harlow:2011ke,Banks:1998dd,Balasubramanian:1998de},
\begin{equation}
    \lim_{z\to 0} z^{-\Delta}\,\hat{\Phi}(z,\mathsf{x})
\ = \ \hat{\mathcal{O}}(\mathsf{x})\, .
    \label{eq:extrapolate}
\end{equation}
Thus the boundary operator is obtained by taking the appropriately rescaled boundary limit of the bulk field.\footnote{As a technical note, throughout we will employ a normalization of the free bulk field $\hat{\Phi}$ in which $\hat{\Phi}$ is dimensionless: $[\hat{\Phi}]=(\textrm{length})^{0}$.
Then, eq.~\eqref{eq:extrapolate} gives the correct mass dimension for the dual primary field:
$[\hat{\mathcal{O}}]=(\textrm{length})^{-\Delta}$.
Finally, the HKLL kernel $K_\Delta$ introduced below has 
$[K_\Delta]=(\textrm{length})^{\Delta-d}$. 
\label{foot:units}
}

Equation \eqref{eq:extrapolate} is a starting point, but it does not by itself give a boundary description of local fields or measurements deep in the bulk. To describe such measurements, one would like to ``back off'' from the boundary limit and represent $\hat{\Phi}(\sfX)$, at finite radial position $z$, directly as an operator in the CFT. This is the problem of bulk reconstruction, and there is a broad literature devoted to this question \cite{Banks:1998dd,Balasubramanian:1998de,Bena:1999jv,Balasubramanian:1999ri,Hamilton:2005ju,Hamilton:2006az,Hamilton:2006fh,Morrison:2014jha,Heemskerk:2012np,Heemskerk:2012mn,Kabat:2011rz,Kabat:2012av,Kabat:2013wga,Bousso:2012mh}. Early work showed that, at least perturbatively in the large-$N$ expansion, one can construct CFT operators corresponding to local bulk fields at finite radial position by solving the bulk equations of motion with boundary conditions fixed by the extrapolate dictionary. This construction is usually referred to as HKLL reconstruction, after the works of Hamilton, Kabat, Lifschytz, and Lowe \cite{Hamilton:2005ju,Hamilton:2006az,Hamilton:2006fh}, who were the first to study this question in detail.

The word ``perturbatively'' is key here, as bulk reconstruction is generally an approximate notion. 
This is because the description of the bulk in terms of local quantum fields on a fixed spacetime is valid only in a semiclassical regime. In practice, this means working at large $N$ (or large central charge) in the boundary theory, or equivalently at small Newton constant $G_N$ in the bulk, where gravitational backreaction is perturbatively small. 
This involves restricting one's attention to a set of states describing small excitations about a fixed semiclassical geometry.
This set of states is often called a code subspace, emphasizing the quantum-error-correction structure of holography \cite{Almheiri:2014lwa,Dong:2016eik,Harlow:2016vwg}. 

Within such a semiclassical sector, the bulk physics is well-approximated by free quantum fields on a fixed asymptotically AdS background. The single-trace boundary operator $\hat{\mathcal{O}}$ hence behaves as a generalized free field, since its correlators are reproduced by those of the dual bulk field $\hat{\Phi}$. Within this setting,
a local bulk field can be represented as a CFT operator smeared over some boundary region, 
\begin{equation}\label{eq:HKLL-schematic}
  \hat{\Phi}(\mathsf{X})\ = \ \int\!\dd^d \mathsf{x}'\;K_\Delta(\mathsf{X}\,|\,\mathsf{x}')\,\hat{\mathcal{O}}(\mathsf{x}')\, .
\end{equation}
Here $\mathsf{X}=(z, \mathsf{x})$ is the bulk point being reconstructed, while $\mathsf{x}'$ is the boundary point over which we integrate. The vertical bar notation in $K_\Delta(\sfX\,|\,\sfx')$ separates the bulk label of the kernel from its boundary argument. 
The operator equality in eq.~\eqref{eq:HKLL-schematic} should be understood within the chosen semiclassical sector or code subspace.
Equivalently, it is an equality inside correlation functions in that regime:
\begin{equation}
    \langle \hat{\Phi}(\sfX_1)\cdots\hat{\Phi}(\sfX_n)\rangle_{\text{bulk}}
\ = \ 
    \int\!\prod_{i=1}^n \dd^d \sfx_i'\;
    K_\Delta(\sfX_i\,|\,\sfx_i')\,
    \langle \hat{\mathcal{O}}(\sfx_1')\cdots
    \hat{\mathcal{O}}(\sfx_n')\rangle_{\text{bdy}}\, .
    \label{eq:hkll-correlator}
\end{equation}
In the following, when we say that a kernel or boundary region reconstructs bulk operators, we mean that it allows a boundary representation of the form~\eqref{eq:HKLL-schematic} which can be used to reproduce bulk correlators as above. These calculations can also be extended perturbatively beyond leading order in the large-$N$ expansion to include bulk interactions \cite{Kabat:2011rz,Kabat:2012av,Kabat:2013wga,Heemskerk:2012mn}.

The form of the smearing kernel $K_\Delta$ is constrained by the defining properties of the bulk field. First, away from the boundary insertion, the reconstructed field must obey the free bulk equation of motion,
\begin{equation}
    (\Box_\sfX - m^2)\hat{\Phi}(\sfX)\ = \ 0\, .
\end{equation}
Second, $\hat{\Phi}(\sfX)$ must obey the extrapolate dictionary~\eqref{eq:extrapolate}. Substituting the HKLL representation into these two requirements gives
\begin{enumerate}
    \item \( (\Box_\sfX-m^2)K_\Delta(\sfX|\sfx')=0 \) for bulk points away from the boundary insertion,
    \item \( \displaystyle \lim_{z\to0} z^{-\Delta}\,K_\Delta(z,\sfx|\sfx') = \delta^{(d)}(\sfx-\sfx') \).
\end{enumerate}

A useful feature, and also a subtlety, is that the smearing kernel is not unique: different boundary integrals can represent the same bulk operator within the code subspace of interest. 
One source of this freedom is the possibility of adding a null kernel,
\begin{equation}
K_\Delta(\sfX|\sfx')\longrightarrow K_\Delta(\sfX|\sfx')+\delta K(\sfX|\sfx').
\end{equation}
A null kernel is defined relative to the chosen code subspace by the condition
\begin{equation}
\int\!\dd^d\sfx'\,\delta K(\sfX|\sfx')\,\hat{\mathcal O}(\sfx')=0
\end{equation}
inside all correlation functions in that code subspace.  
Nontrivial null kernels can exist because the boundary data carried by 
\(\hat{\mathcal O}\) span only a restricted space of modes;
a boundary distribution with zero overlap with every such mode cannot affect the reconstructed operator.
It follows directly that \(K_\Delta\) and \(K_\Delta+\delta K\) give equivalent boundary representations of the same bulk operator.
A second source of freedom is in the boundary subregion: the same bulk point may be reconstructible from more than one boundary subregion. 
If $\sfX$ can be reconstructed from two different boundary regions $A$ and $B$, with kernels $K_A$ and $K_B$, then
\begin{equation}
    \int_A\!\dd^d \sfx'\,K_A(\sfX|\sfx')\,\hat{\mathcal{O}}(\sfx')
\ = \
    \int_B\!\dd^d \sfx'\,K_B(\sfX|\sfx')\,\hat{\mathcal{O}}(\sfx')
    \label{eq:differKAB}
\end{equation}
as operators within the relevant code subspace. Thus the same bulk operator may admit different boundary representatives, whose kernels differ in their support and their detailed form.

\subsection{Causal diamond kernel}
\label{sec:ads-rindler}

The original HKLL construction~\cite{Hamilton:2005ju,Hamilton:2006az} represents a free field at a bulk point  as a boundary operator smeared over an extended region on the AdS boundary which includes an entire Cauchy slice. 
For example, in Poincar\'e coordinates \eqref{eq:metric}
and when $d+1$ is odd,
the smearing kernel has support over the full \(d\)-dimensional Minkowski spacetime at the conformal boundary. However, bulk reconstruction can also be formulated using data on a finite boundary region \cite{Hamilton:2006fh,Morrison:2014jha}, and this approach is more natural for producing boundary descriptions of bulk RQI protocols, as we consider here. We therefore focus on the case in which the boundary support is restricted to a finite causal diamond. 
This section reviews the corresponding smearing kernel, following~\cite{Morrison:2014jha}. We note, however, that the causal diamond kernel \(K_D\) associated with reconstruction at a fixed bulk point is distributional and must be understood in a weak sense.
Motivated by this, in section~\ref{sec:smeared-hkll}, we also consider the kernel obtained by reconstructing a smeared bulk field, which is an ordinary smooth function in the interior of the diamond.

Let \(D\) denote a causal diamond on the boundary, defined as the domain of dependence of a spatial region on a boundary Cauchy slice. 
The associated causal wedge \(W[D]\) is the bulk region that can both send signals to, and receive signals from, the boundary region \(D\) \cite{Hubeny:2012wa,Hubeny:2013gba}. Equivalently, we can write
\begin{equation}
    W[D]\ =\ J^+(D)\cap J^-(D)\,,
\end{equation}
where the causal future and past are computed in the bulk spacetime. Causal wedge
reconstruction  says that if a bulk point \(\sfX\) lies in \(W[D]\), then the bulk
operator \(\hat{\Phi}(\sfX)\) can be represented as an integral of boundary operators with support only in \(D\) \cite{Hamilton:2006az,Harlow:2018fse}.\footnote{In general, the bulk region reconstructible from a fixed boundary region may extend beyond the causal wedge to the so-called entanglement wedge \cite{Almheiri:2014lwa,Dong:2016eik,Harlow:2016vwg}. However,
for the ball-shaped boundary regions we will consider in the AdS vacuum geometry, the entanglement wedge coincides with the causal wedge.}

\begin{figure}[t]
  \centering
  \includegraphics[width=.9\textwidth]{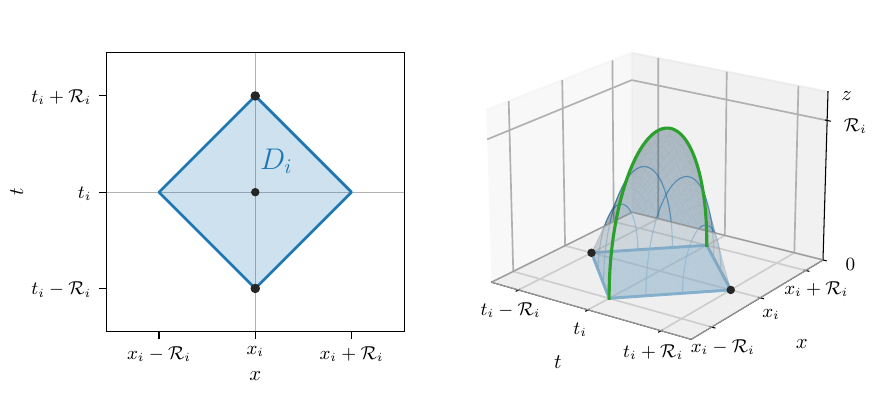}
  \caption{Left panel: the support of the kernel $K_{D_i}$ is the boundary causal diamond $D_i=\{(t,x): |t-t_i|+|x-x_i|\le \mR_i\}$.
  Right panel: the set of bulk points reconstructible from this boundary diamond is its causal wedge $W[D_i]$. The intersection of the two lightcones as given in eq.~\eqref{eq:bifur} is shown in green.}
  \label{fig:diamond-support}
\end{figure}

For our explicit calculations, we will work in \(d=2\), but it is straightforward to extend the qualitative content of our discussion to general boundary dimensions.
In $d=2$, the bulk metric \eqref{eq:metric} for the vacuum AdS$_3$ spacetime reduces to
\begin{equation}\label{eq:metric-d2}
  \dd s^2
\ = \ \frac{\ell^2}{z^2}\,
  \bigl(\dd z^{2} - \dd t^{2}
  + \dd x^{2}\bigr)\, ,
\end{equation}
and we keep the notation \(\sfX=(z,t,x)\) for bulk points and \(\sfx=(t,x)\) for boundary points. Let
\(D_i\) denote a boundary causal diamond centered at $c_i=(t_i,x_i)$
with size \(\mR_i\), i.e.,
\begin{equation}
    D_i
\ = \
    \{(t,x): |t-t_i|+|x-x_i|\leq \mR_i\}\, ,
    \label{eq:diamond-general}
\end{equation}
as shown in figure~\ref{fig:diamond-support}. 
The $i$ label is a placeholder which becomes useful when we wish to refer to distinct causal diamonds in the same equation, e.g.~$D_a$ and $D_b$.
In general, the diamond $D_i$ is the boundary domain of dependence of the interval \(-\mR_i\leq x-x_i\leq \mR_i\) on the \(t=t_i\) slice.
The corresponding causal wedge \(W_i=W[D_i]\) is the bulk region bounded by two bulk lightcones emanating from the past and future tips of the diamond at
\((t,x)=(t_i-\mR_i,x_i)\) and \((t,x)=(t_i+\mR_i,x_i)\). These lightcones meet at
\begin{equation}
    t=t_i\, ,
    \qquad
    (x-x_i)^2+z^2=\mR_i^2\, ,\label{eq:bifur}
\end{equation}
shown by a green curve in the right panel of figure~\ref{fig:diamond-support}.

The construction of \(K_{D_i}\)  proceeds by writing the causal wedge \(W[D_i]\) in
AdS-Rindler coordinates.
In Rindler coordinates, the wedge has a simple coordinate description and $D_i$ becomes the complete boundary of the Rindler patch. 
One constructs an HKLL kernel \(K_R\) in these Rindler coordinates, and then maps it back to the Poincar\'e boundary coordinates \((t,x)\)
on the diamond \(D_i\). 

To simplify the formulae in what follows, we introduce the dimensionless Poincar\'e coordinates
\begin{equation}
  \tilde z_i = \frac{z}{\mR_i}\,,\qquad \tilde t_i=\frac{t-t_i}{\mR_i}\, ,
    \qquad
    \tilde x_i=\frac{x-x_i}{\mR_i}\, .
    \label{eq:dimensionless-diamond-coords}
\end{equation}
Following \cite{Casini:2011kv},
we may now introduce the (dimensionless) AdS-Rindler coordinates \((Z,T,X)\) with
\begin{equation}
    \frac{1}{\tz_i}\left(1,\ttt_i,\tx_i\right)
\ = \
    \frac{1}{Z}
    \left(
    \cosh X+\sqrt{1-Z^2}\,\cosh T,\,
    \sqrt{1-Z^2}\,\sinh T,\,
    \sinh X
    \right)\, .
    \label{eq:poincare-to-rindler}
\end{equation}
In these coordinates, the AdS$_3$ metric \eqref{eq:metric-d2} reads
\begin{equation}
    \dd s^2
\ = \
    \frac{\ell^2}{Z^2}
    \left[
    -(1-Z^2)\,\dd T^2
    + \frac{\dd Z^2}{1-Z^2}
    + \dd X^2
    \right]\, .
    \label{eq:ads-rindler-metric}
\end{equation}
The bulk causal wedge \(W[D_i]\) is simply the region \(0<Z<1\), with \(T,X\in(-\infty,\infty)\). 
Meanwhile, the two null boundaries of the causal wedge 
are 
the Rindler-like horizons at \((Z,T)\to(1,\pm\infty)\), and one can easily verify that they intersect along the surface in eq.~\eqref{eq:bifur}.

Restricting eq.~\eqref{eq:poincare-to-rindler} to the conformal boundary, where \(z\to0\) and \(Z\to0\), gives
\begin{equation}
\label{eq:rindler-poincare}
    (t-t_i,\,x-x_i)
\ = \
    \frac{\mR_i}{\cosh T+\cosh X}\,
    \left(\sinh T,\,\sinh X\right)\, .
\end{equation}
This map identifies the full Rindler boundary, parametrized by \(T,X\in\mathbb{R}\), with the interior of the causal diamond \(D_i\).
The null edges of the diamond are approached only at infinite Rindler coordinate distance.

Crucially, while eq.~\eqref{eq:poincare-to-rindler} is only a bulk coordinate change, 
the Poincar\'e and AdS--Rindler radial coordinates naturally select different representatives of the boundary conformal class.
Indeed, near the boundary, the first component of eq.~\eqref{eq:poincare-to-rindler} gives
\begin{equation}
    \frac{Z}{z}\longrightarrow\frac{\cosh T+\cosh X}{\mR_i}\equiv \Omega(t,x)\,.
\end{equation}
Consequently, the flat boundary representatives selected by the two radial coordinates are related as follows:
\begin{equation}
  -\dd T^2+\dd X^2
\ = \
  \Omega^2(t,x)\,
  \bigl(-\dd t^2+\dd x^2\bigr)\, ,
  \label{eq:boundary-conformal-factor}
\end{equation}
where the Weyl factor can be independently computed as\footnote{Alternatively, we can introduce boundary lightcone coordinates:
$\tilde u_i\equiv \tilde t_i+\tilde x_i= \tanh\!\big(\frac{T+X}{2}\big)$ and
$\tilde v_i\equiv \tilde t_i-\tilde x_i
    = \tanh\!\big(\frac{T-X}{2}\big)$.
Then the Weyl factor becomes $\Omega^{-2}(\tilde u_i,\tilde v_i)=\mR_i^2 (1-\tilde u_i^{\,2})(1-\tilde v_i^{\,2})/4$.}

\begin{equation}\label{eq:jacobian-omega}
\begin{split}
\Omega^2(t,x)
\ \equiv& \
  \left|\frac{\partial(T,X)}{\partial(t,x)}\right|
\ = \
  \frac{(\cosh T+\cosh X)^2}{\mR_i^2}\\
\ =& \
  \frac{4\,\mR_i^2}{
  \bigl((\mR_i-x+x_i)^2-(t-t_i)^2\bigr)
  \bigl((\mR_i+x-x_i)^2-(t-t_i)^2\bigr)}\, .     
\end{split}  
\end{equation}
Note that \(\Omega\) diverges at the four null boundaries of the diamond $D_i$.
This divergence reflects the fact that the
boundary of \(D_i\) lies at infinite distance in the Rindler frame.

We now construct the Rindler kernel. The first step is to expand the bulk field in a Fourier basis of Klein-Gordon solutions:
\begin{equation}\label{eq:mode-expansion-rindler}
  \hat{\Phi}_R(\sfX_R)
\ = \
  \int\!\frac{\dd\omega\,\dd k}{4\pi^2}\ 
  V_{\omega k}^{R}(Z)\,
  e^{-\ii\omega T+\ii kX}\,
  \hat{a}_{\omega k}\, ,
\end{equation}
where $\sfX_R=(Z,T,X)$ denotes the Rindler coordinates of a bulk point,
and throughout this construction, both $\omega$ and $k$ run over the full real line.
The radial profiles $V_{\omega k}^{R}(Z)$ are fixed by the bulk equation of motion and the
near-boundary behaviour \(V_{\omega k}^{R}(Z)\to Z^\Delta\) as \(Z\to0\).
They read \cite{Morrison:2014jha}
\begin{equation}\label{eq:V-rindler}
  V_{\omega k}^{R}(Z)
\ = \
  Z^{\Delta}\,
  (1-Z^2)^{-\ii\omega/2}\,
  {}_2F_1\left(
    \frac{\Delta-\ii\omega+\ii k}{2},
    \frac{\Delta-\ii\omega-\ii k}{2};
    \Delta;\,Z^2
  \right)\, .
\end{equation}
Taking the $Z\to0 $ limit on both sides of eq.~\eqref{eq:mode-expansion-rindler},
using the extrapolate dictionary \eqref{eq:extrapolate},
and using
\(V_{\omega k}^{R}\to Z^\Delta\) as $Z\to 0$ identifies \(\hat{a}_{\omega k}\) as the Fourier 
coefficients of the Rindler boundary operator: 
\begin{equation}\label{eq:rindler-mode-coefficients}
  \hat{a}_{\omega k}
  \ = \
  \int\!\dd T'\,\dd X'\;
  e^{\ii\omega T'-\ii kX'}\,
  \hat{\mathcal{O}}_R(T',X')\, .
\end{equation}
Following eq.~\eqref{eq:HKLL-schematic-recap}, the HKLL kernel $K_R$ is defined as follows:
\begin{equation}\label{eq:HKLL-diamond-2}
  \hat{\Phi}_R(\sfX_R)
\ = \
  \int_{\mathbb{R}^{1,1}}\dd T'\,\dd X'\;
  K_R(\sfX_R|T',X')\,
  \hat{\mathcal{O}}_R(T',X')\, .
\end{equation}
Substituting eq.~\eqref{eq:rindler-mode-coefficients}  into \eqref{eq:mode-expansion-rindler} gives the formal expression
\begin{equation}\label{eq:K-rindler}
  K_R(\sfX_R|T',X')
  \ = \
  \int\!\frac{\dd \omega\,\dd k}{4\pi^2}\,
  e^{-\ii\omega(T-T')+\ii k(X-X')}\,
  V_{\omega k}^{R}(Z)\, 
\end{equation}
for the Rindler HKLL kernel.

At fixed \(\omega\) and \(0<Z<1\), the
large-\(k\) behaviour of the hypergeometric function in eq.~\eqref{eq:V-rindler} carries a non-oscillatory exponential envelope\footnote{See
\cite{Hamilton:2006az,Bousso:2012mh}, and our derivation in appendix \ref{app:WKB}.}
\begin{equation}\label{eq:V-rindler-asymptotic}
  V_{\omega k}^{R}(Z)
  \ \sim \
  e^{|k|\sin^{-1}(Z)}
  \qquad
  \text{as } |k|\to\infty\, .
\end{equation}
Thus, the integral defining \(K_R\) does not converge to an ordinary pointwise function.
The reconstruction should instead be understood in a weaker sense
\cite{Morrison:2014jha}.\footnote{Alternatively, the authors of \cite{Hamilton:2006az,Hamilton:2006fh} obtain a well-defined smearing function by analytically continuing to complexified boundary coordinates. \label{footy1}}   
Namely, for a smooth, compactly supported
bulk test function \(f\) in the AdS-Rindler wedge $W_R$, one first defines the
induced boundary smearing
\begin{equation}
  F_f(T',X')
  \equiv
  \int_{W_R}\!\dd^3 \sfX_R \,\sqrt{-g}\,
  f(\sfX_R)\,K_R(\sfX_R|T',X') .
  \label{morrsmear}
\end{equation}
Morrison showed that the smeared boundary operators 
\(\hat{\mathcal O}_R[F_f]\) and their correlation functions are
well-defined whenever the boundary state is quasi-free  (Gaussian) and obeys the
microlocal spectrum condition. 
The reconstruction formula is
therefore an equality of operator-valued distributions, or,
equivalently, an equality after insertion into suitably smeared
correlation functions.

With these qualifications in mind,
we may now derive the diamond kernel $K_D$ from the Rindler kernel $K_R$.
Given a causal diamond $D_i$,
the defining property of $K_{D_i}$ is the reconstruction property
\begin{equation}\label{eq:HKLL-diamond}
  \hat{\Phi}(\sfX)
\ = \
  \int_{D_i} \dd t'\,\dd x'\,
  K_{D_i}(\sfX|\sfx')\,\hat{\mathcal{O}}(\sfx')\, ,
\end{equation}
with \(\sfX=(z,t,x)\) in $W[D_i]$ and \(\sfx'=(t',x')\) in $D_i$. 
The bulk field transforms as a scalar under the change to Rindler coordinates \eqref{eq:poincare-to-rindler}, and so we have \(\hat{\Phi}(\sfX)=\hat{\Phi}_R(\sfX_R)\),
where $\hat{\Phi}_R(\sfX_R)$ is given in eq.~\eqref{eq:HKLL-diamond-2}.
Hence, to derive $K_{D_i}$, it remains only to express the integral over $(T',X')$ as an integral over $\sfx'=(t',x')$,
where these are related by \eqref{eq:rindler-poincare}.
The measure and the primary operator transform nontrivially under this rewriting:
\begin{equation}
    \dd T'\,\dd X'
\ = \
    \Omega(\sfx')^2\,\dd t'\,\dd x'\qquad{\rm and}
    \qquad
    \hat{\mathcal{O}}_R(T',X')
\ = \
    \Omega(\sfx')^{-\Delta}\,
    \hat{\mathcal{O}}(\sfx')\, .
    \label{eq:mp-transform}
\end{equation}
Substituting eq.~\eqref{eq:mp-transform} into eq.~\eqref{eq:HKLL-diamond-2} and comparing the result with eq.~\eqref{eq:HKLL-diamond}, we find\footnote{Note that while $K_D$ has dimensions of $(\textrm{length})^{\Delta-d}$,
the calculations in this section implicitly require that $K_R$ is dimensionless.
This is consistent with the fact that $\Omega$ has units of inverse length from eq.~\eqref{eq:jacobian-omega}.
}
\begin{equation}\label{eq:K-diamond}
  K_{D_i}(\sfX|\sfx')
\ = \
  \Omega(\sfx')^{2-\Delta}\,
  K_R(\sfX_R|T',X').
\end{equation}
Thus the causal diamond kernel differs from the Rindler kernel by an overall prefactor set by the conformal factor \eqref{eq:jacobian-omega} raised to the power of $2-\Delta$.
For marginal boundary operators ($\Delta=2$),  the power vanishes and this factor disappears. For relevant operators ($\Delta<2$), this factor enhances the kernel near the edges of the diamond $D_i$, where $\Omega$ diverges. 
Meanwhile, for irrelevant operators ($\Delta>2$), it suppresses the kernel near the edges of $D_i$. While this conformal factor affects the boundary behaviour of the kernel, it does not change its support.

Finally, we comment that $K_{D_i}$ inherits the distributional pathologies of $K_R$ and hence must be interpreted in the weak sense of Morrison \cite{Morrison:2014jha}.
By contrast, in RQI applications, 
detector protocols are typically defined using ordinary smearing functions.
This motivates considering bulk operators smeared over a small spacetime region in the next section.

\subsection{Reconstructing a smeared operator}
\label{sec:smeared-hkll}

For the moment, let us return to a general number of spacetime dimensions $d$. So far, we have seen that a bulk field operator at a point can be reconstructed as
\begin{equation}\label{eq:HKLL-schematic-recap}
  \hat{\Phi}(\mathsf{X})\ = \ \int_A\!\dd^d \sfx'\;K_A(\mathsf{X}|\mathsf{x}')\,\hat{\mathcal{O}}(\mathsf{x}'),
\end{equation}
where $K_A(\mathsf{X}|\mathsf{x}')$ is an HKLL kernel with support in the boundary region $A$. In this section, we consider instead the reconstruction of field operators smeared over a region of the bulk spacetime,
\begin{equation}
    \hat{\Phi}_f(\mathsf{X}_0)\ = \ \int \dd^{d+1}\mathsf{X}\, \sqrt{|g|} \, f(\mathsf{X},\mathsf{X}_0) \hat{\Phi}(\mathsf{X})\,,
    \label{eq:smear-field}
\end{equation}
where $f(\mathsf{X},\mathsf{X}_0)$ is a function localized around the bulk point $\mathsf{X}=\mathsf{X}_0$, \eg a Gaussian centered at $\mathsf{X}_0$. 
Following \cite{Bousso:2012mh}, we  substitute eq.~\eqref{eq:HKLL-schematic-recap} in for $\hat{\Phi}(\mathsf{X})$ and exchange the order of integration to find
\begin{equation}
    \hat{\Phi}_f(\sfX_0)\ = \ \int_A\!\dd^d \sfx'\, \kfA(\sfX_0|\sfx')\, \hat{\mathcal{O}}(\sfx')
    \label{eq:smearing}
\end{equation}
where we define
\begin{equation}
    \kfA(\sfX_0|\sfx')
\ = \ 
    \int \dd^{d+1}\sfX\, \sqrt{|g|} \, f(\mathsf{X},\mathsf{X}_0)
    \;K_A(\sfX|\sfx')\,.
    \label{eq:K-F-def}
\end{equation}
Here, we have assumed that the support of the smearing function $f(\sfX,\sfX_0)$ lies entirely inside the causal wedge of the boundary region $A$.
The new kernel \eqref{eq:K-F-def} is a convolution which averages the original kernel $K_{A}(\sfX|\sfx')$ over bulk points $\sfX$, weighted by the profile $f(\sfX,\sfX_0)$, while the boundary point $\sfx'$ remains fixed.

As an aside, we comment that in certain cases, the bulk average in eq.~\eqref{eq:K-F-def} can be rewritten as a convolution in boundary coordinates. 
This requires several conditions which, although restrictive, are all  satisfied for Rindler reconstruction.
First, we require that moving the bulk point within some $n$-dimensional subspace of the boundary directions simply shifts the corresponding boundary dependence of the pointwise reconstruction kernel, \iee
\begin{equation}\label{eq:kernel-translation-property}
    K_A(\sfX_0+\sfy|\sfx')\ = \ K_A(\sfX_0|\sfx'-\sfy)\,,
\end{equation}
where $\sfy$ denotes a displacement in the $n$-dimensional subspace. 
Further, we require that the bulk measure $\sqrt{|g|}$ and the boundary region $A$ are invariant under these translations. Now suppose that the profile $f(\sfX,\sfX_0)$ smears only over bulk points with the form $\sfX_0+\sfy$, where again $\sfy$ lies in the $n$-dimensional subspace.
Schematically, 
suppressing the covariant normalization of the transverse delta distribution and the associated measure factors, 
we write $f(\sfX,\sfX_0)=\delta(\sfX_\perp-\sfX_{\perp,0})\,\tilde{f}(\sfX_\parallel-\sfX_{\parallel,0})$. Then substituting into eq.~\eqref{eq:K-F-def} yields
\begin{equation}
    \kfA(\sfX_0|\sfx')
\ = \ 
    \int\!\dd^{n}\sfy\,\sqrt{|g(\sfX_0)|}\,\tilde f(\sfy)\,
    K_A(\sfX_0|\sfx'-\sfy)\,.
    \label{eq:K-F-boundary-conv}
\end{equation}
Hence, we have rewritten eq.~\eqref{eq:K-F-def} using a convolution in boundary coordinates.
Further, eq.~\eqref{eq:K-F-boundary-conv} can be substituted into eq.~\eqref{eq:smearing}, allowing us to express the smeared bulk operator \eqref{eq:smear-field} in terms of smeared boundary operators
\begin{equation}
\hat{\Phi}_f(\sfX_0)\ = \ \int_A\!\dd^d \sfx'\, \sqrt{|g(\sfX_0)|} \,  K_A(\sfX_0|\sfx')\, \hat{\mathcal{O}}_{\tilde f}(\sfx')
    \label{eq:so-what}
\end{equation}
with
\begin{equation}
\hat{\mathcal{O}}_{\tilde f}(\sfx')\ = \ 
    \int\!\dd^{n}\sfy\,\tilde f(\sfy-\sfx')\,\hat{\mathcal{O}}(\sfy)\,.
\label{eq:so-what2}
\end{equation}
Again, while this construction may appear somewhat contrived, all of the required
conditions are satisfied if $\tilde f(\sfy)$ is chosen to be a Gaussian in the Rindler coordinates $T$ and $X$ in eq.~\eqref{eq:ads-rindler-metric}.
In this case, the Rindler kernel $K_R$ depends only on the differences $T-T'$ and $X-X'$, as shown in eq.~\eqref{eq:K-rindler}. 
It would be interesting to evaluate the usefulness of this boundary convolution perspective in future work.


We now use eqs.~\eqref{eq:smearing} and \eqref{eq:K-F-def} to improve the behaviour of the causal diamond kernel, \iee to smooth out its distributional structure. We return our focus to $d=2$, for which explicit expressions were obtained in the previous subsection. Naively, one might try to apply the convolution in eq.~\eqref{eq:K-F-def} directly to the diamond kernel \eqref{eq:K-diamond}, however, constructing an appropriate smearing function with support adapted to the causal wedge $W[D]$ is rather involved.  Instead, it is simpler to choose a Gaussian profile in the Rindler directions $T$ and $X$,\footnote{Implicitly, we are choosing the smearing function in eq.~\eqref{eq:smear-field} to have a delta function in  $Z$.}
\begin{align}
    K_{D}^{(f)}(z,t_0,x_0|t',x')\ \equiv& \ \Omega(t',x')^{2-\Delta}\,\int \dd T\dd X\, f_0(T-T_0)\,f_1(X-X_0)
    \label{eq:newK}\\
    &\qquad\times\ \   K_R(Z,T,X|T'(t',x'),X'(t',x'))\,.
    \nonumber
\end{align}
The support of this smearing is fully contained within the Rindler wedge, or equivalently, the causal wedge $W[D]$.
Here $(Z,T_0,X_0)$ are the Rindler coordinates corresponding to the bulk point $(z,t_0,x_0)$ under the change of coordinates \eqref{eq:poincare-to-rindler}, and we have introduced two functions
\begin{equation}
    f_{i}(Y)\ \equiv \ \frac{1}{\sqrt{2\pi}\,\sigma_{i}}\,e^{-{Y^2}/{2\sigma_{i}^2}}\,, \qquad i\ = \ 0,1\,
\end{equation}
with widths $\sigma_0$ and $\sigma_1$ in the $T$ and $X$ directions, respectively.
Substituting eq.~\eqref{eq:K-rindler} into the above, we find
\begin{align}
  K_{D}^{(f)}
&\ =\ \Omega^{2-\Delta} \! \int\!\frac{\dd k \, \dd \omega}{4\pi^2}\,
  e^{-\ii \omega(T_0- T')+ \ii k(X_0- X')} e^{-\frac{1}{2}(\sigma_0^2 \omega^2+\sigma_1^2 k^2)}\,V_{\omega k}^{R}(Z)\,. \label{eq:KF-3}
  \end{align}
  
Recall that the expression \eqref{eq:K-rindler} for the Rindler kernel  diverges by itself because $V_{\omega k}^{R}$ grows exponentially at large $|k|$ (eq.~\eqref{eq:V-rindler-asymptotic}). Hence, both the Rindler and diamond kernels must be interpreted in a weak sense.
By contrast, in eq.~\eqref{eq:KF-3}, the Gaussian suppression factor $e^{-\sigma_1^2 k^2/2}$ overrides this large-$|k|$ growth. Therefore, the Gaussian smearing in eq.~\eqref{eq:newK} turns both $\kfr$ and $\kfd$ into well-behaved functions. 

\begin{figure}[t]
    \centering   \includegraphics[width=0.9\linewidth]{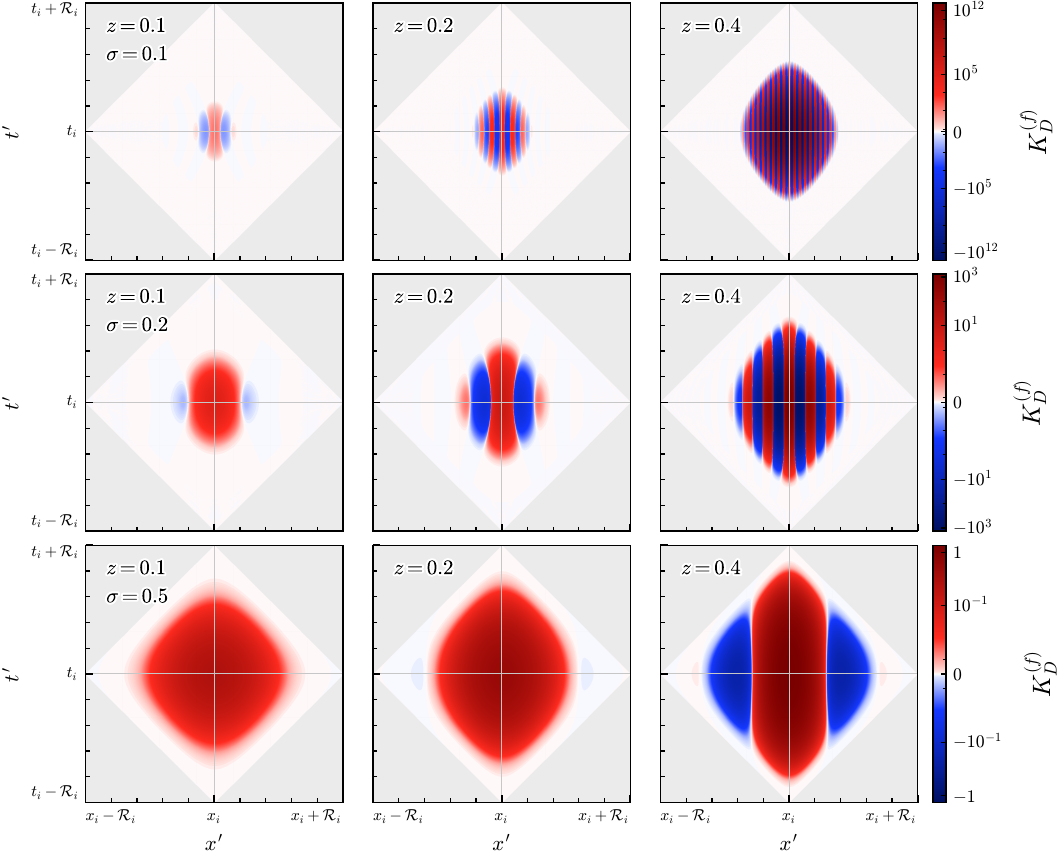}
    \caption{Numerical evaluation of the smeared kernel $\kfd$ in eq.~\eqref{eq:KF-3}, for $\Delta=1$ in $d=2$. Columns show $z=0.1$, $z=0.2$, and $z=0.4$ from left to right in units of $\mathcal{R}_i$. Rows show $\sigma_0=\sigma_1=0.1$, $0.2$, and $0.5$ from top to bottom. The plots are shown on a general boundary diamond $D_i$, with outer tick labels indicating $x'=x_i-\mR_i,x_i,x_i+\mR_i$ along the bottom and $t'=t_i-\mR_i,t_i,t_i+\mR_i$ along the left.
    Each row has its own colour scale, shared across the three panels in that row, showing the raw value of $\kfd$.
    }
    \label{fig:K_D_Hkll_full_numeric}
\end{figure}
    
Figure \ref{fig:K_D_Hkll_full_numeric} shows numerical evaluations of $\kfd$ from eq.~\eqref{eq:KF-3} at three bulk depths and for three smearing widths. In the figure and throughout the remainder of this subsection, we set $(t_0,x_0)=(t_i,x_i)$ (and hence, $T_0=0=X_0$), which aligns the bulk operator at the center of the causal diamond. In Rindler coordinates, the operator lies along the line $(Z,T=0,X=0)$, where $Z=2\mR_iz/(z^2+\mR_i^2)$ from eq.~\eqref{eq:poincare-to-rindler}.

The oscillations visible in figure~\ref{fig:K_D_Hkll_full_numeric}
can be understood from the large-\(|k|\) part of the kernel on the
horizontal time slice. 
We therefore set \(T_0=X_0=T'=0\) and define
\begin{equation}
    \theta\equiv\arcsin Z\,,
    \qquad \tau_\theta \equiv\tan\theta\,,
    \qquad 0<\theta<\frac{\pi}{2}\,.
\end{equation}
For fixed \(0<Z<1\), fixed \(\omega\), and fixed \(\Delta\), the WKB
analysis in appendix~\ref{app:WKB} gives
\begin{equation}
\begin{split}
    V_{\omega k}^{R}(Z)
    ={}& C_\Delta(\theta)\,|k|^{1/2-\Delta}
    \exp\!\left[
        |k|\theta
        -\frac{a_\Delta(\theta)+\omega^2\tau_\theta}{2|k|}
        +\mathcal{O}(|k|^{-2})
    \right],\\
    C_\Delta(\theta)
    \equiv{}&
    \frac{2^{\Delta-1}\Gamma(\Delta)}{\sqrt{2\pi}}
    \sqrt{\tan\theta}\,,\\
    a_\Delta(\theta)
    \equiv{}&
    \left((\Delta-1)^2-\frac14\right)\cot\theta
    +\frac14\tan\theta\,.
\end{split}
\label{eq:large-k-approx}
\end{equation}
In particular, eq.~\eqref{eq:V-rindler-asymptotic} retains only the
leading exponential envelope in this expression. 
The more detailed \eqref{eq:large-k-approx} is a fixed-\(\omega\), fixed-\(Z\)
statement; it is not uniform, for example, as \(Z\to0\). We give the
conditions needed to use it inside the frequency integral below.

At the order displayed in eq.~\eqref{eq:large-k-approx}, the
\(\omega\)-dependence on the slice \(T'=0=T_0\) is Gaussian. Thus
\begin{equation}
\begin{split}
    &\int_{-\infty}^{\infty}\!\dd\omega\,
    e^{-\frac12\sigma_0^2\omega^2}V_{\omega k}^{R}(Z)\;\;
\simeq\;\;
    \sqrt{2\pi}\,C_\Delta(\theta)\,
    \frac{|k|^{1/2-\Delta}}
    {\sqrt{\sigma_0^2+\tau_\theta/|k|}}
    \exp\!\left[
        |k|\theta-\frac{a_\Delta(\theta)}{2|k|}
    \right].
\end{split}
\label{eq:omega-integrated-large-k}
\end{equation}
Here and below, \(\simeq\) denotes the stated asymptotic approximation,
not an exact equality up to an unspecified normalization. At a fixed
positive \(k\), the frequency distribution in the (approximated) integrand is
centered at the origin and has standard deviation
\begin{equation}
    \omega_s=0\,,
    \qquad
    \delta\omega(k)
    =\left(\sigma_0^2+\frac{\tau_\theta}{k}\right)^{-1/2}\,.
    \label{eq:omega-width}
\end{equation}

Ignoring slowly-varying prefactors, the
envelope in $k$ for positive $k$ is proportional to
\(\exp(-\sigma_1^2k^2/2+\theta k)\). Its center and width are
\begin{equation}
    k_0=\frac{\theta}{\sigma_1^2}\,,
    \qquad
    \delta k=\frac{1}{\sigma_1}\,.
    \label{eq:k-envelope}
\end{equation}
The Fourier factor \(e^{-\ii kX'}\) moves the 
saddle into the complex plane:
\begin{equation}
    k_s=\frac{\theta-\ii X'}{\sigma_1^2}\,.
    \label{eq:k-sad}
\end{equation}
The prefactors of the exponential in eq.~\eqref{eq:omega-integrated-large-k} shift this
saddle only at subleading order, under the conditions stated next.

Let \(\delta\omega_0\equiv\delta\omega(k_0)\),
and assume the true saddle $k_s$ is close to $k_0$:
\begin{equation}
    |X'| \ll \theta\,.
    \label{truesad}
\end{equation}
Then, a convenient set of
sufficient conditions for a saddle approximation in $k$ is the following:
\begin{equation}
    \frac{\delta k}{k_0}\ll1\,,
    \qquad
    k_0\sin\theta\gg \sqrt{|(\Delta-1)^2-\frac{1}{4}|}\,,
    \qquad
    k_0\cos\theta\gg
    \sqrt{\delta\omega_0^{\,2}+\frac14}\,.
    \label{eq:large-k-validity}
\end{equation}
The first inequality separates the saddle from the endpoint at \(k=0\) which arises when we split the $k$ integral into $k>0$ and $k<0$ parts.
The second and third ensure that,
on the scales set by $k_0$ and $\delta \omega_0$,
the WKB approximation yielding \eqref{eq:large-k-approx} makes sense 
(see eq.~\eqref{eq:wkb-start}).
These conditions also serve to control the frequency width so that higher $\omega$-dependent terms in the WKB exponent are suppressed.

For equal widths, \(\sigma_0=\sigma_1=\sigma\), and fixed
\(\Delta=\mathcal{O}(1)\), a concise sufficient condition is
\begin{equation}
    \frac{\theta\sin\theta\cos\theta}{\sigma^2}\gg1\,.
    \label{eq:equal-width-global-validity}
\end{equation}
This reduces near the boundary ($\theta\ll1$) to
\begin{equation}
    \theta\gg\sigma\,.
    \label{eq:equal-width-validity}
\end{equation}
Consequently, at small-\(Z\), the saddle formula must be taken in
the regime
\begin{equation}
    \sigma\ll\theta\ll1\,.
    \label{eq:small-Z-overlap}
\end{equation}
To satisfy eq.~\eqref{truesad}, we can take $X'$ to be of order $\sigma$.
In words, our approximation works best  near the center of the diamond.

Up to an overall positive factor, the
positive-\(k\), high-momentum contribution is
\begin{equation}
\begin{split}
    J(X')\;\simeq\;
    \int_{k_{\rm as}}^\infty\!\dd k\,
    &\frac{k^{1-\Delta}}
    {\sqrt{\sigma_0^2 k+\tau_\theta}}
    \exp\!\left[-\frac{a_\Delta(\theta)}{2k}\right]\;
    \exp\!\left[-\frac12\sigma_1^2k^2
    +(\theta-\ii X')k\right],
\end{split}
\label{eq:KD-high-k-positive}
\end{equation}
where $k_{\rm as}$ is chosen in an overlap region where WKB is valid and yet \(k_{\rm as}\ll k_0\).
Within our domain of validity, the
leading saddle contribution is independent of the particular choice of
\(k_{\rm as}\).
The negative-\(k\) contribution is
the complex conjugate of eq.~\eqref{eq:KD-high-k-positive}, so 
\begin{equation}
    K_{D}^{(f)}\simeq
    \mathcal{N}_\Delta(Z,t',x')\,2\operatorname{Re}J(X')\,,
    \qquad \mathcal{N}_\Delta(Z,t',x')>0\,.
    \label{eq:KD-high-k-saddle}
\end{equation}
The positive factor \(\mathcal{N}_\Delta\) includes the Weyl factor
\(\Omega(t',x')^{2-\Delta}\) and does not affect the zeros.

We now evaluate the phase in the saddle approximation. 
The \(a_\Delta/k\) factor in eq.~\eqref{eq:KD-high-k-positive} produces only a subleading correction. 
Evaluating the remaining parts of the integrand at $k_s$ (eq.~\eqref{eq:k-sad}) gives
\begin{equation}
    K_{D}^{(f)}
    \simeq \mathcal{A}_\Delta(X')
    \cos\bigl(\Phi_\Delta(X')\bigr)\,,
    \qquad \mathcal{A}_\Delta(X')>0\,,
    \label{eq:KD-saddle-phase}
\end{equation}
where
\begin{equation}
\begin{split}
    \Phi_\Delta(X')={}&
    -\frac{\theta X'}{\sigma_1^2}
    +(\Delta-1)\arctan\!\left(\frac{X'}{\theta}\right)+\frac12\arctan\!\left(
    \frac{X'}{\theta+(\sigma_1^2/\sigma_0^2)\tan\theta}
    \right).
\end{split}
\label{eq:general-saddle-phase}
\end{equation}

\begin{figure}
    \centering
    \includegraphics[width=1\linewidth]{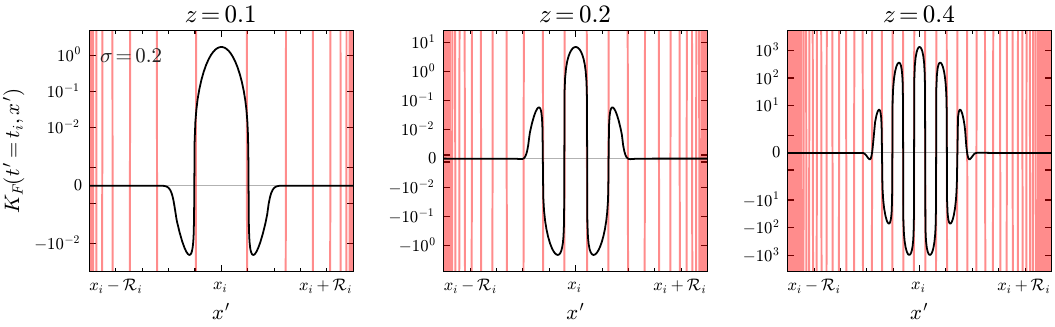}
    \caption{Eq.~\eqref{eq:cosine} gives a good fit to the phase behaviour of $K_D^{(f)}$.
    The zeros of eq.~\eqref{eq:cosine} are shown as vertical red lines.
    Meanwhile, the $K_D^{(f)}$ profile along the $t'=t_i$ slice is shown in black,
    and its zeros agree well with the locations of the red lines.
    Here we have set $\sigma=0.2$, but a similar matching holds for the other cases in figure \ref{fig:K_D_Hkll_full_numeric}.
    }
    \label{fig:slice}
\end{figure}

For the parameters used in figure~\ref{fig:K_D_Hkll_full_numeric},
\(\Delta=1\) and \(\sigma_0=\sigma_1=\sigma\), the phase reduces to
\begin{equation}
    K_{D,\mathrm{sadd}}^{(f)}
    \simeq \mathcal{A}(X')
    \cos\!\left[
        -\frac{\theta X'}{\sigma^2}
        +\frac12\arctan\!\left(
            \frac{X'}{\theta+\tan\theta}
        \right)
    \right]\,.
    \label{eq:cosine}
\end{equation}
The zeros predicted by eq.~\eqref{eq:cosine} can be compared directly
with the numerical kernel in figure~\ref{fig:slice}.

The leading full period
in Rindler distance is 
\begin{equation}
    \lambda_X
    =\frac{2\pi\sigma^2}{\theta}
    \left[1+\mathcal{O}\!\left(\frac{\sigma^2}{\theta^2}\right)\right].
    \label{eq:Rindler-wavelength}
\end{equation}
To go back to Poincar\'e coordinates, note that 
at \(t'=t_i\), the boundary coordinate transformation is
\begin{equation}
    X'=2\operatorname{arctanh}\!\left(
        \frac{x'-x_i}{\mathcal{R}_i}
    \right),
    \qquad
    \theta=\arcsin\!\left(
        \frac{2\mathcal{R}_i z}{\mathcal{R}_i^2+z^2}
    \right)
    =2\arctan\!\left(\frac{z}{\mathcal{R}_i}\right),
    \label{eq:central-slice-theta-X}
\end{equation}
where the final equality uses \(0<z<\mathcal{R}_i\), as appropriate
inside the central axis of the causal wedge. The leading phase in
eq.~\eqref{eq:cosine} gives the Poincar\'e-coordinate period
\begin{equation}
    \lambda_x(x')
    \simeq
    \frac{\pi\sigma^2\mathcal{R}_i}{\theta}
    \left(1-\left(\frac{x'-x_i}{\mathcal{R}_i}\right)^2\right).
    \label{eq:local-Poincare-wavelength}
\end{equation}
This period is not constant across the diamond; at the center \(x'=x_i\), it reads
\begin{equation}
    \lambda_x(x_i)
    \simeq
    \frac{\pi\sigma^2\mathcal{R}_i}
    {2\arctan(z/\mathcal{R}_i)}
    \simeq
    \frac{\pi\sigma^2\mathcal{R}_i^2}{2z}
    \qquad
    \left(\sigma\ll\frac{2z}{\mathcal{R}_i}\ll1\right).
    \label{eq:central-Poincare-wavelength}
\end{equation}
Adjacent zeros are separated by one half of this value.
Similar methods can also be used to estimate the amplitude  (and not just the phase).

\begin{figure}[t]
    \centering
    \includegraphics[width=0.95\linewidth]{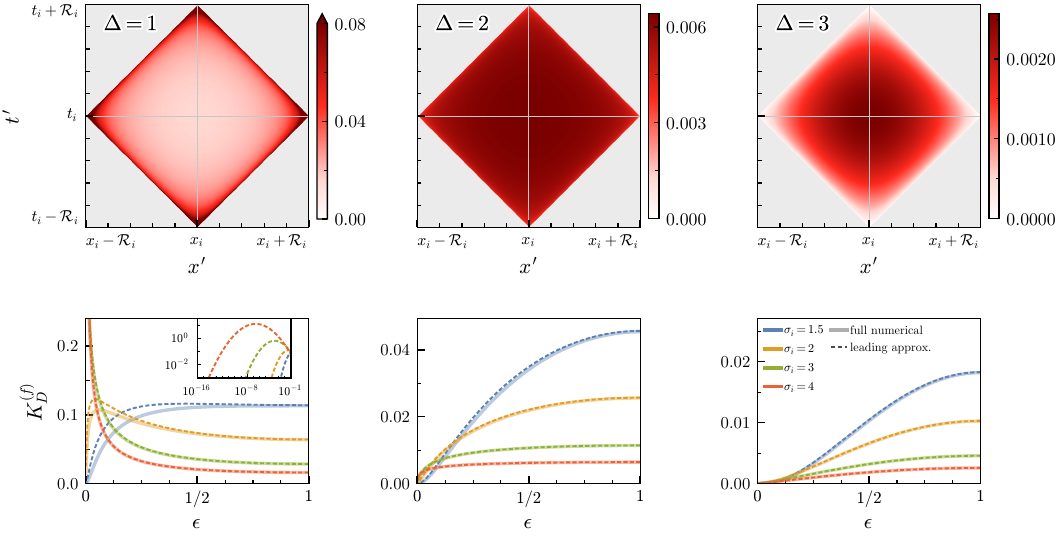}
    \caption{
    Large-smearing behaviour of $\kfd$.  From left to right,
$\Delta=1,2,3$, with $z/\mR_i=0.4$.  The top row shows the leading
approximation~\eqref{eq:KD_smeared_final} for
$\sigma_0=\sigma_1=4$.  The bottom row shows the corresponding
$t'=t_i$ cross-sections as functions of $\epsilon$, comparing
$\sigma_0=\sigma_1=1.5,2,3,4$.  Thick translucent curves show the
full numerical integral, while thin dashed curves show the
approximation.  The bottom-left inset resolves the decay near the
diamond edge.
    }
    \label{fig:K_D_Hkll_plot}
\end{figure}

For now, we instead turn to a different asymptotic parameter regime in which the integral can be evaluated in closed form at leading order. 
In particular, we will consider the limit of large smearings $\sigma_0,\sigma_1\gg 1$. In this case, the Gaussian factor $e^{-\frac{1}{2}(\sigma_0^2\omega^2+\sigma_1^2 k^2)}$ strongly localizes the momentum integrals near $\omega=k=0$, 
and the oscillatory behaviour exhibited in figures \ref{fig:K_D_Hkll_full_numeric} and \ref{fig:slice} goes away. 
To obtain the leading large-smearing behaviour, we only need the leading behaviour of the radial mode for small $\omega$ and $k$. In this regime, the leading approximation for the Rindler profiles \eqref{eq:V-rindler} becomes
\begin{equation}\label{eq:V-rindler-F0}
  V_{\omega k}^{R}(Z)
\ = \ Z^{\Delta}\,
 F_0+\mathcal{O}(\omega,k^2)\qquad{\rm with}\ \ 
  F_0={}_2F_1\left(
    \frac{\Delta}{2},
    \frac{\Delta}{2};
    \Delta;\,Z^2
  \right)\,.
\end{equation}
The linear term in $k$ is absent because the hypergeometric function is symmetric under $k\to-k$. 
With this, we can perform the $k$ and $\omega$ integrals in eq.~\eqref{eq:KF-3} and obtain, to leading order,
\begin{equation}
\begin{aligned}
\kfd(z,t_i,x_i|t',x')&\ \simeq \ \Omega(t',x')^{2-\Delta}\,Z^{\Delta}\,F_0\,
  \frac{e^{-T'^2/2\sigma_0^2}\; e^{-X'^2/2\sigma_1^2}}{2\pi\, \sigma_0\,\sigma_1}\,.
\end{aligned}
\label{eq:KD_smeared_final}
\end{equation}
This  is plotted in figure \ref{fig:K_D_Hkll_plot} for $\Delta = 1,2,3$, with $z=0.4$ and $\sigma_0=\sigma_1=4$. We see there is good agreement between this approximation and the full numerical integral in most of the instances shown there.  

Eq.~\eqref{eq:KD_smeared_final} is a fixed-coordinate
large-smearing approximation and is
not uniform as the boundary of the diamond is approached.
Nevertheless, as shown in figure \ref{fig:K_D_Hkll_plot},
it does capture the qualitative  behaviour of the kernel near the spatial tip of the diamond at $t'=t_i$ and $x'=\mR_i$ (again, we are momentarily setting $x_i$ to zero). 
Define a small dimensionless parameter $\epsilon$ with
\begin{equation}
\frac{x'}{\mR_i}=1-\epsilon\,,
\label{eq:epsilon-i}
\end{equation}
where $0<\epsilon\ll1$. From eq.~\eqref{eq:jacobian-omega}, we  have
\begin{equation}
    \Omega^2
    \ = \ \frac{4}{\mR_i^2\,\epsilon^2(2-\epsilon)^2}
    \ \simeq \ \frac1{\mR_i^2\,\epsilon^{2}}\,.
\end{equation}
Hence the conformal factor in the smeared kernel \eqref{eq:KD_smeared_final} yields
\begin{equation}
    \kfd\ \supset \ \Omega(t',x')^{2-\Delta}\ \sim \ \epsilon^{\Delta-2}\,.
    \label{eq:factor22}
\end{equation}
For $\Delta>2$, this factor causes the smeared kernel to approach zero near the boundary of the diamond, as is  shown in the bottom-right panel of figure~\ref{fig:K_D_Hkll_plot} with $\Delta=3$. 
For $\Delta=2$, this factor plays no role; from the bottom-center panel of figure~\ref{fig:K_D_Hkll_plot}, we see that there must be other factors which cause the smeared kernel to vanish near the boundary.
In contrast, for $\Delta<2$,  this factor diverges at the boundary, which produces the growth shown in the bottom-left panel as $\epsilon\to0$.
However, a careful examination reveals that the kernel still vanishes at the boundary of the diamond, again indicating that other factors must become important there.

Hence we turn our attention to the Gaussian factor $e^{-X'^2/2\sigma_1^2}$ in eq.~\eqref{eq:KD_smeared_final}. Using the boundary map \eqref{eq:rindler-poincare} with $T'=0$ and $x_i=0$, we find $\tfrac{x'}{\mR_i}=\tanh(X'/2)$ or alternatively
\begin{equation}
    X'\, =  \, \log\left(\frac{1+\frac{x'}{\mR_i}}{1-\frac{x'}{\mR_i}}\right)\,= \, \log\left(\frac{2-\epsilon}{\epsilon}\right)
    \, = \, -\log\frac{\epsilon}{2}+\mathcal{O}(\epsilon)\,.
\end{equation}
Therefore, near the corner, we have
\begin{equation}
    e^{-{X'^2}/{2\sigma_1^2}}
    \, = \, \exp\left[-\frac{(\log(\epsilon/2))^2}{2\sigma_1^2}
    +\mathcal{O}(\epsilon\,|\log\epsilon|)\right]\,\simeq
    \,  \Big(\frac{\epsilon}2 \Big)^{\frac{|\log(\epsilon/2)|}{2\sigma_1^2}}
    \,.\label{eq:factor33}
\end{equation}
The final form shows $\epsilon$ to a (positive) power, similar to the conformal factor in eq.~\eqref{eq:factor22}; however, in this case, the power diverges as $\epsilon\to0$. 
Therefore, this Gaussian factor will always dominate very close to the corner.
Hence, even if we have $\Delta<2$, $\kfd$ will vanish at the boundary of the causal diamond.

We can estimate the regime where the Gaussian suppression becomes important by comparing the powers in eqs.~\eqref{eq:factor22} and \eqref{eq:factor33}. That is, we compare
\begin{equation}
\frac{|\log(\epsilon/2)|}{2\sigma_1^2} \sim |\Delta-2| \qquad\implies\qquad
\epsilon \sim 2\,\exp[-2\sigma_1^2|\Delta-2|]\,.
\label{eq:compare22}
\end{equation}
For the two left panels of figure~\ref{fig:K_D_Hkll_plot} with $\Delta=1$ and $\sigma_i=4$, we find $\epsilon\sim 10^{-14}$.
For $\Delta=2$, the contribution  coming from the Weyl factor is constant,
and it is the Gaussian factor which forces the kernel to vanish at the boundary. 
Meanwhile, for $\Delta>2$,
the Weyl factor already suppresses the kernel near the boundary of the causal diamond.
The estimate in eq.~\eqref{eq:compare22} then indicates the regime where the suppression from the Gaussian factor begins to dominate and drives the kernel to zero much faster than the Weyl factor. In any event, this analysis shows that smearing the bulk fields and the diamond kernel (as in eq.~\eqref{eq:newK}) not only cures the distributional nature of $K_D$, but also renders the kernel finite at the boundaries of the diamond.

To close here, we note that we have included a high-resolution inset in the bottom-left panel of figure~\ref{fig:K_D_Hkll_plot}. This confirms that the growth of the smeared kernel $\kfd$ does indeed turn over and reach zero at the boundary of the diamond. 
In this regime with $\Delta<2$, the estimates in eqs.~\eqref{eq:factor22} and \eqref{eq:factor33} allow us to determine the turnover point by simply evaluating the derivative,
\begin{equation}
   \frac{d\ }{d\epsilon}\Big[ e^{(\Delta-2)\log \epsilon-\frac{1}{2\sigma_1^2}(\log \frac{\epsilon}{2})^2}\Big]=0
    \quad\implies\quad
    \epsilon_\mt{max}= 2\,e^{\sigma_1^2(\Delta-2)}\,.
    \label{eq:turnover}
\end{equation}
With $\Delta=1$ and $\sigma_1=4$ again, we find $\epsilon_\mt{max}= 2.25\times 10^{-7}$, which agrees quite well with the inset in figure~\ref{fig:K_D_Hkll_plot}. 
Note that this turnover estimate \eqref{eq:turnover} differs from the former in eq.~\eqref{eq:compare22} because the former was concerned with the equality of exponents rather than the saddle-point value.

\section{Detector setup with HKLL}
\label{sec:detector-setup}

In this section, we first briefly review Unruh--DeWitt detector models.
We then explain how smeared detectors in a holographic CFT can be used to represent pointlike detectors in the bulk using HKLL reconstruction.  Finally, we introduce two different detector protocols for entanglement harvesting in holography which will be analyzed in section~\ref{sec:harvesting-geometries}.
The initial detector and HKLL discussion applies in general boundary dimension $d$.
Beginning with our introduction of HKLL detector protocols, we specialize to a two-dimensional boundary.

\subsection{Unruh--DeWitt detectors}
\label{sec:pointlike-detector}

In its canonical form, an Unruh--DeWitt detector is a two-level system with energy gap $\Omega$, coupled along its worldline $\sfx(\tau)$ to an operator in a quantum field theory, such as the field amplitude. In the following, we take the detector to couple to a scalar operator $\hat{\Psi}$. We denote the ground and excited states of the detector Hilbert space $\mathcal{H}_{\sA}$ by $\ket{g}$ and $\ket{e}$, respectively, and denote the QFT Hilbert space by $\mathcal{H}_{\rm QFT}$. 

To study the evolution of the detector, or of a pair of detectors as in the next section, we work in the interaction picture. The free Hamiltonians of the detector(s) and the quantum field theory generate the evolution of operators on their respective Hilbert spaces. 
Meanwhile, the interaction Hamiltonian $\hat{H}_{\rm int}$ generates the evolution of states in the joint Hilbert space.

When the detector is treated as an idealized pointlike system, the detector-field coupling is supported directly on the detector worldline $\sfx(\tau)$, and the time-integrated interaction Hamiltonian is defined as
\begin{equation}
\int\!\dd \tau\;\hat{H}_{\rm int}(\tau)
  \ = \ \lambda \int\!\dd \tau\;\chi(\tau)\,
  \hat{\mu}(\tau;\Omega)\otimes
  \hat{\Psi}(\sfx(\tau))\,,
\label{eq:point}
\end{equation}
where $\lambda$ is the coupling constant,\footnote{
The dimensions of $\lambda$ are given by $\lambda \sim (\textrm{length})^{-a-1}$,
given $\hat{\Psi}\sim (\textrm{length})^{a}$.
For instance, with a conformal primary operator $\hat{\Psi}=\hat{\mathcal{O}}$, one has
$\lambda \sim (\textrm{length})^{\Delta-1}$.
Meanwhile, for a free scalar field with the normalization established in footnote \ref{foot:units},  one has
$\lambda \sim (\textrm{length})^{-1}$.
} $\chi(\tau)$ is the switching function, which controls the time dependence of the interaction (\iee when the detector coupling is turned on and for how long), and $\hat{\Psi}$ is the scalar operator in the quantum field theory introduced above. 
Meanwhile, $\hat{\mu}$ is the detector monopole operator,
which in the Schr\"odinger picture simply reads $\hat{\mu}=\hat{\sigma}_x=\hat{\sigma}^+ + \hat{\sigma}^-$, 
where $\hat{\sigma}^+$ raises the detector from $\ket{g}$ to $\ket{e}$, and $\hat{\sigma}^-$ lowers it from $\ket{e}$ to $\ket{g}$.
We are working in the interaction picture, 
so in eq.~\eqref{eq:point}
we instead have a time-dependent monopole operator
\begin{equation}
    \hat{\mu}(\tau;\Omega)
    \ = \ 
    e^{\ii\Omega \tau}\,\hat{\sigma}^+
    +
    e^{-\ii\Omega \tau}\,\hat{\sigma}^-\,.
    \label{eq:monopole-operator}
\end{equation}
The operator \eqref{eq:monopole-operator} is simply $\hat\sigma_x$, appropriately evolved in the interaction picture with respect to the detector's free Hamiltonian,
\begin{equation}
    \hat{H}_{\sA} = \frac{\Omega}{2}\,\hat{\sigma}_z\,,
    \label{eq:freeH-detect}
\end{equation}
consistently with \(\Omega\) being the energy gap between the two detector states.

One can describe a pointlike detector using 
eq.~\eqref{eq:point};
however, in a more realistic scenario,
the detector will have a finite size.
In this case, the detector-field coupling is supported over a spacetime region, typically a timelike tube centered on the detector's putative worldline.
The covariant object
is the spacetime-smeared interaction 
\cite{Schlicht:2003iy,Pozas-Kerstjens:2015gta,Martin-Martinez:2018gzb,
Martin-Martinez:2020pss,Perche:2023nde} 
\begin{equation}\label{eq:smeared-folded-1}
   \lambda \int_{\mathcal{M}}\!\dd V'\;
  \hat{\Lambda}(\sfx')\otimes\hat{\Psi}(\sfx')\,,
\end{equation}
where $\lambda$ is the coupling constant, $\dd V'$ is the invariant
volume element of the spacetime on which $\hat{\Psi}$ lives, and
$\hat{\Lambda}$ is an operator-valued function supported on the timelike tube of the detector, with length dimension $1-D$, where $D$ is the spacetime dimension.

In an appropriate detector-adapted coordinate system $\sfx'=(\tau,\vec{x}')$, 
$\hat{\Lambda}$ may be chosen to factorize as
\begin{equation}
    \hat{\Lambda}(\sfx')=\Lambda(\sfx')\,\hat{\mu}(\tau(\sfx'); \Omega)\,,
    \label{eq:lambda-factorizes}
\end{equation}
where $\Lambda(\sfx')$ is the scalar spacetime profile.
In the spirit of eq.~\eqref{eq:point},
we can then express eq.~\eqref{eq:smeared-folded-1} as a time-integrated interaction:
\begin{equation}\label{eq:smeared-folded}
  \int\!\dd \tau\;\hat{H}_{\rm int}(\tau)
  \ = \ \lambda \int_{\mathcal{M}}\!\dd V'\;
  \hat{\Lambda}(\sfx')\otimes\hat{\Psi}(\sfx')\,,
\end{equation}
with
\begin{equation}
  \hat{H}_{\rm int}(\tau)
  \ = \ 
  \lambda\,
  \hat{\mu}(\tau;\Omega)\otimes
  \!\int_{\Sigma_\tau}\!\dd \Sigma'\;
  \Lambda(\tau,\vec{x}')\,
  \hat{\Psi}(\tau,\vec{x}')\,,
\end{equation}
and where $\dd\Sigma'$ is the induced volume element on the constant-$\tau$ slice $\Sigma_{\tau}$.\footnote{For simplicity, we have assumed the metric has unit lapse in the detector-adapted coordinates, so that the spacetime volume element factorizes as $\dd V'=\dd \tau \,\dd\Sigma'$.}
When the detector is rigid, the spacetime smearing $\Lambda(\sfx')$ further factorizes into a switching function $\chi(\tau)$ that controls the time dependence of the interaction (as above), and a spatial smearing profile $f(\vec{x}')$ prescribed on each constant-$\tau$ slice $\Sigma_\tau$. 
Then, the above equation becomes
\begin{equation}\label{eq:smeared-hamiltonian}
  \hat{H}_{\rm int}(\tau)\ = \ \lambda\,\chi(\tau)\,
  \hat{\mu}(\tau; \Omega)\otimes \!\int_{\Sigma_\tau}\!\dd \Sigma'\;
f(\vec{x}')\,\hat{\Psi}\bigl(\tau,\vec{x}')\,.
\end{equation}
We recover the pointlike detector with a $\delta$-function smearing profile, \iee $f(\vec{x}') \ = \ \delta_{\Sigma_\tau}\bigl(\vec{x}'- \vec{x}(\tau)\bigr)$, where $\delta_{\Sigma_\tau}$ is the delta distribution on the spatial slice $\Sigma_\tau$. The interaction is then supported on the single timelike curve $\sfx(\tau)$, as in eq.~\eqref{eq:point}.

Although the spacetime integral in eq.~\eqref{eq:smeared-folded-1} is manifestly covariant, 
we note that detectors with nontrivial spatial extent can suffer from frame-dependence ambiguities at the quantum level \cite{Martin-Martinez:2020lul}.
This arises from the Dyson time-ordering $\mathcal{T}$ 
used to define the evolution operator:
\begin{equation}
    \hat{U}_I=
     \mathcal T
  \exp\left[
    -i \int_{\mathcal{M}}\!\dd V'\;\hat h_I(\sfx')
  \right]\,,
  \qquad
  \hat h_I(\sfx)
=
\lambda\,\Lambda(\sfx)\,
\hat{\mu}(\tau;\Omega)\otimes\hat{\Psi}(\sfx)\,.
\label{dyson}
\end{equation}
In particular, two interaction events
$\sfx=(\tau,\vec{x})$ and $\sfx'=(\tau',\vec{x}')$ within the detector's worldtube may be spacelike separated, in which case different admissible time-ordering prescriptions $\mathcal{T}$ and $\widetilde{\mathcal{T}}$ may order them differently. Such a reversal would be harmless if the interaction Hamiltonian density were microcausal. For the interaction density given above,
however, we have
\begin{align}
\bigl[\hat h_I(\sfx),\hat h_I(\sfx')\bigr]
&=
2\ii\lambda^2\Lambda(\sfx)\Lambda(\sfx')\,
\sin\!\bigl(\Omega(\tau-\tau')\bigr)\,
\hat{\sigma}_z\otimes
\hat{\Psi}(\sfx)\hat{\Psi}(\sfx')
\label{eq:time-order-mess}
\end{align}
at spacelike separation,
where we used eq.~\eqref{eq:monopole-operator} and  microcausality of the quantum field.
This commutator is generically nonzero, so changing the foliation which defines the time-ordering can change the  evolution operator $\hat{U}_I$. 
Pointlike detectors are free of this ambiguity:  the interaction events on a single timelike worldline are pairwise timelike separated, so every admissible foliation preserves their ordering.
More generally, 
failures of detector covariance are suppressed by the size of the detector \cite{Martin-Martinez:2020lul}.

Below, we will work with both pointlike detectors and detectors with spatial extent.
We will always take the switching function to be a Gaussian with switching timescale $\mathcal{T}$, \iee
\begin{equation}
    \chi(\tau; \mathcal{T})\ = \ e^{-{\tau^2}/{\mathcal{T}^2}}\,.
    \label{eq:chi-gaussian}
\end{equation}
Our detector model therefore has three tunable  parameters: the detector gap $\Omega$, the interaction time $\mathcal{T}$, and the coupling $\lambda$.
Note that the parameter $\mathcal{T}$ is unrelated to the time ordering operator $\mathcal{T}$ in eq.~\eqref{dyson}.

In the above equations we have taken care to use a general scalar operator $\hat{\Psi}$ acting on a general Hilbert space $\mathcal{H}_{\rm QFT}$. 
So, the above discussion applies equally well for describing detectors in anti-de Sitter spacetime, 
which couple to the amplitude of a bulk scalar field $\hat{\Psi}=\hat{\Phi}$, and for describing detectors in the dual CFT, which couple to a scalar conformal primary $\hat{\Psi}=\hat{\mathcal{O}}$. In the former case, $\mathcal{H}_{\rm QFT}=\mathcal{H}_{\rm bulk}$ is the Hilbert space of the free scalar field in the bulk, while in the latter case, $\mathcal{H}_{\rm QFT}=\mathcal{H}_{\rm bdy}$ is the associated Hilbert space of the dual generalized free field theory.
Of course, in the large $N$ limit, the holographic dictionary identifies $\mathcal{H}_{\rm bulk}$ with $\mathcal{H}_{\rm bdy}$ in the appropriate sense.

\subsection{HKLL kernels as bulk detectors}

We now use HKLL reconstruction to express a pointlike bulk detector in boundary variables. As explained with eq.~\eqref{eq:point}, a
pointlike detector following a bulk worldline $\sfX(\tau)$ has the following interaction Hamiltonian:
\begin{equation}
  \hat{H}_{\rm int}(\tau)
  \ = \ \lambda\;\chi(\tau; \mathcal{T})\;
  \hat{\mu}(\tau; \Omega)\otimes\hat{\Phi}(\sfX(\tau))\,,
  \label{eq:bulk-detector-coordinate}
\end{equation}
where we are coupling the detector to the amplitude of a (free) bulk scalar field, denoted $\hat{\Psi}=\hat{\Phi}$.
HKLL (eq.~\eqref{eq:HKLL-schematic}) rewrites the field operator at each point of the worldline as a boundary spacetime integral,
\begin{equation}
  \hat{\Phi}(\sfX(\tau))\ = \ \int_A\!\dd^d \mathsf{x}'\;K_A(\sfX(\tau)|\mathsf{x}')\,\hat{\mathcal{O}}(\mathsf{x}')\,,
    \label{eq:boundary-detector}
\end{equation}
where we have specified the particular boundary region $A$ associated with the smearing kernel $K_A$. 
So, at a fixed value of $\tau$, one can insert eq.~\eqref{eq:boundary-detector} into eq.~\eqref{eq:bulk-detector-coordinate},
to obtain
\begin{equation}
  \hat{H}_{\rm int}(\tau)
  \ = \ \lambda\,\chi(\tau; \mathcal{T})\,
  \hat{\mu}(\tau; \Omega)\otimes
  \int_A\!\dd^d \mathsf{x}'\;K_A(\sfX(\tau)|\mathsf{x}')\,\hat{\mathcal{O}}(\mathsf{x}')
\label{eq:boundary-int-hkll}
\end{equation}
as the expression for a pointlike bulk detector in terms of boundary variables. This boundary dual of a local bulk detector defines what we call an ``HKLL detector.''

Throughout this construction, we treat the detector qubit as an auxiliary degree of freedom external to the bulk quantum field theory, rather than as a composite excitation of the bulk fields. 
Accordingly, HKLL reconstruction is applied only to the bulk-field operator $\hat{\Phi}$; the detector Hilbert space and its monopole operator $\hat{\mu}$ are carried unchanged into the boundary description. 
Constructing a boundary representation of the microscopic degrees of freedom comprising the detector would be a separate problem beyond the scope of the present analysis.

Returning to eq.~\eqref{eq:boundary-int-hkll}, we observe that there is a fair degree of freedom implicit in this expression. Recall from eq.~\eqref{eq:differKAB} that the bulk field $\hat{\Phi}(\sfX(\tau))$ can in principle be reconstructed with many different smearing kernels associated with different boundary regions. 
Hence, to complete our prescription for the interaction Hamiltonian \eqref{eq:boundary-int-hkll}, we must specify the smearing kernel $K_A$ and the corresponding boundary region $A$ associated with each point $\sfX(\tau)$ along the detector's bulk worldline. Two simple examples would be as follows: if the entire worldline remains within the causal wedge of a certain boundary diamond $D$, we could use the same causal diamond kernel $K_D$ and boundary region $D$ for all values of $\tau$. Otherwise, we may want to use a scheme where for each point $\sfX(\tau)=(z(\tau),t(\tau),\vec{x}(\tau))$ along the bulk worldline, a new boundary causal diamond $D_\tau$ centered at $c_\tau=(t(\tau),\vec{x}(\tau))$ is chosen with some width $\mR_\tau > z(\tau)$.

At first glance, the interaction \eqref{eq:boundary-int-hkll} which characterizes an  HKLL detector resembles our previous expression for smeared  detectors in eq.~\eqref{eq:smeared-hamiltonian}.
However, there is a crucial difference owing to the fact that $K_A(\sfX(\tau)|\mathsf{x}')$ is integrated over a nontrivial space{\it time} region, whereas $f(\vec{x}')$ is supported only on the fixed time-slice $\Sigma_\tau$.
In the latter case, we are simply modelling a detector with finite spatial extent, while in the former case, the interpretation is subtle: the detector, at a fixed detector time $\tau$, is sensitive to the behaviour of the boundary quantum fields over a nontrivial spacetime subregion. Despite this coupling 
to a generally nonlocal boundary operator, 
in a slight abuse of terminology, we will still say that the HKLL detectors in eq.~\eqref{eq:boundary-int-hkll} are ``smeared boundary detectors.''

The spacetime smearing introduced here raises a subtle but important question concerning the time-ordering of the interaction in eq.~\eqref{eq:boundary-int-hkll} in harvesting calculations. Here we are considering a pointlike bulk detector \eqref{eq:bulk-detector-coordinate} following the worldline $\sfX(\tau)$, and hence its evolution is described by
\begin{equation}
  \hat{U}_I
  =
  \mathcal T_\tau
  \exp\left[
    -\ii\lambda\int \dd\tau\,
    \chi(\tau;\mathcal{T})\,\hat\mu(\tau;\Omega)
    \otimes\hat{\Phi}(\sfX(\tau))
  \right]\, ,
  \label{eq:HKLL-evolution1}
\end{equation}
where the time ordering is defined with respect to the detector's proper time. Using eq.~\eqref{eq:boundary-detector}, the boundary
representation of the same evolution operator becomes
\begin{equation}
  \hat{U}_I
  =
  \mathcal T_\tau
  \exp\left[
    -\ii\lambda\int \dd \tau\,
    \chi(\tau;\mathcal{T})\,\hat\mu(\tau;\Omega)
    \otimes
    \int_A\!\dd^d\mathsf{x}'\,
    K_A(\sfX(\tau)|\mathsf{x}')\,
    \hat{\mathcal O}(\mathsf{x}')
  \right]\, .
  \label{eq:HKLL-evolution2}
\end{equation}
Hence the boundary evolution operator inherits the time ordering of
the bulk detector. More precisely, 
one can Taylor expand the exponential in eq.~\eqref{eq:HKLL-evolution2}.
In each term,
$\mathcal T_\tau$ orders operators
according to the detector time $\tau$, which labels the bulk point $\sfX(\tau)$ appearing in the kernel, and therefore the entire HKLL-smeared boundary operator. In particular, after inserting the HKLL representation, one does not separately reorder the local CFT operators according to the boundary times $t'$ appearing inside the smearing integral. Doing so would define a different boundary protocol, rather than the boundary representation of the original bulk detector.\footnote{For protocols involving more than one detector, the corresponding
Dyson expansion must instead be ordered with respect to a common bulk
time function. In the examples considered below, this role is played
by the Poincar\'e time $t$, which is monotonic along both stationary
worldlines. }
We return to this issue and its relation to covariance in section~\ref{sec:discussion}. But we observe that it does not alter our explicit calculations in section~\ref{sec:harvesting-geometries}.

Next, note that if we integrate the HKLL interaction \eqref{eq:boundary-int-hkll} over $\tau$, we may recast the result in the covariant form given by eq.~\eqref{eq:smeared-folded},
namely
\begin{equation}\label{eq:smeared-folded-2}
  \int\!\dd \tau\;\hat{H}_{\rm int}(\tau)
  \ = \ \lambda^{\textrm{CFT}} \int_{\mathcal{M}}\!\dd V'\;
  \hat{\Lambda}(\sfx')\otimes\hat{\Psi}(\sfx')\,,
\end{equation}
with the field $\hat{\Psi}(\sfx')=\hat{\mathcal{O}}(\sfx')$ at a single spacetime point coupled to the following nonlocal operator-valued smearing on the detector Hilbert space:
\begin{equation}
    \hat{\Lambda}(\mathsf{x}')
    =
    L_0^{-\Delta}
    \int_{\mathcal{I}_{\sfx'}} \!\!d\tau\;\chi(\tau; \mathcal{T})\,
  \hat{\mu}(\tau; \Omega)
  \!\;
  K_A(\sfX(\tau)|\mathsf{x}')\,,
  \label{eq:operator-smearing}
\end{equation}
if we define $\lambda^{\textrm{CFT}}= L_0^{\Delta}\lambda$ for some length scale $L_0$.
Here, with the range $\mathcal{I}_{\sfx'}$, we are indicating that the integral runs over all values of $\tau$ for which the domain of the HKLL kernel contains $\sfx'$.
Further, recall from our discussion above that the form of the smearing kernel $K_A$ itself may also be a function of $\tau$.
In this presentation, the quantum field at a fixed spacetime point $\mathsf{x}'$ is coupled to detector operators evaluated over a nontrivial interval of detector time. While eqs.~\eqref{eq:smeared-folded-2} and \eqref{eq:operator-smearing} are equivalent to eq.~\eqref{eq:boundary-int-hkll} as expressions
for the interaction integrated over $\tau$, this rewriting makes the time-ordering prescription less transparent. To define the same evolution operator, the spacetime-smeared form must retain the ordering inherited from $\tau$, as specified in eq.~\eqref{eq:HKLL-evolution2}. 
We will therefore use the presentation in eq.~\eqref{eq:boundary-int-hkll}, in which the detector clock and the associated ordering prescription remain explicit.

Below, we consider different detector protocols involving HKLL detectors, restricting our attention to two spacetime dimensions on the boundary and hence a three-dimensional bulk.
Each HKLL detector that we consider is associated with a stationary bulk worldline:
\begin{equation}
    \sfX(\tau)\ = \ (z, z \tau/\ell, x)\,,
    \label{eq:diamond-detector}
\end{equation}
where the depth $z$ and transverse location $x$ are both fixed in time,
and we are using that the proper time $\tau$ along the worldline is related to the Poincar\'e coordinate time $t$ as $t=z \tau/\ell$.
To reconstruct this point on the boundary, we use the diamond kernel described in eqs.~\eqref{eq:HKLL-diamond} and \eqref{eq:K-diamond}. 
Specifically, following the second case described above, we specify a family of boundary diamonds $D_\tau$ each centered at $c_\tau=(z\tau/\ell,x)$ and with a fixed
radius $\mathcal{R}$, which is sufficiently large, \iee $\mR> z$.
Then in eq.~\eqref{eq:boundary-int-hkll}, we choose $K_{A}$ to be the diamond kernel $K_{D_\tau}$.
We emphasize that in this prescription, the boundary regions $A=D_\tau$ are moving in time. This is required for the worldline \eqref{eq:diamond-detector}, since for a fixed $\tau$, the diamond $D_{\tau}$ can only be used to reconstruct the bulk scalar along the segment given by $|t- z\tau/\ell|<\mathcal{R}-z$, while the full worldline runs over all $t$. Hence eq.~\eqref{eq:boundary-int-hkll} becomes
\begin{equation}
  \hat{H}_{\rm int}(\tau)
  \ = \ \lambda\,\chi(\tau; \mathcal{T})\,
  \hat{\mu}(\tau; \Omega)\otimes
  \int\dd^d \mathsf{x}'\;K_{D_\tau}(\sfX(\tau)|\mathsf{x}')\,\hat{\mathcal{O}}(\mathsf{x}')\,,
\label{eq:boundary-int-hkllv2}
\end{equation}
where the stationary bulk trajectory is specified by eq.~\eqref{eq:diamond-detector},
and we have left the integration region $D_\tau$ implicit above
for notational simplicity.

As a technical aside, note that 
in standard detector models, it is conventional for the detector profile to be characterized by a function rather than a distribution.
However, as discussed in section \ref{sec:ads-rindler}, the kernel $K_D$ is distributional.
This issue can be sidestepped by convolving $K_D$ with a Gaussian in Rindler coordinates, as in eq.~\eqref{eq:K-F-def} of section~\ref{sec:smeared-hkll}. 
In this case, the bulk dual of the HKLL detector will not be pointlike; instead, it will have a nontrivial smearing as in eq.~\eqref{eq:smear-field}.
In this paper, we will prefer to use the distributional form, since computations are simpler (and still well-defined, when carefully understood \cite{Morrison:2014jha}) with pointlike detectors in the bulk.

Finally, we comment on 
an important distinction between bulk and boundary detectors.
Note that a boundary detector following a fixed worldline has proper time $\tau$ equal to the Minkowski coordinate time $t$.
In contrast, a stationary detector in Poincar\'e coordinates \eqref{eq:metric-d2} has the proper time and coordinate time related by $\dd\tau= (\ell/z)\dd t$. 
In this case, the integrated interaction is easily rewritten in terms of Poincar\'e time with
\begin{equation}
\label{eq:integrated-h-again}
  \int\!\dd\tau\;\hat{H}_{\rm int}(\tau)
  \ = \ \int\!\dd t\;\frac{\partial\tau}{\partial t}\;
  \hat{H}_{\rm int}(\tau(t))\,,
\end{equation}
and accordingly, one can define an interaction Hamiltonian $\hat{H}_{\rm int}^{t}(t)$ with respect to  Poincar\'e time as 
\begin{equation}
  \hat{H}_{\rm int}^{t}(t)
  \ = \ \frac{\partial\tau}{\partial t}\;
  \hat{H}_{\rm int}(\tau(t))
  \ = \ \frac{\ell}{z}\,
  \hat{H}_{\rm int}(t \ell/z)\,,
  \label{eq:assume-const}
\end{equation}
or more explicitly, for a pointlike detector,\footnote{This relation can also be stated at the level of the full Schr\"odinger-picture Hamiltonian.
Let
\begin{equation}
     \hat{H}_D^{(\tau)}=\frac{1}{2}\Omega \hat{\sigma}_z
\end{equation}
denote the detector Hamiltonian generating proper-time translations,
and let $\hat{B}_z$ be a boundary representative of $\hat{\Phi}(z,0,x)$.
Then, with identity operators on the complementary factors understood, the boundary-time Hamiltonian in the Schr\"odinger picture is
\begin{equation}
\hat H_{\mathrm{bdy},S}(t)
=
\hat H_{\mathrm{CFT}}
+\frac{\ell}{z}\hat H_D^{(\tau)}
+\frac{\ell}{z}\lambda\chi\left(\frac{\ell}{z}t;\mathcal T\right)
\hat\sigma_x\otimes\hat B_z .
\end{equation}
In the interaction picture defined by
\(\hat H_{\mathrm{CFT}}+\frac{\ell}{z}\hat H_D^{(\tau)}\), this becomes
\begin{equation}
\hat H_{\mathrm{bdy},I}(t)
=
\frac{\ell}{z}\lambda\chi\left(t;\frac{z}{\ell}\mathcal T\right)\,
\hat\mu\left(t;\frac{\Omega}{z/\ell}\right)\otimes\hat B_z(t),
\end{equation}
where
$\hat B_z(t)=e^{i\hat H_{\mathrm{CFT}}t}\hat B_z
e^{-i\hat H_{\mathrm{CFT}}t}$
is a boundary representative of $\hat\Phi(z,t,x)$. This makes explicit that the rescaled parameters in eq.~\eqref{eq:H-int-00} arise from expressing the same detector dynamics in terms of boundary Poincar\'e time.
}
\begin{equation}
    \hat{H}_{\rm int}^t(t)\ = \ \lambda_{\rm lab}\; \chi(t; \mathcal{T}_{\rm lab})\; \hat{\mu}(t; \Omega_{\rm lab})\,\otimes\, \hat{\Phi}(\sfX(t))\,,
    \label{eq:H-int-0}
\end{equation}
with
\begin{equation}
 \lambda_{\rm lab}\equiv\frac{\lambda}{(z/\ell)}\,,\quad 
    \mathcal{T}_{\rm lab}\equiv \frac{z}{\ell}\,\mathcal{T}\,,\quad
    \Omega_{\rm lab}\equiv\frac{\Omega}{(z/\ell)}\,.
    \label{eq:H-int-00}
\end{equation}
That is, for the simple worldlines \eqref{eq:diamond-detector} considered here, the redshift factor between the boundary time and the proper time can be absorbed via constant multiplicative rescalings of the parameters defining the detector. So far, this is just a formal rewriting of the same bulk interaction \eqref{eq:integrated-h-again}. However, it also indicates that there is a simple relation between the parameters specifying a pointlike bulk detector following a simple stationary trajectory \eqref{eq:diamond-detector} and those specifying the dual HKLL detector \eqref{eq:boundary-int-hkll} in the boundary CFT, since the latter is naturally presented in Poincar\'e coordinates; this fact will be useful in our harvesting setups below.

Physical differences can arise depending on the treatment of lab (or ``boundary'') parameters versus those in the detector's rest frame.
In particular, if $\lambda$, $\mathcal{T}$ and $\Omega$ are treated as constants as we vary $z$, then with respect to Poincar\'e time, the corresponding parameters vary with appropriate factors of $z/\ell$, as indicated in eq.~\eqref{eq:H-int-0}.
On the other hand, if the lab parameters are held fixed while \(z\) is varied, the proper-frame parameters vary across the resulting family of stationary detectors, although they remain constant along each individual stationary worldline.
Both perspectives will be useful in our harvesting setups below.

Of course, for a pointlike boundary detector following a general trajectory $\sfx(\tau)=(t(\tau),\vec{x}(\tau))$, there will not be a simple relation between the detector's proper time $\tau$ and Minkowski time $t$. 
Hence, there will not be a simple relation between the detector parameters defined with respect to the two different time coordinates.
More explicitly, consider a pointlike bulk detector following a nontrivial trajectory
$\sfX(\tau)=(z(\tau), t(\tau),\vec{x}(\tau))$, \eg a bulk detector falling towards the Poincar\'e horizon or towards a black hole. Here again, we need only specify the coupling $\lambda$, the switching time $\cal T$ and gap $\Omega$, which are naturally constants in the detector's rest frame. However, there will not be a set of simple (constant) parameters which specify the dual HKLL detector. For example, the gap between the detector levels would appear to be a function of time (\iee $t$) from the boundary perspective. Of course, this should not be viewed as a problem, but rather as an interesting feature of our construction that deserves further investigation in future work.

\subsection{Two-detector setups: Examples 1 and 2} \label{sec:exam12}

We now introduce two entanglement harvesting protocols in a holographic CFT$_2$.
We  start from the total Hilbert space of  two detectors and the quantum field theory: $\mathcal{H}_{\sA}\otimes \mathcal{H}_{\sB} \otimes \mathcal{H}_{\rm QFT}$,
and an interaction of the general form
\begin{equation}
\begin{aligned}
\hat H_{\rm int}(t)
={}&
\lambda_\A\chi(t;\mathcal T_\A)\,
\hat\mu(t;\Omega_\A)\otimes\mathbf 1_\B
\otimes\hat\Psi_\A(t)
\\
&+
\lambda_\B\chi(t;\mathcal T_\B)\,
\mathbf 1_\A\otimes\hat\mu(t;\Omega_\B)
\otimes\hat\Psi_\B(t)\,.
\end{aligned}
\end{equation}
Going forward, we will typically use the following convention:
\begin{equation}
    \hat\Psi_I(t)=
\begin{cases}
\hat\Phi(\mathsf X_I(t)), & \text{bulk detector},\\
\hat{\mathcal O}(\mathsf x_I(t)), & \text{boundary detector}
\end{cases}
\end{equation}
and suppress identity factors $\mathbf 1$ in tensor products.
As our discussion above indicates,
many possible detector protocols can be constructed depending on the choice of detector smearings, as well as detector parameters: $\mathcal{T}_{\sA}$, $\Omega_{\sA}$, $\lambda_{\sA}$, $\mathcal{T}_{\sB}$, $\Omega_{\sB}$, and $\lambda_{\sB}$.

One natural protocol was already considered in \cite{Wurtz:2026ofi}, where harvesting was examined with two pointlike detectors in the boundary CFT. 
The total interaction Hamiltonian is given by\footnote{Note that Ref.~\cite{Wurtz:2026ofi} also introduced a time shift between the switching functions for the two detectors; it would be interesting to extend our present analysis by introducing similar time shifts.}
\begin{equation}\label{eq:Hint-prior}
  \hat{H}_{\rm int}(t)
  \ = \ \lambda\,\chi(t; \mathcal{T})\,
    \hat{\mu}(t; \Omega)\otimes\hat{\mathcal{O}}(\sfx_{\sA}(t))
  \ + \ \lambda\,\chi(t; \mathcal{T})\,
    \hat{\mu}(t; \Omega)\otimes \hat{\mathcal{O}}(\sfx_\sB(t))\,,
\end{equation}
where $\lambda\sim (\textrm{length})^{\Delta-1}$.
We will generalize this setup in our examples 1 and 2 below by replacing one or both of these pointlike detectors by a smeared HKLL detector. 

As our first example, we will consider the case where the first boundary detector A is an HKLL detector, while the second boundary detector B remains pointlike. Following the logic of eq.~\eqref{eq:boundary-int-hkllv2}, we define the total interaction as
\begin{equation}
\begin{aligned}
 \;\hat{H}_{\rm int}(t)
 \ = \
  &\frac{\lambda}{z_{\sA}^{\Delta}}\,\chi(t; \mathcal{T})\,
    \hat{\mu}(t; \Omega)\otimes\!\int\!\dd^{d}\sfx'\,
 K_{D_t}(\sfX_{\sA}(t)|\sfx')\,\hat{\mathcal{O}}\bigl(\sfx'\bigr)
  \\
  &\qquad+ \ \lambda\,\chi(t; \mathcal{T})\,
    \hat{\mu}(t; \Omega)\otimes \hat{\mathcal{O}}(\sfx_\sB(t))\,.
\end{aligned}
  \label{eq:mixed1}
\end{equation}
For detector A, we take the bulk worldline to be
\begin{equation}
    \sfX_\sA(t)\ = \ (z_{\sA},\,t,\,0)\,,
    \label{eq:lazy33}
\end{equation}
while for detector~B, we have $\sfx_\sB(t)=(t,L)$. That is, for a fixed $t$, the pointlike detector B sits at a fixed distance $L$ along the spatial direction in the boundary and on the same time slice as the center of the diamond $D_t$ defining the HKLL smearing for detector A. The above prescription is completed by specifying the size of the boundary causal diamonds $D_t$ with a fixed $\mR$.

The HKLL interaction specifying detector A in eq.~\eqref{eq:mixed1} differs from eq.~\eqref{eq:boundary-int-hkllv2} in two important ways. First, each of the factors in eq.~\eqref{eq:mixed1} is defined with respect to the boundary time rather than the proper time of the dual bulk detector, \eg both detectors have the same switching function $\chi(t; \mathcal{T})=e^{-t^2/\mathcal{T}^2}$.
The second difference is that the coupling in the first line of eq.~\eqref{eq:mixed1} depends on $z_\sA$. The reason behind this choice is the following: In the limit $z_\sA\to0$ which pushes detector A towards the asymptotic boundary, the boundary condition on the HKLL kernel (discussed below eq.~\eqref{eq:extrapolate}) gives
\begin{equation}
    K_{\Delta}(z_\sA,\sfx_\sA|\sfx') \to z_\sA^{\Delta} \delta^{(d)}(\sfx_\sA-\sfx')\,.
    \label{eq:gone-to-bdy}
\end{equation}
We see that in order to recover the interaction Hamiltonian \eqref{eq:Hint-prior} for two pointlike boundary detectors in this limit, the coupling must be allowed to depend on $z_{\sA}$ as $\lambda z_{\sA}^{-\Delta}$ to cancel the prefactor in eq.~\eqref{eq:gone-to-bdy}.

Crucially, HKLL reconstruction tells us that the interaction \eqref{eq:mixed1} can equivalently be written as
\begin{equation}
  \hat{H}_{\rm int}(t)
  \ = \ \frac{\lambda}{z_{\sA}^{\Delta}}\,\chi(t; \mathcal{T})\,
    \hat{\mu}(t; \Omega)\otimes\hat{\Phi}(\sfX_{\sA}(t)) + \ \lambda\,\chi(t; \mathcal{T})\,
    \hat{\mu}(t; \Omega)\otimes \hat{\mathcal{O}}(\sfx_\sB(t))\,.
  \label{eq:mixed2}
\end{equation}
Using eq.~\eqref{eq:H-int-0}, we see that detector A admits a dual bulk interpretation as a detector whose proper-time parameters are given by $\lambda\, z_{\sA}^{-\Delta}(z_\sA/\ell)$, $\frac{\mathcal{T}}{(z_\sA/\ell)}$ and $\Omega\,(\zA/\ell)$.
Since we are treating $\mathcal{T}$ and $\Omega$ as constants,
eq.~\eqref{eq:H-int-0} tells us that with respect to bulk proper time, detector A has a parametrically long switching time and a parametrically small gap as $z_\sA$ approaches the asymptotic boundary.
Again, these features are chosen in order to recover 
the ordinary interaction Hamiltonian \eqref{eq:Hint-prior} for two pointlike boundary detectors in the $z_{\sA}\to 0$ limit. 

Our example 2 will be the modification of eq.~\eqref{eq:Hint-prior} in which both of the pointlike boundary detectors are replaced by HKLL detectors. Specifically, we define the total interaction as
\begin{equation}
\begin{aligned}
  \hat{H}_{\rm int}(t)
&\ = 
  \frac{\lambda}{\tilde{z}_{\sA}}\,\chi(t; \tilde{z}_{\sA}\mathcal{T})\,
    \hat{\mu}(t; \Omega/\tilde{z}_{\sA})\otimes\!\int\!\dd^{d}\sfx'\;
 K_{D_{\A,t}}(\sfX_{\sA}(t)|\sfx')\,\hat{\mathcal{O}}\bigl(\sfx'\bigr)
  \ \\
  &\qquad+ \ \frac{\lambda}{\tilde{z}_{\sB}}\,\chi(t; \tilde{z}_{\sB}\mathcal{T})\,
    \hat{\mu}(t; \Omega/\tilde{z}_{\sB})\otimes \!\int\!\dd^{d}\sfx'\;
 K_{D_{\B,t}}(\sfX_{\sB}(t)|\sfx')\,\hat{\mathcal{O}}\bigl(\sfx'\bigr)
\,,
  \label{eq:bb1}
\end{aligned}
\end{equation}
where to avoid clutter we temporarily define (here and in eq.~\eqref{eq:bb2} below)
\begin{equation}
    \tilde{z}_\sA = \zA/\ell\qquad 
    \tilde{z}_\sB = \zB/\ell\,,
\end{equation}
and the two bulk worldlines are chosen as
\begin{equation}
  \sfX_\sA(t)\ = \ (\zA,\,t,\,0),
  \qquad
  \sfX_\sB(t)\ = \ (\zB,\,t,\,L)\,.
  \label{eq:bulkworldline}
\end{equation}
To avoid redundancy, we will take $z_\sB \ge z_\sA$ in our explicit calculations in the next section.

Notice that unlike the HKLL detector A in example 1 (see eq.~\eqref{eq:mixed1}), we have chosen the couplings, switching times and gap parameters with specific $z$-dependencies motivated by eq.~\eqref{eq:H-int-0}. 
To explain this more clearly, we use HKLL reconstruction to re-express eq.~\eqref{eq:bb1} as an interaction between pointlike bulk detectors:
\begin{equation}
  \hat{H}_{\rm int}(t)
\ = \ \frac{\lambda}{\tilde{z}_{\sA}}\,\chi(t; \tilde{z}_{\sA}\mathcal{T})\,
    \hat{\mu}(t; \Omega/\tilde{z}_{\sA})\otimes\,\hat{\Phi}(\sfX_{\sA}(t)) + \ \frac{\lambda}{\tilde{z}_{\sB}}\,\chi(t; \tilde{z}_{\sB}\mathcal{T})\,
    \hat{\mu}(t; \Omega/\tilde{z}_{\sB})\otimes \, \hat{\Phi}(\sfX_{\sB}(t))\,.
  \label{eq:bb2}
\end{equation}
Comparing to eq.~\eqref{eq:H-int-0}, we see that this interaction Hamiltonian describes a standard detector protocol in the bulk where  the coupling, switching time, and gap are constants $\lambda$, $\Omega$, $\mathcal{T}$ with respect to the detectors' proper times.

To summarize so far: In example 1, we fine-tune the choices of the bulk detector parameters in order to recover a nice near-boundary limit.
Meanwhile, in example 2, we fine-tune the choices of the boundary detector parameters in order to recover standard bulk detector physics.

In the next section, we will analyze entanglement harvesting in examples 1 and 2.
For this purpose, it is convenient to recast the Hamiltonians in \eqref{eq:mixed2} and \eqref{eq:bb2} in the following general form:
\begin{equation}
  \hat{H}_{\rm int}(t)
  \ = \ \lambda_{\sA}\,\chi(t; \mathcal{T}_\sA)\,
    \hat{\mu}(t; \Omega_\sA)\otimes\hat{\Psi}_\sA(t)
  \ + \ \lambda_{\sB}\,\chi(t; \mathcal{T}_\sB)\,
    \hat{\mu}(t; \Omega_\sB)\otimes \hat{\Psi}_\sB(t)\,,
  \label{eq:h-int-general}
\end{equation}
where $\hat{\Psi}_{A/B}$ stands in for either the bulk scalar $\hat{\Phi}(\sfX)$ or the dual boundary operator $\hat{\mathcal{O}}(\sfx)$, and the parameters $\lambda_{\sA}, \mathcal{T}_{\sB}, \Omega_{\sA}$, etc. should be chosen to comply with eqs.~\eqref{eq:mixed2} or \eqref{eq:bb2}. Specifically, for example 1, we have $\hat{\Psi}_{\sA} =\hat{\Phi}$, $\hat{\Psi}_{\sB} = \hat{\mathcal{O}}$, and
\begin{equation}
\begin{aligned}
    \mathcal{T}_{\sA}&\ = \ \mathcal{T}\,, \qquad\quad \  \mathcal{T}_{\sB}\ = \ \mathcal{T}\,,\\
    \qquad 
    \Omega_{\sA}&\ = \ \Omega\,, \qquad \quad \ \,\Omega_{\sB}\ = \ \Omega\,,
    \qquad
    \lambda \sim (\textrm{length})^{\Delta-1}\,.
    \\
    \lambda_{\sA}&\ = \ \lambda z_{\sA}^{-\Delta}\,,\qquad\lambda_{\sB}\ = \ \lambda\,
\end{aligned}
\label{eq:case1}
\end{equation}
Meanwhile for example 2, we have $\hat{\Psi}_{\sA}=\hat{\Psi}_{\sB}=\hat{\Phi}$ and
\begin{equation}
\begin{aligned}
    \mathcal{T}_{\sA}&\ = \ \frac{z_{\sA}}{\ell} \mathcal{T} \qquad \quad \quad\  \mathcal{T}_{\sB}\ = \ \frac{z_{\sB}}{\ell}\mathcal{T}\,,\\
    \qquad
    \Omega_{\sA}&\ = \ \frac{\Omega}{(z_{\sA}/\ell)} \qquad
    \quad\ \,\Omega_{\sB}\ = \ \frac{\Omega}{(z_{\sB}/\ell)} \,,
    \qquad \lambda \sim (\textrm{length})^{-1}\,.
    \\
    \lambda_{\sA}&\ = \ \frac{\lambda}{(z_{\sA} /\ell)} \qquad\quad\ \, \lambda_{\sB}\ = \ \frac{\lambda}{(z_{\sB}/\ell)}\,
\end{aligned}
\label{eq:case2}
\end{equation}

As we review in the following section, the key ingredients in entanglement harvesting calculations are correlators of the schematic form $\langle \hat{\Psi}_{\sA} \hat{\Psi}_{\sA}\rangle$, $\langle \hat{\Psi}_{\sB} \hat{\Psi}_{\sB}\rangle$, and $\langle \hat{\Psi}_{\sA} \hat{\Psi}_{\sB}\rangle$. Accordingly, we conclude this section by summarizing the relevant bulk-to-bulk, boundary-to-boundary, and bulk-to-boundary two-point functions for AdS$_{d+1}$ with  Poincar\'e coordinates, \eg see \cite{Freedman:1998tz,DHoker:2002nbb}. 

As described at the beginning of section \ref{sec:setup}, we are considering a free bulk scalar $\hat{\Phi}(\sfX)$ with mass $m$ and the corresponding dual conformal primary $\hat{\mathcal{O}}(\sfx)$ with conformal dimension $\Delta$ given by eq.~\eqref{eq:confdimen}.
The boundary-to-boundary Wightman function (\iee $\langle \hat{\mathcal{O}}(\sfx)\, \hat{\mathcal{O}}(\sfx') \rangle$) reads
\begin{equation}\label{eq:Gbdybdy}
  G^+_{\text{bdy-bdy}}(\sfx, \sfx')
\ = \ \frac{1}
    {\bigl[-(t-t'-\ie)^2 + (\vec{x}-\vec{x}')^2\bigr]^{\Delta}},
\end{equation}
where we are using the $\ie$ prescription appropriate for the positive-frequency Wightman function.
The bulk-to-boundary Wightman function (\iee $\langle \hat{\Phi}(\sfX)\, \hat{\mathcal{O}}(\sfx')\rangle$) reads
\begin{equation}\label{eq:Gbulkbdy}
  G^+_{\text{bulk-bdy}}\bigl(\sfX,\,\sfx'\bigr)
\ = \ \frac{z^{\Delta}}
    {\bigl[- (t-t'-\ie)^2 + (\vec{x}-\vec{x}')^2 +z^2\bigr]^{\Delta}}\,.
\end{equation}
Finally, the bulk-to-bulk Wightman function (\iee $\langle \hat{\Phi}(\sfX) \,\hat{\Phi}(\sfX')\rangle$) in $d+1$ bulk dimensions is given by
\begin{equation}\label{eq:Gbb}
\begin{aligned}
  G^+_{\text{bulk-bulk}}(\sfX, \sfX')
\ 
 &= \ 2^{-\Delta}\,\xi^{\Delta}\;
     {}_2F_1\Bigl(\frac{\Delta}{2},\;\frac{\Delta +1}{2};\;
     \Delta - \frac{d}{2} + 1;\;\xi^2\Bigr)\\
     &\overset{d=2}=
    \frac{\xi}{2\sqrt{1-\xi^2}}\,\bigg(\frac{\xi}{1+\sqrt{1-\xi^2}}\bigg)^{\Delta-1}
    \,,
\end{aligned}
\end{equation}
where we introduced the AdS-invariant chordal distance (with $\ie$ prescription) 
\begin{equation}
    \xi\ = \ \frac{2zz'}{z^2 + z'^2 + (\vec{x}-\vec{x}')^2 -
(t-t'-\ie)^2}\,.
\label{eq:invardistance}
\end{equation}
The boundary dimension $d$  appears only implicitly in the two propagators \eqref{eq:Gbdybdy} and \eqref{eq:Gbulkbdy}. In section \ref{sec:harvesting-geometries}, we will focus on $d=2$, where the spatial position vectors $\vec{x}$ reduce to a single coordinate $x$. Further, as we have shown in the second line of eq.~\eqref{eq:Gbb}, the expression for the bulk-to-bulk propagator is greatly simplified for $d=2$.
From this expression, we see the
bulk-to-bulk propagator has branch point singularities at $\xi=1$ and $\xi=-1$, which physically  correspond to the direct and image
bulk lightcones.\footnote{That is, for $\xi=1$, the two points, call them $A$ and $B$, are null separated, while for $\xi=-1$, the two points $A$ and $\bar{B}$ are null separated, where $\bar{B}$ is the reflection of $B$ via $\zB\to -\zB$.} Denoting $L\equiv |x-x'|$, these are located at
\begin{align}
  |t-t'|&\ = \ \sqrt{(\zA - \zB)^2 + L^2}
    &&(\xi\ = \ 1), \label{eq:v1}\\
  |t-t'|&\ = \ \sqrt{(\zA+\zB)^2+ L^2}
    &&(\xi\ = \ -1). \label{eq:v2}
\end{align}

The Feynman propagators can be assembled using the standard time-ordering prescription. Schematically, we have
\begin{equation}
    \GF{ }(\sfX,\sfX') = \theta(t-t')\, G^+(\sfX,\sfX') + \theta(t'-t)\, G^-(\sfX,\sfX')\,.
    \label{eq:Feyn}
\end{equation}
The relative normalizations of the correlators are fixed by the extrapolate dictionary. That is, the coefficient $2^{-\Delta}$ in eq.~\eqref{eq:Gbb} ensures that extrapolating either $\zA$ or $\zB$ to the boundary yields eq.~\eqref{eq:Gbulkbdy} and then extrapolating the second bulk point to the boundary while multiplying by $z^{-\Delta}$ reproduces eq.~\eqref{eq:Gbdybdy}.

\section{Harvesting results with HKLL detectors}
\label{sec:harvesting-geometries}

We now analyze entanglement harvesting with two detectors, for the two protocols described in section \ref{sec:exam12}. 
All results below are evaluated for $d=2$ boundary dimensions and hence three bulk spacetime dimensions. 

\subsection{The two-detector density matrix}
\label{sec:detector-bridge}

We may write the Hilbert space of the combined system (\iee the two detectors and the quantum field theory) as $\mathcal{H}_{\sA}\otimes \mathcal{H}_{\sB} \otimes \mathcal{H}_{\rm QFT}$. 
We begin the evolution with an initial state which corresponds to the vacuum in all three subsystems:
\begin{equation}
    \ket{g_\sA}\otimes\ket{g_\sB}\otimes \ket{0}\ \in \ \mathcal{H}_{\sA}\otimes\mathcal{H}_{\sB}\otimes \mathcal{H}_{\rm QFT}\,.
\end{equation}
To determine the entanglement which develops between two detectors resulting from the interaction given in  eq.~\eqref{eq:h-int-general}, we compute the late-time reduced density matrix of the two-detector system to leading order in the coupling constants $\lambda_{\sA}$ and $\lambda_{\sB}$.
We will use the detector basis
\begin{equation}
	\{|g_\sA g_\sB\rangle, |e_\sA g_\sB\rangle,
    |g_\sA e_\sB\rangle,
    |e_\sA e_\sB\rangle\}\,.
\end{equation}
Then, at leading order, the reduced density matrix takes the standard entanglement-harvesting form~\cite{Pozas-Kerstjens:2015gta},\footnote{To avoid $\lambda^3$ corrections, we implicitly assumed that the QFT is a free  theory (or generalized free theory) and that the state  is zero-mean Gaussian.}
\begin{equation}\label{eq:rho2}
  \hat{\rho}_{\sA\sB}\ = \ 
  \begin{pmatrix}        
    1-\cL_{\sA\sA}-\cL_{\sB\sB} & 0 & 0 & \cM^* \\
    0 & \cL_{\sA\sA} & \cL_{\sA\sB}^* & 0 \\
    0 & \cL_{\sA\sB} & \cL_{\sB\sB} & 0 \\
    \cM & 0 & 0 & 0
  \end{pmatrix}
  \ + \ O(\lambda^4),
\end{equation}
where the matrix elements $\cL_{\sA\sA}$, $\cL_{\sB\sB}$, $\cL_{\sA\sB}$, and $\cM$ are all of order $\lambda^2$. Physically, $\cL_{\sA\sA}$ and $\cL_{\sB\sB}$ are the local self-excitation probabilities (or the self-noise terms) induced by vacuum fluctuations within each detector's respective coupling region. 
That is, the diagonal part of  $\hat{\rho}_{\sA\sB}$ describes a mixed state with the following measurement probabilities: 
\begin{equation}
\begin{aligned}
    P(|g_{\sA} g_{\sB}\rangle)
    &=
    1-\mathcal{L}_{\sA\sA}-\mathcal{L}_{\sB\sB}
    \\
    P(|e_{\sA} g_{\sB}\rangle)
    &=\mathcal{L}_{\sA\sA}
    \\
    P(|g_{\sA} e_{\sB}\rangle)&=
    \mathcal{L}_{\sB\sB}\,.
\end{aligned}
\end{equation}
Meanwhile, the off-diagonal terms capture the effect of field correlations between A and B. The $\cM$ term mixes the $|e_{\sA}e_{\sB}\rangle$ state with detector ground state $|g_{\sA}g_{\sB}\rangle$ and is called the nonlocal coherence. Meanwhile, $\cL_{\sA\sB}$ is called the one-excitation coherence; it mixes $|e_{\sA}g_{\sB}\rangle$ and $|g_{\sA}e_{\sB}\rangle$. 

Given $\hat{\rho}_{\sA\sB}$, one can measure the entanglement between the detectors by computing the negativity $\cN(\hat{\rho}_{\sA\sB})$. This is defined by taking the partial transpose of $\hat{\rho}_{\sA\sB}$ with respect to one of the detectors, say detector B, and writing 
\begin{equation}
  \cN(\hat{\rho}_{\sA\sB})
\ = \ \frac{\|\hat{\rho}_{\sA\sB}^{\intercal_{\sB}}\|_1 - 1}{2}\,,
\label{eq:nega}
\end{equation}
where $\|\cdot\|_1$ denotes the trace norm.
Equivalently,
$\cN(\hat{\rho}_{\sA\sB})$ is the sum of the absolute values of the negative eigenvalues of $\hat{\rho}_{\sA\sB}^{\intercal_{\sB}}$ \cite{Vidal:2002zz}. 
For a density matrix of the form \eqref{eq:rho2}, the negativity reads (to leading order)
\begin{equation}\label{eq:neg}
  \cN\ = \ \max\left(0,\;
    \frac{\sqrt{(\cL_{\sA\sA}-\cL_{\sB\sB})^2 + 4|\cM|^2}
          - \cL_{\sA\sA} - \cL_{\sB\sB}}{2}
  \right)\,.
\end{equation}
Since the two detectors form a $2\otimes2$ system,  the joint state is entangled if and only if $\cN(\hat{\rho}_{\sA\sB})$ is nonzero \cite{Peres:1996dw,Horodecki:1996nc}. 
Hence given eq.~\eqref{eq:neg}, the threshold to have nonzero entanglement simply reads
\begin{equation}
    |\cM|\ > \ \sqrt{\cL_{\sA\sA}\,\cL_{\sB\sB}}\,.
    \label{eq:harvest-threshold}
\end{equation}
The generation of entanglement between the detectors therefore comes down to a competition between the nonlocal coherence $\mathcal{M}$ and the self-excitation probabilities.

At $O(\lambda^2)$, the elements entering this competition are explicitly given by
\begin{align}
  \cL_{\sA\sA}
&\ = \ \lambda_{\sA}^2 \int\!\dd t\,\dd t'\;
    \chi(t; \mathcal{T}_\sA)\,\chi(t'; \mathcal{T}_\sA)\;
    e^{-\ii\Omega_\sA(t-t')}\;
    \langle \hat{\Psi}_{\sA}(t)\hat{\Psi}_{\sA}(t')\rangle\,,
  \label{eq:LAA-general}\\[6pt]
  \cM
&\ = \ - \lambda_{\sA} \lambda_{\sB}\int\!\dd t\,\dd t'\;
    \chi(t; \mathcal{T}_\sA)\,\chi(t'; \mathcal{T}_\sB)\;
    e^{\ii(\Omega_\sA t + \Omega_\sB t')}\;
    \GF{ }(\hat{\Psi}_{\sA}(t),\hat{\Psi}_{\sB}(t')),
  \label{eq:M-general}
\end{align}
and with $\mathcal{L}_{\sB\sB}$ defined analogously to $\mathcal{L}_{\sA\sA}$.\footnote{For the coherence $\mathcal M$, the time ordering in $G^{(F)}$ is inherited from the common
evolution parameter $t$ appearing in the interaction Hamiltonian. Following the discussion below eq.~\eqref{eq:HKLL-evolution2}, when either operator is expanded using a boundary HKLL representation, this ordering is not replaced by chronological ordering of the boundary time which is integrated over in the corresponding kernels.} 
Assuming time-translation invariance, the correlators entering these equations depend only on $v\equiv t-t'$, in which case we can change variables and integrate over $u=t+t'$ to reduce the above expressions to one-dimensional integrals:
\begin{equation}
\begin{aligned}
  \frac{\cL_{\sA\sA}}{\lambda_{\sA}^2}
&\ = \ \sqrt{\frac{\pi}{2}}\,
  \mathcal{T}_{\sA}  \int\!\dd v\; e^{-\frac{v^2}{2\mathcal{T}_{\sA}^2}}\;
    e^{-\ii\Omega_\sA v}\;
    \langle \hat{\Psi}_{\sA}(v)\hat{\Psi}_{\sA}(0)\rangle\,,\\
\frac{\cM}{\lambda_{\sA} \lambda_{\sB}}
&\ = \, -
  \sqrt{\frac{\pi}{2a_+}}\,e^{ -\frac{\Omega_+^2}{2a_+}}
  \int\!\dd v\;
    e^{-\beta v^2+\ii \gamma v}\;
    \GF{}(\hat{\Psi}_{\sA}(v),\hat{\Psi}_{\sB}(0)),
\end{aligned}\label{eq:almost-done}
\end{equation}
where we have defined
\begin{eqnarray}
  a_{\pm}\ &=& \ \frac{1}{2}\left(\frac{1}{\mathcal{T}_\sA^2}
  \pm 
\frac{1}{\mathcal{T}_\sB^2}\right)\,,
  \qquad\quad\qquad
  \beta \  = \ \frac{1}{\mathcal{T}_\sA^2+\mathcal{T}_\sB^2}\,,
\nonumber\\
    \Omega_{\pm}\ &=& \ \frac{1}{2}\left(\Omega_\sA \pm\Omega_\sB\right)\,,
\qquad\quad\quad\quad\;\;\; \
    \gamma \ = \ \Omega_- - \frac{a_-}{a_+}\Omega_+\,.
    \label{eq:lots-of-parm}
\end{eqnarray}
These are the key formulas that we must evaluate in our two protocols.

\subsection{Example 1: One HKLL detector, one pointlike detector }
\label{sec:bulk-boundary-harvesting}

We consider the first setup described in section \ref{sec:exam12}, in which detector A is an HKLL detector, 
which admits a dual representation as a pointlike bulk detector, while
detector B is a pointlike detector on the boundary. The bulk worldline for
detector A is given by eq.~\eqref{eq:lazy33}, while detector B follows the boundary worldline $\sfx_\sB(t)=(t,L)$.
Plugging in the parameters for example 1 from eq.~\eqref{eq:case1}, we have
\begin{equation}
\begin{aligned}
    a_-\ = \ \Omega_-\ = \ \gamma\ = \ 0\qquad \Omega_+\ = \ \Omega\qquad a_+\ = \ 1/\mathcal{T}^2\ = \ 2\beta
\end{aligned}
\end{equation}
as well as $\hat{\Psi}_{\sA} =\hat{\Phi}$ and $\hat{\Psi}_{\sB} = \hat{\mathcal{O}}$. Hence the density matrix elements in eq.~\eqref{eq:almost-done} become
\begin{equation}
\label{eq:ogcornball}
\begin{aligned}
  \frac{\cL_{\sA\sA}}{\lambda^2}
&\ = \ \ z_{\sA}^{-2\Delta}\,\sqrt{\frac{\pi}{2}}\,
  \mathcal{T}  \int\!\dd v e^{-\frac{v^2}{2\mathcal{T}^2}}\;
    e^{-\ii\Omega v}\;
    G^+_{\text{bulk-bulk}}(v)\,,\\
    \frac{\cL_{\sB\sB}}{\lambda^2}
&\ = \ \ \ \qquad \sqrt{\frac{\pi}{2}}\,
  \mathcal{T}  \int\!\dd v e^{-\frac{v^2}{2\mathcal{T}^2}}\;
    e^{-\ii\Omega v}\;
    G^+_{\text{bdy-bdy}}(v)\,,\\
\frac{\cM}{\lambda^2 }
&\ =\, -
z_{\sA}^{-\Delta}\,
  \sqrt{\frac{\pi}{2}}\,\mathcal{T} e^{ -\frac{1}{2}\Omega^2\mathcal{T}^2}
  \int\!\dd v\;
    e^{-\frac{v^2}{2\mathcal{T}^2}}\;
    \GF{bulk-bdy}(v)\,,
\end{aligned}
\end{equation}
where $G^+_{\text{bulk-bulk}}(v)$ is given by the second line in eq.~\eqref{eq:Gbb} with $\xi(v)=2\zA^2/[2\zA^2-(v-\ie)^2]$, while $G^+_{\text{bdy-bdy}}(v)=(-(v-\ie)^2)^{-\Delta}$ from eq.~\eqref{eq:Gbdybdy}. Finally combining eqs.~\eqref{eq:Gbulkbdy} and \eqref{eq:Feyn}, we have
\begin{equation}\label{eq:GF2}
  \GF{bulk-bdy}(v)
\ = \ \frac{z_\sA^{\Delta}}
    {\bigl[z_{\sA}^2 + L^2 - v^2 +i \epsilon |v|\bigr]^{\Delta}}.
\end{equation}
By construction, near the boundary ($\zA\to 0$), the above matrix elements are finite and match the corresponding matrix elements in \cite{Wurtz:2026ofi}.\footnote{Note however that Ref.~\cite{Wurtz:2026ofi} prefers to work with  a dimensionless coupling $\bar{\lambda}=\lambda/\mathcal{T}^{\Delta-1}$.}
That is, in $\mathcal{M}$, the factor of $z_\sA^{-\Delta}$ in the coupling cancels the falloff of the propagator, with $\zA^{-\Delta}\,\GF{bulk-bdy}(v) \to 1/(L^2-v^2)^{\Delta}$, which is the boundary-to-boundary Feynman propagator at separation~$L$. The self-noise recovers its boundary value in a similar way, \iee by~\eqref{eq:Gbb}, we have $\zA^{-2\Delta}\,G^+_{\text{bulk-bulk}}(v) \to (-(v-\ie)^2)^{-\Delta}$, which is the Wightman function in the  CFT.

We now examine each of the matrix elements in turn.

\paragraph{Self-noise for A:}
Many of the integrals we will consider here, including the $\cL_{\sA\sA}$ integral, are of the general form
\begin{equation}
    I = \int_{\mathbb{R}}\dd v\, e^{-S(v)}\,G(v)\,,
    \label{eq:schem-int}
\end{equation}
where $S(v)$ is a quadratic function with saddle point $v_s\in \mathbb{C}$, 
while $G(v)$ is analytic away from (possibly singular) branch points $v_i \in \mathbb{R}$.
Since the branch points lie along the contour of integration,
an appropriate Feynman or Wightman $\ie$ prescription is essential for 
making sense of the integral \eqref{eq:schem-int}, as we will see in examples below.

\begin{figure}
    \centering
    \includegraphics[width=0.7\linewidth]{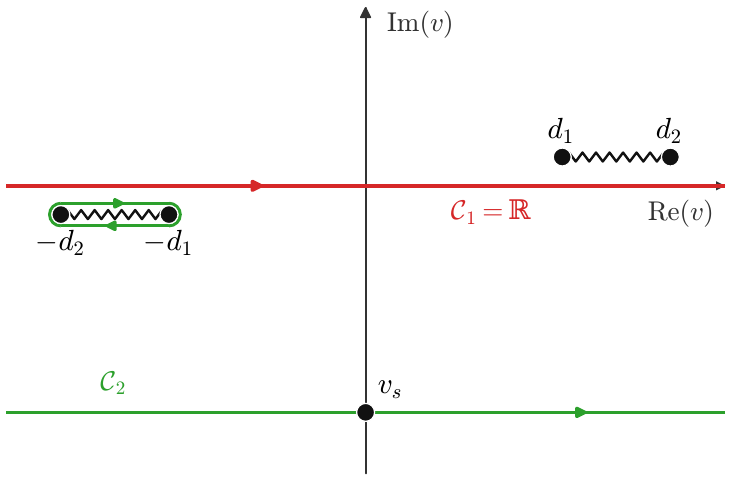}
    \caption{To evaluate an integral of the form \eqref{eq:schem-int}, we push the original integration contour $\mathcal{C}_1=\mathbb{R}$ to the steepest-descent contour through the saddle point $v_s$; however,
    the resulting contour $\mathcal{C}_2$ may contain branch cut contributions in addition to the steepest descent contour.}
    \label{fig:complex_v_plane}
\end{figure}

To approximate the integral,
one can deform the original real-axis contour to a steepest descent contour through $v_s$.
Then, in the saddle point approximation, the integral along the steepest-descent contour is approximated by $I_s$, with
\begin{equation}
    I_s=\sqrt{\frac{2\pi}{S''(v_s)}}\,e^{-S(v_s)}\,G(v_s)\,.
    \label{eq:saddle-contribution}
\end{equation}
In order to deform the original contour to the steepest-descent contour, 
it may be forced to wrap branch cuts in $G(v)$ --- see the example shown in figure \ref{fig:complex_v_plane}. 
In this case, the branch cut contribution
\begin{equation}
    I_{\rm br}
    =
    \int_{\shortstack{$\scriptstyle\mathrm{branch}$\\$\scriptstyle\mathrm{cut}$}}\dd v\, e^{-S(v)}\,\operatorname{Disc}G(v)\,,
    \label{eq:disc}
\end{equation}
may compete with the contribution in eq.~\eqref{eq:saddle-contribution}.

For the integrals considered below,
we will assume that $G(v)$ is slowly varying near $v_s$
whenever $v_s$ is far (in units of the Gaussian width) from the singularities of $G(v)$.\footnote{We will not take care to justify this assumption. However, the resulting predictions give a good fit to numerical results.
}
This assumption controls the local saddle approximation to the steepest-descent contribution. It does not, however, guarantee that $I_s$ dominates the full integral. 
Neglecting the branch contribution additionally requires, roughly,  $\textrm{Re}(S(v_i)-S(v_s))\gg1$.

For the $\mathcal{L}_{\sA\sA}$ integral in eq.~\eqref{eq:ogcornball}, the $e^{-S(v)}$ factor is a Gaussian of width $\mathcal{T}$.
Its saddle point lies at
\begin{equation}
    v_s = - i \mathcal{T}^2 \Omega\,,
\end{equation}
and the associated steepest descent contour is $\mathbb{R}- i \mathcal{T}^2 \Omega$.
Meanwhile, the $G(v)$ factor has branch point singularities at
\begin{equation}
    v_1= \ie \qquad
    v_2 = 2z_{\sA}+\ie
    \qquad
    v_3 =- 2z_{\sA}+\ie\,.
    \label{eq:humpty}
\end{equation}
These branch points arise from the lightcone singularities of the bulk-to-bulk propagator.
As indicated, the Wightman  $\ie$ prescription  pushes the branch cut slightly above the real axis 
meaning  we can freely deform the original contour to the steepest descent contour.
In the regime $|v_s|\gg \mathcal{T}$,
or equivalently $\Omega \mathcal{T}\gg 1$,
the singularities of $G(v)$ are far from the saddle point, and indeed, the saddle approximation agrees well with the numerical results shown in figure~\ref{fig:boundary-recovery} below, where we have set $\Omega\mathcal{T}=10$.

The full form of $I_{s}$ is complicated, so let us simply present $I_{s}$ in extreme regimes of $z_{\sA}$.
When $z_{\sA}$ is small compared to $|v_s|$, the saddle  contribution \eqref{eq:saddle-contribution} becomes
\begin{equation}
  \frac{\cL_{\sA\sA}}{\lambda^2}
\ = 
\frac{\pi \mathcal{T}^{\,2-2\Delta}}{(\mathcal{T}\Omega)^{2\Delta}} e^{-\frac{1}{2}\mathcal{T}^2\Omega^2}\,,
\label{eq:self-noise-bdy-ref}
\end{equation}
up to corrections of order $z_{\sA}^2/|v_s|^2$.
This says in particular that $\mathcal{L}_{\sA\sA}$ is approximately flat near $z_{\sA}=0$.
Meanwhile, when $z_{\sA}$ is large compared to  $|v_s|$, the saddle contribution reads
\begin{equation}
     \frac{\cL_{\sA\sA}}{\lambda^2}= 
     \frac{\pi z_{\sA}^{1-2\Delta}}{2\Omega}
  e^{-\frac{1}{2}\Omega^2 \mathcal{T}^2}\,,
  \label{eq:LAA-estimate}
\end{equation}
with relative corrections of order $|v_s|/z_{\sA}$.
Hence, the self-noise decays as $z_{\sA}^{1-2\Delta}$ at large $z_{\sA}$.
Again, this behaviour is indeed reflected in figure~\ref{fig:boundary-recovery} below.

Note that by definition (eq.~\eqref{eq:rho2}), the self-noise matrix elements are always real.
This reality property is manifest in eqs.~\eqref{eq:almost-done} and \eqref{eq:ogcornball}, since complex conjugation of $\mathcal{L}_{\sA\sA}$ (or $\mathcal{L}_{\sB\sB}$) 
can be absorbed into a redefinition $v\to-v$ of the integration variable.
The same is not true of $\mathcal{M}$, as we will see below.

\paragraph{Self-noise for B:}
The $\mathcal{L}_{\sB \sB}$ integral in eq.~\eqref{eq:ogcornball} is also of the schematic form \eqref{eq:schem-int}, with Gaussian width $\mathcal{T}$ and saddle point $v_s = -i \mathcal{T}^2 \Omega$, 
except that it has only a single branch-point singularity at $v_*=\ie$.
Applying the saddle approximation in the regime $\Omega \mathcal{T}\gg 1$, where $v_*$ lies well outside the Gaussian's primary support region,
simply reproduces eq.~\eqref{eq:self-noise-bdy-ref}.

In fact, the $\mathcal{L}_{\sB\sB}$ integral admits a closed form in terms of the confluent hypergeometric function \cite{Wurtz:2026ofi}
\begin{equation}\label{eq:LBB-closedform}
  \frac{\cL_{\sB\sB}}{\lambda^2}
\ = \ \frac{\pi^{3/2}\,\mathcal{T}^{\,2-2\Delta}}{2^\Delta}
  \left[
    \frac{{}_1F_1\left(\frac{1}{2}-\Delta;\,\frac{1}{2};\,-\frac{\mathcal{T}^2\Omega^2}{2}\right)}
         {\Gamma\left(\frac{1}{2}+\Delta\right)}
    \;-\;\frac{\sqrt{2}\,\mathcal{T}\Omega}{\Gamma(\Delta)}\,
    {}_1F_1\left(1-\Delta;\,\frac{3}{2};\,-\frac{\mathcal{T}^2\Omega^2}{2}\right)
  \right],
\end{equation}
which is valid for any $\Delta\in\mathbb{R}_+$ and compatible with the approximation \eqref{eq:self-noise-bdy-ref} in the regime $\Omega \mathcal{T}\gg 1$.

\paragraph{Nonlocal coherence:}
The $\mathcal{M}$ integral is also of the schematic form \eqref{eq:schem-int}, with
\begin{equation}
    v_s=0\,, \qquad 
    v_1=\sqrt{z_{\sA}^2 + L^2}+\ie\,,
    \qquad\textrm{and}\qquad
    v_2=-\sqrt{z_{\sA}^2 + L^2}-\ie\,,
\end{equation}
where the singularities at $v_1$ and $v_2$ come from lightcone singularities in the Feynman bulk-to-boundary propagator \eqref{eq:GF2}.\footnote{
To derive the $\ie$'s,
recall 
$\GF{ }(v) = \theta(v)\, G(v-\ie) + \theta(-v)\, G(v+\ie)$.}
In this case, the original contour already runs along the steepest descent contour through $v_s$.
Moreover, the singularities of $G(v)$, and both branch cuts, are far from $v_s$ (in units of the Gaussian width $\mathcal{T}$) in the regime
\begin{equation}\label{eq:v0-sigma-I}
  \sqrt{\zA^2 + L^2}\gg \mathcal{T}\,.
\end{equation}
Hence, we expect the integral to be well approximated by the saddle approximation in this regime.
There are two ways for the above inequality to be satisfied: either $z_{\sA}$ is large, or $L$ is large (or both).
In Figure~\ref{fig:boundary-recovery} below, we have $L \gg \mathcal{T}$ (specifically, $L/\mathcal{T}=5$) and the saddle approximation,
\begin{equation}\label{eq:M-saddle-I}
 \begin{aligned}
- \frac{\cM}{\lambda^2 }
&\ = \ 
  \pi \mathcal{T}^2 e^{ -\frac{1}{2}\Omega^2\mathcal{T}^2}
    \frac{1}
    {\bigl[z_{\sA}^2 + L^2\bigr]^{\Delta}}\,,
\end{aligned}
\end{equation}
is indeed a good fit to the numerical results.
In particular, the numerical values show $\cM$ has a negligible imaginary part, as required for the real function \eqref{eq:M-saddle-I} to be a good approximation.
$\mathcal{M}$ is roughly constant for $z_\A\ll L$, 
with the constant value matching the results of Ref.~\cite{Wurtz:2026ofi},
and monotonically decreases at large $\zA$, decaying as $\zA^{-2\Delta}$.

\begin{figure}[ht]
  \centering
  \includegraphics{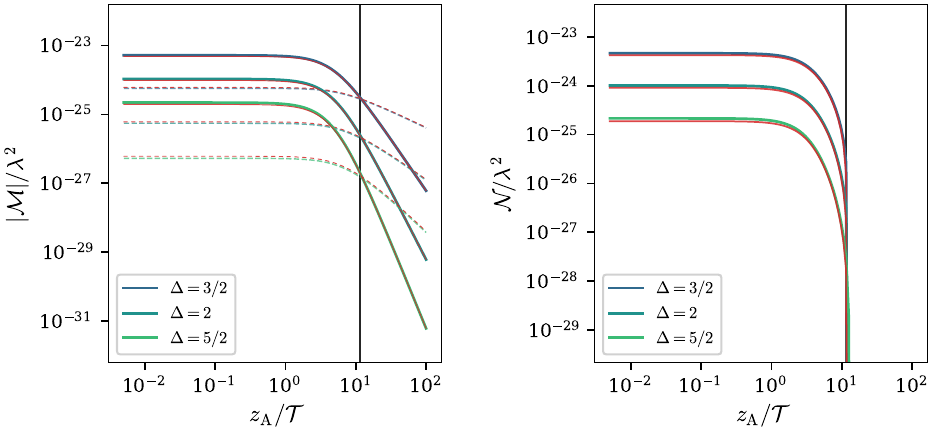}
  \caption{
  Numerical harvesting results for the single-HKLL-detector protocol with $\Omega=10$ and $L=5$, in units where $\mathcal{T}=1$.
  In each plot, we consider $\Delta = 1.5$,  $2$, and $2.5$.
  \emph{Left:} The nonlocal coherence $|\cM|/\lambda^2$ (solid) and the harvesting threshold
    $\sqrt{\cL_{\sA\sA}\cL_{\sB\sB}}/\lambda^2$ (dashed).
    Red  curves show the corresponding saddle-point approximations derived in this section and we see that they are a good fit to the numerical results. 
    In the limit $\zA \to 0$, the matrix elements plateau to the boundary-to-boundary values found in \cite{Wurtz:2026ofi}, while at large $z_\A$ we find that they decay rapidly in accord with eqs.~\eqref{eq:LAA-estimate} and \eqref{eq:M-saddle-I}.
    \emph{Right:} The negativity~$\cN/\lambda^2$ vs.\ $\zA$.
    Harvesting persists from the boundary down to $\zA/\mathcal{T} \approx 12$--$13$ in the numerical curves.
    The vertical line marks a heuristic estimate $z_{\sA}/\mathcal{T}=2^{\frac{1}{2\Delta+1}}\,\Omega\,\mathcal{T}$ for the depth at which the negativity vanishes, evaluated for $\Delta=2$; see the end of the subsection.}
  \label{fig:boundary-recovery}
\end{figure}

\paragraph{Numerical results}
Figure~\ref{fig:boundary-recovery} shows the numerical evaluation of the nonlocal coherence $\mathcal{M}$, the harvesting threshold $\sqrt{\cL_{\sA\sA}\cL_{\sB\sB}}$ and the negativity $\mathcal{N}$ for $\Delta=3/2$, 2 and $5/2$, as well as the various saddle point approximations derived above.\footnote{
Let us elaborate on our numerical methods in figure \ref{fig:boundary-recovery}
Firstly, rather than evaluating the singular integrands at a finite $i\epsilon$ and extrapolating as $\epsilon\to0^+$, we deformed the integration contours appropriately into the complex $v$ plane, 
and varied the contour displacements  to verify that the results were deformation-independent.
Secondly, we truncated Gaussian tails of the integrals conservatively, 
with the integration domain for \({\mathcal L}_{\sA\sA}\) extending at least \(12\mathcal T\) from the origin and, for \(\mathcal M\), at least 
\(6\mathcal T\) beyond the bulk lightcone singularities.
Thirdly, 
the integrals for \(\mathcal M\) and \(\mathcal L_{AA}\) were evaluated with adaptive numerical integration, which automatically refined the integration grid in difficult regions. Intermediate calculations retained 50 and 60 decimal digits, respectively, to control errors from oscillatory behaviour.
}
In practice, when evaluating the matrix elements numerically, we set $\mathcal{T}=1$ and hence say that all dimensionful quantities are expressed in units of $\mathcal{T}$.
We find that all of the matrix elements are flat near the boundary, \iee for $z_\sA\ll L$ and $z_\sA\ll\mathcal{T}^2\Omega$. 
As we go into the bulk, 
the negativity vanishes near $\zA \approx 12$--$13\mathcal{T}$, at the depth where $|\cM|$ intersects $\sqrt{\cL_{\sA\sA}\cL_{\sB\sB}}$.
The value of $z_\sA$ where this occurs can be heuristically estimated from the saddle approximation discussed above eq.~\eqref{eq:self-noise-bdy-ref} and that in eq.~\eqref{eq:M-saddle-I}.
The closed-form expression is complicated, but simplifies in the large $\zA$ limit  ($\zA\gg L$, $\zA \gg \mathcal{T}^2 \Omega$) to $z_{\sA}=2^{\frac{1}{2\Delta+1}}\,\Omega\,\mathcal{T}^2$.
We indicate the $\Delta=2$ value of this expression with a vertical black line in figure \ref{fig:boundary-recovery}.

\subsection{Example 2: Two HKLL detectors}
\label{sec:bulk-bulk-harvesting}

We now consider the example 2 setup for harvesting, in which both boundary detectors are smeared using HKLL kernels
associated with the two bulk worldlines in eq.~\eqref{eq:bulkworldline}.
For simplicity, we will take $z_\sB > z_\sA$ throughout the following.

Substituting the parameters in eq.~\eqref{eq:case2} for example 2, the parameters in eq.~\eqref{eq:lots-of-parm} become
\begin{eqnarray}
  a_{\pm}\ &= \ \frac{1}{2\mathcal{T}^2}\left(\frac{\ell^2}{z_\sA^2}\pm\frac{\ell^2}{z_\sB^2}\right)\,
  \qquad \beta\ &= \ \frac{\ell^2}{\mathcal{T}^2(z_{\sA}^2 + z_{\sB}^2)}\,,
\nonumber\\
  \Omega_{\pm} &= \ \frac{\Omega}{2}\left(\frac{\ell}{z_{\sA}} \pm\frac{\ell}{z_{\sB}}\right)\,,
  \qquad \gamma &= \ \Omega\frac{\ell(z_{\sA}-z_{\sB})}{z_{\sA}^2 + z_{\sB}^2}\,,
\end{eqnarray}
and we have $\hat{\Psi}_{\sA}=\hat{\Psi}_{\sB}=\hat{\Phi}$. Hence the density matrix elements \eqref{eq:almost-done} read 
\begin{equation}
\begin{aligned}
\frac{\cL_{\sA\sA}}{\lambda^2}
&\ = \ \ \sqrt{\frac{\pi}{2}}\,
  \mathcal{T}  \int\!\dd \tv\; e^{-\frac{\tv^2}{2\mathcal{T}^2}}\;
    e^{-\ii\Omega \tv}\;
    G^+_{\text{bulk-bulk}}(\tv)\,,\\
\frac{\cM}{\lambda^2}
&\ = -
  \frac{\ell^2}{z_{\sA}z_{\sB}}\sqrt{\frac{\pi}{2a_+}}e^{ -\frac{\Omega_+^2}{2a_+}}
  \int\!\dd v\;
    e^{-\beta v^2+\ii \gamma v}\;
    \GF{bulk-bulk}(v)\,\,.
\end{aligned}  
\label{eq:cornball}
\end{equation}
and $\cL_{\sB\sB}$ has the same form as $\cL_{\sA\sA}$.
Note that
in the self-noise term $\cL_{\sA\sA}$, we have changed the integration variable as $v=(z_{\sA}/\ell) \tv$.
Hence the corresponding Wightman function $G^+_{\text{bulk-bulk}}(\tv)$ is given by eq.~\eqref{eq:Gbb} with $\xi = {2\ell^2}/(2\ell^2-(\tv-\ie)^2)$. In the expression for the coherence, the time-ordered propagator $\GF{bulk-bulk}(v)$ takes the same form as eq.~\eqref{eq:Gbb} but with
\begin{equation}
    \xi=\frac{2\,z_\sA\,z_\sB}{z_\sA^2+z_\sB^2+L^2-v^2+2\ie |v|}\,.
    \label{eq:stretch}
\end{equation}
We now discuss the different matrix elements in turn.

\paragraph{Self-noise terms:} The self-noise in eq.~\eqref{eq:cornball} is  independent of $z_\sA$, the depth of the corresponding worldline. 
This reflects the symmetries of the bulk AdS geometry (\iee the translation and dilatation isometries of the Poincar\'e patch), which can be used to map the static worldline at one depth and $x$ position to any other while preserving the proper-frame detector parameters. Consequently, we have  $\cL_{\sB\sB} = \cL_{\sA\sA} \equiv \cL_{\text{bulk-self}}$. Since the two self-noise terms are equal, the negativity~\eqref{eq:neg} simplifies to
\begin{equation}\label{eq:neg-bb}
  \cN\ = \ \max\Bigl(0,\;|\cM| - \cL_{\text{bulk-self}}\Bigr),
\end{equation}
and the harvesting threshold is $|\cM| > \cL_{\text{bulk-self}}$.

\begin{figure}[t]
  \centering
  \includegraphics[width=\textwidth]{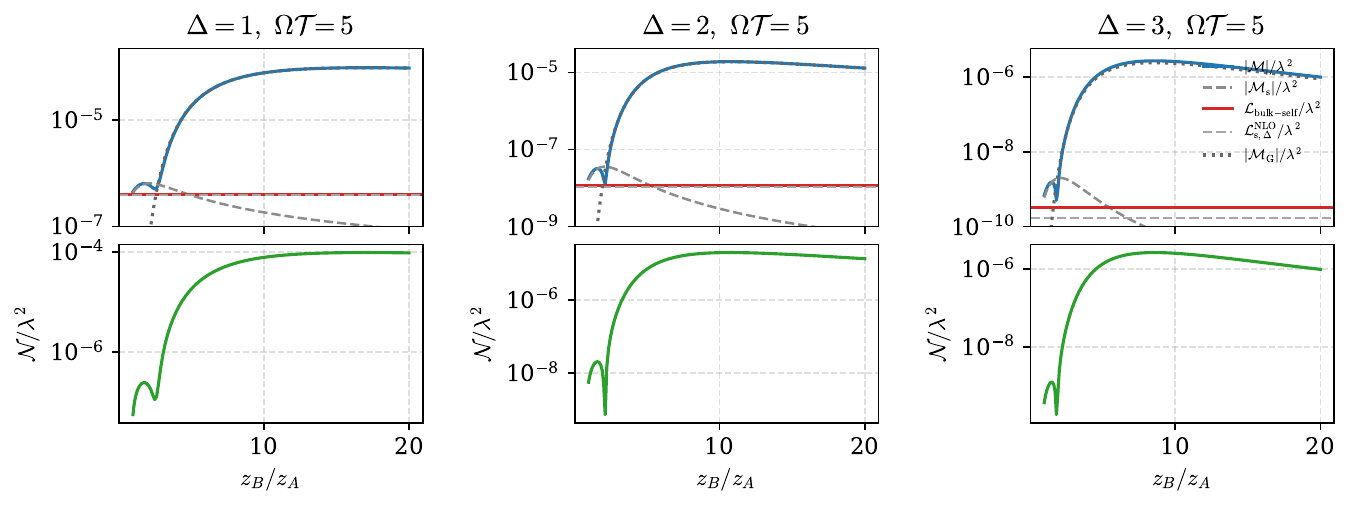}
  \caption{Numerical harvesting results for the two-HKLL-detector setup,
  with $\zA=1$ and
    $\zB$ varied.
    We set $\ell=1$ and
  $L=5=\Omega$ (again, we use units where $\mathcal{T}=1$),
  and we vary $\Delta=1,2,3$.
    \textbf{Top:} 
    $|\cM|/\lambda^2$
    (solid, blue) with the saddle contribution~\eqref{eq:M-saddle-bb} (dashed, gray),
    and the branch-cut contribution~\eqref{eq:M-branch-bb} (dotted, gray).
    The data tracks the saddle contribution when $\zB/\zA$ is order 1, and the branch-cut contribution at
    large~$\zB/\zA$.
    We have also shown 
    the constant self-noise reference $\cL_{\text{bulk-self}}/\lambda^2$ (red,
    dashed) and its approximation \eqref{eq:L-sad} (appropriately generalized to $\Delta \ne 1$). Note that increasing $\Omega$ improves the saddle approximations relative to the exact answer.
    \textbf{Bottom:} The negativity
    $\cN/\lambda^2 = \max(0,|\cM|-\cL_{\text{bulk-self}})$.
   }
  \label{fig:bulk-bulk-sweep}
\end{figure}

The integral for $\cL_{\textrm{bulk-self}}$ in the first line of eq.~\eqref{eq:cornball} 
is of the schematic form \eqref{eq:schem-int}, with 
\begin{equation}
    \tilde{v}_s = -i \mathcal{T}^2 \Omega
    \qquad 
    \tilde{v}_1 = \ie
    \qquad
    \tilde{v}_2 = 2\ell + \ie
    \qquad
    \tilde{v}_3 =-  2\ell + \ie\,,
\end{equation}
and Gaussian width $\mathcal{T}$. 
As above, we expect the saddle approximation is valid in the limit $|\tilde{v}_s|\gg \mathcal{T}$ or equivalently
$\Omega\mathcal{T}\gg1$. 
At leading order we obtain 
\begin{equation}
    \frac{\cL_{\textrm{bulk-self}}^{(0)}}{\lambda^2} = \pi
  \,\mathcal{T}^2 
  e^{-\frac{1}{2}\Omega^2\mathcal{T}^2}\;
    G^+_{\textrm{bulk-bulk}}\Big(\xi = \frac{2}{2+q^2}\Big)\,,
    \qquad \qquad q \equiv \frac{|\tilde{v}_s|}{\ell} =\frac{\mathcal{T}^2 \Omega}{\ell} \,.
\end{equation}
If we Taylor expand $G^+_{\textrm{bulk-bulk}}$ near the saddle point,
we can also extract a subleading correction which improves the match to the numerical results in figure \ref{fig:bulk-bulk-sweep}.
For instance, the corrected answer reads 
\begin{equation}
    \frac{\cL_{\textrm{bulk-self}}}{\lambda^2} = 
    \frac{\cL_{\textrm{bulk-self}}^{(0)}}{\lambda^2}
  \left[1-\frac{\mathcal{T}^2}{\ell^2}\frac{3q^4 + 10q^2 + 16}{q^2(q^2+4)^2}\right]\,
  \label{eq:L-sad}
\end{equation}
for $\Delta=1$.

\paragraph{Nonlocal coherence:} 
To simplify the expression for the coherence, 
we change integration variable as follows:
\begin{equation}
  v\ =\   \bar v/\sqrt{\beta}\ =  \ 
    \mathcal{T}\sqrt{\zA^2+\zB^2}\,\bar v/\ell \,.
    \label{eq:vbar}
\end{equation}
Note that while $v$ had units of length, $\bar{v}$ is dimensionless.
Then, the integral for $\mathcal{M}$ in eq.~\eqref{eq:cornball}  becomes
\begin{equation}
    -\frac{\cM}{\lambda^2}
\ = \ 
  \sqrt{\pi}\,\mathcal{T}^2\,e^{ -\frac{\Omega^2 \mathcal{T}^2}{2(1+\alpha^2)}}
  \int\!\dd \bar{v}\;
    e^{-\bar{v}^2-\ii \omega \bar{v}}\;
    \GF{bulk-bulk}(v=\mathcal{T}\sqrt{\zA^2 + \zB^2}\,\bar{v}/\ell)\,,
    \label{eq:M-bb-rescaled}
\end{equation}
where we define 
\begin{equation}\label{eq:alpha-def}
  \alpha\ \equiv \ \frac{\zB - \zA}{\zA + \zB}
\ \in \ [0,\,1)\,,
  \qquad
  \omega\ \equiv \
\Omega\mathcal{T}\,\frac{z_{\sB}-z_{\sA}}{\sqrt{z_{\sA}^2 + z_{\sB}^2}}\
\,,
\qquad
\begin{cases}
    d_1 \equiv \sqrt{L^2 + (\zB-\zA)^2}\,,\\
    d_2 \equiv \sqrt{L^2 + (\zB+\zA)^2}\,.
\end{cases}
\end{equation}

The integral in eq.~\eqref{eq:M-bb-rescaled} has the schematic form \eqref{eq:schem-int}, with
\begin{equation}
 \label{eq:v1v2-bar}
    \bar v_s = -\ii\omega/2
    \qquad
  \bar v_1\ = \frac{\ell d_1}{\mathcal{T}\sqrt{\zA^2 + \zB^2}} ,
  \qquad
  \bar v_2\ = \ \frac{\ell d_2}{\mathcal{T}\sqrt{\zA^2 + \zB^2}} \,.
\end{equation}
In the regime $\max\{ |\omega|, \sqrt{2}\bar{v}_1\}\gg 1$, these singularities  lie well outside the Gaussian's primary support region, and so we expect the local saddle expansion is valid.\footnote{Note that unlike in the previous examples,
where
the Gaussian width was set by $\mathcal{T}$, here
it is simply $1/\sqrt{2}$.
Note also that validity of the local saddle expansion does not imply that the saddle dominates the full contour.
Dominating over the branch cut contribution (to be discussed momentarily) requires an additional condition $\bar{v}_1^2-\omega^2/4\gg 1$.
} 
Applying the saddle approximation gives
\begin{equation}
    -\frac{\cM_s}{\lambda^2}
\ = \ 
  \pi\,\mathcal{T}^2\,e^{ -\frac{\Omega^2 \mathcal{T}^2}{2}}
    \;
    \GF{bulk-bulk}\left(v=\frac{i \mathcal{T}^2 \Omega}{2}\frac{\zA-\zB}{\ell}\right)\,.
    \label{eq:M-saddle-bb}
\end{equation}
In figure \ref{fig:bulk-bulk-sweep}, the numerical results are approximately given by this real function, but only for small $z_\sB$.  
The approximation \eqref{eq:M-saddle-bb} is valid in figure \ref{fig:bulk-bulk-sweep} because at fixed nonzero depth asymmetry $\zB-\zA$,
our choice of large 
$\Omega \mathcal{T}$ (namely $\Omega \mathcal{T}=5$) guarantees $|\omega|\gg 1$.
On the other hand, as $\zB$ approaches $\zA$, $|\omega|$ becomes small but our choice $L\gg \zA$ guarantees $\sqrt{2}\bar{v}_1\gg 1$.

To understand what's going on at larger values of $z_\sB$,
we must take a closer look at the branch structure of the integrand of eq.~\eqref{eq:M-bb-rescaled}. 
To do so, note that in terms of $d_1$, $d_2$,  and $v$,
the AdS-invariant distance reads simply $\xi = 4\zA\zB/(d_1^2 + d_2^2-2v^2)$,
and the bulk-to-bulk propagator reads
\begin{equation}
    G^{(F)}_{\textrm{bulk-bulk}}(v)
    =
    \frac{\zA \zB}{\sqrt{(v^2-d_1^2)(v^2-d_2^2)}}
    \left(
    \frac{4\zA\zB}{d_1^2+d_2^2 -2 v^2 + 2 \sqrt{(v^2-d_1^2)(v^2-d_2^2)}}
    \right)^{\Delta-1}
    \,.
\end{equation}
Assuming that $\Delta$ is an integer,\footnote{
For non-integer $\Delta$, 
additional branch discontinuities will contribute to $\mathcal{M}_{\textrm{br}}$.
However, our numerical results show that the qualitative behaviour of harvesting remains the same.
Indeed, in many cases, a rough approximation to $\mathcal{M}_{\textrm{br}}$ can be obtained using eq.~\eqref{eq:M-branch-bb} below,
but replacing the Bessel function with the left-hand side of eq.~\eqref{bessel} divided by $\pi F(\pi/2)$.
} 
all of the interesting branch structure in $G^{(F)}_{\textrm{bulk-bulk}}$ comes from the prefactor outside the parentheses.
Specifically, the bulk-to-bulk Feynman propagator has branch cuts along two intervals 
\begin{equation}
    v \in (-d_2,-d_1) -\ie
    \qquad 
    v\in(d_1,d_2)+\ie \,.
\end{equation}
We see that in order to deform the original real-axis contour to reach the steepest descent contour through $\bar{v}_s=-i \omega/2$ with $\omega>0$, 
the contour must wrap the branch cut along the interval $v\in(-d_2,-d_1)$.
This is the example shown in figure \ref{fig:complex_v_plane}.

To evaluate the contribution from the branch cut, we must evaluate
\begin{equation}
    -\frac{\cM_{\rm br}}{\lambda^2}
\ = \ 
  \sqrt{\pi}\,\mathcal{T}^2\,e^{ -\frac{\Omega^2 \mathcal{T}^2}{2(1+\alpha^2)}}
  \int_{-\bar{v}_2}^{-\bar{v}_1}\!\dd \bar{v}\;
    e^{-\bar{v}^2-\ii \omega \bar{v}}\;
    \textrm{Disc}\GF{bulk-bulk}(v)\,.
\end{equation}
We can equivalently present this equation as an integral over the branch cut on the positive real axis, as follows:
\begin{equation}
  - \frac{\cM_{\rm br}^*}{\lambda^2}
\ = \ 
\sqrt{\pi}\,\mathcal{T}^2\,e^{ -\frac{\Omega^2 \mathcal{T}^2}{2(1+\alpha^2)}}
  \int_{\bar{v}_1}^{\bar{v}_2}\!\dd \bar{v}\;
    e^{-\bar{v}^2-\ii \omega \bar{v}}\;
    \textrm{Disc}\GF{bulk-bulk}(v)\,,
\end{equation}
where we have used the complex conjugation identity $\textrm{Disc}(f^*)=-\textrm{Disc}(f)^*$,
and that here we have $\textrm{Disc}f(-x)=-\textrm{Disc}f(x)$.\footnote{We use the standard definition $\textrm{Disc}f(x)=f(x+i0)-f(x-i0)$.}
To evaluate this integral, we parameterize the branch cut by $\theta$:
\begin{equation}
\cos\theta =  \frac{D-v^2}{2 \zA \zB}\,,\qquad D\equiv 2 \zA \zB+d_1^2\,.
\end{equation}
The endpoints of the branch are at $\theta=0$ ($v=d_1$) and $\theta=\pi$ ($v=d_2$).
It follows that
\begin{equation}
    G^{(F)}_{\textrm{bulk-bulk}}(v(\theta)\pm i 0)
    =
    \frac{\pm i e^{\pm i(\Delta-1)\theta}}{2\sin\theta}
    \,.
    \label{eq:G-bb-vd1d2}
\end{equation}
Hence,
\begin{equation}
    \textrm{Disc}\GF{bulk-bulk}(v(\theta))
    =
    \frac{ \ii \cos\left((\Delta-1)\theta\right)}{\sin\theta}\,.
\end{equation}
Meanwhile, there is a Jacobian from changing variables from $v$ to $\theta$:
\begin{equation}
    \frac{\dd v}{\dd \theta}=\frac{\zA\zB\sin\theta}{v(\theta)}\,,
\end{equation}
and with this the branch cut integral becomes
\begin{equation}
  - \frac{\cM_{\rm br}^*}{\lambda^2}
\ = \ 
\ii\sqrt{\pi}\,\mathcal{T}\,e^{ -\frac{\Omega^2 \mathcal{T}^2}{2(1+\alpha^2)}}
\frac{\ell\zA\zB}{\sqrt{\zA^2+\zB^2}}
  \int_{0}^{\pi}\!\dd \theta\,
  F(\theta)
    \,\cos\left((\Delta-1)\theta\right)\,,
     \label{eq:m-brrrrr}
\end{equation}
where $v(\theta)$, $\bar{v}(\theta)$, and $F(\theta)$ read
\begin{equation}
   v(\theta)=\sqrt{D} \sqrt{1 -\rho \cos \theta}\,,
   \qquad
   \bar{v}(\theta)
    =
    \sqrt{\eta} \sqrt{1 -\rho \cos \theta}
     \qquad
   F(\theta)= \frac{e^{-\bar{v}(\theta)^2-\ii \omega \bar{v}(\theta)}}{v(\theta)}\,,
\end{equation}
and we have defined
\begin{equation}
   \rho \equiv \frac{2\zA \zB}{D}\,,
   \qquad
   D\equiv 2\zA \zB+d_1^2\,,
    \qquad
    \eta \equiv
    \frac{\ell^2 D }{\mathcal{T}^2(\zA^2+\zB^2)}\,,
    \qquad 
    \chi = \omega\sqrt{\eta}\,.
\end{equation}
In terms of these parameters, $F(\theta)$ can be expanded as
\begin{equation}
    F(\theta)= F(\pi/2)\left[\frac{e^{\eta \rho\cos\theta} e^{-\ii\chi (\sqrt{1-\rho\cos\theta}-1)}}{\sqrt{1-\rho\cos\theta}}\right]\,,
    \qquad
    F(\pi/2)=\frac{e^{-\eta-\ii\chi }}{\sqrt{D}} \,.
\end{equation}

We now evaluate the integral \eqref{eq:m-brrrrr} in a small-$\rho$ approximation.\footnote{For the parameter choices in figure \ref{fig:bulk-bulk-sweep},  we have $\rho<0.2$.
Note that this approximation is subtle  because the angular integral against $\cos((\Delta-1)\theta)$ in eqs.~\eqref{eq:m-brrrrr} and \eqref{bessel} can project out naively leading terms in the $\rho$ expansion.
For $\Delta\ge 3$, a more careful analysis would therefore require expanding to higher orders in $\log(F(\theta)/F(\pi/2))$ than what is presented here.
}
When $\rho$ is small with $\chi$ fixed, we have $\log(F(\theta)/F(\pi/2))\approx \Gamma \cos \theta$, with 
\begin{equation}
    \Gamma\equiv (\eta + \frac{1}{2} + \frac{\ii\chi}{2})\rho\,.
\end{equation}
Hence, we can approximate the $\theta$-integral as a simple Bessel integral:
\begin{equation}
  F(\pi/2) \int_{0}^{\pi}\!\dd \theta\,
  e^{\Gamma\cos\theta}\cos((\Delta-1)\theta)
  =
  \pi F(\pi/2) I_{|\Delta-1|}(\Gamma)\,,
  \label{bessel}
\end{equation}
where we have used that $\Delta$ is an integer.
All in all, we have
\begin{equation}
 -  \frac{\cM_{\rm br}^*}{\lambda^2}
\ \approx \ 
i\pi^{3/2}\,\mathcal{T}\,e^{ -\frac{\Omega^2 \mathcal{T}^2}{2(1+\alpha^2)}}
\frac{\ell\zA\zB e^{-\eta-i\chi }}{\sqrt{(\zA^2+\zB^2)(L^2+\zA^2+\zB^2)}}
  I_{|\Delta-1|}
  \left((\eta + \frac{1}{2} + \frac{i\chi}{2})\rho\right)\,.
   \label{eq:M-branch-bb}
\end{equation}
This gives a good fit to the numerics, as shown in figure \ref{fig:bulk-bulk-sweep} for the cases $\Delta=1,2,3$.
Note that unlike in  eqs.~\eqref{eq:M-saddle-I} and~\eqref{eq:M-saddle-bb},
the approximation to $\mathcal{M}$ in eq.~\eqref{eq:M-branch-bb} has a nontrivial imaginary part.
Consequently, the crossover from saddle-dominated to branch-cut dominated regimes is accompanied by a rapid increase in the imaginary part of $\mathcal{M}$.

Two phases within the branch-dominated regime must be distinguished: an initial rapid growth phase and an ultimate decay phase.
In the first phase, $|\mathcal{M}|$ grows rapidly  with $\zB$.
This increase is due to the exponential factor $e^{ -\frac{\Omega^2 \mathcal{T}^2}{2(1+\alpha^2)}}$ in eq.~\eqref{eq:M-branch-bb}.
At larger $\zB$, however,
this exponential factor saturates to its maximum value $e^{-\Omega^2\mathcal{T}^2/4}$
and 
the remaining factors in eq.~\eqref{eq:M-branch-bb} become relevant,
leading to a phase where $|\mathcal{M}|$ decays again.
Explicitly, in the limit $\zB\gg \zA$
in eq.~\eqref{eq:M-branch-bb}, the leading term is of order
$(\zA/\zB)^\Delta$, 
representing polynomial decay.
One sees the onset of this decay towards the right side of the $\Delta=3$ plots in figure \ref{fig:bulk-bulk-sweep}.

It is natural to ask whether the rapid increase in $|\mathcal{M}|$ corresponds to the onset of appreciable causal signalling between the detectors.
We can address this question following the arguments of \cite{Tjoa:2021jgo} (see also \cite{Caribe:2023fhr,Teixido-Bonfill:2024phf}) and the answer to our question appears to be yes. 
We emphasize that the following analysis relies on the fact that we are working with a free scalar field in the bulk. 
To characterize causal signalling, we begin with a standard commutator-anticommutator decomposition of the Wightman function
\begin{equation}
    G^+(\sfX_1,\sfX_2)
    =
   H(\sfX_1,\sfX_2)
   +
   C(\sfX_1,\sfX_2)\,,
   \label{eq:rockandroll}
\end{equation}
with
\begin{equation}
    H(\sfX_1,\sfX_2)=\frac{1}{2}\langle \{ 
    \hat{\Phi}(\sfX_1),
     \hat{\Phi}(\sfX_2)
    \}
    \rangle
    \qquad
    \textrm{and}\qquad
    C(\sfX_1,\sfX_2)=\frac{1}{2}\langle [ 
    \hat{\Phi}(\sfX_1),
     \hat{\Phi}(\sfX_2)
    ]
    \rangle\,.
\end{equation}
One may observe that
the symmetric part, \(H\), should contain  information about correlations in the state which are independent of signalling, since \(H\) can be nonzero between spacelike-separated points, where communication is impossible. Hence
$H$ is associated with the genuine extraction --- or harvesting --- of field correlations.
Meanwhile, one says the commutator $C$ measures signalling because it vanishes for spacelike-separated field operators.
We can make this more precise for a free theory (or generalized free theory), in which case the commutator $C$ is state-independent and therefore insensitive to modifying the entanglement structure of the state. With these assignments, we substitute eq.~\eqref{eq:rockandroll} into eq.~\eqref{eq:M-general} which allows for a decomposition of $\mathcal{M}$ as
\begin{equation}
    \mathcal{M}=\mathcal{M}_H+\mathcal{M}_C\,.
\end{equation}
Given the previous discussion, the idea here is that $\mathcal{M}_H$ captures the correlations arising from harvesting, and $\mathcal{M}_C$ captures the correlations arising from communication.

\begin{figure}
    \centering
    \includegraphics[width=0.9\linewidth]{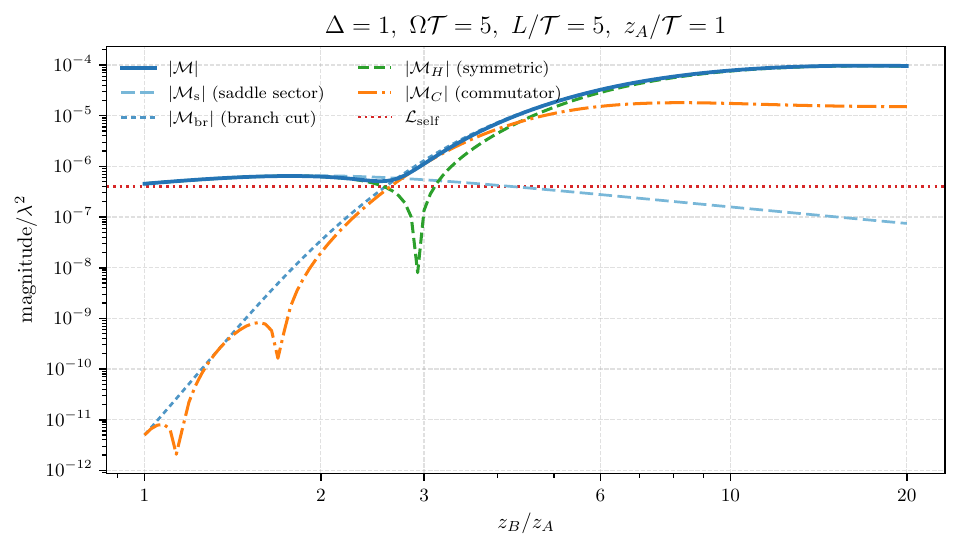}
    \caption{Harvesting/communication decomposition for $\mathcal{M}$.
    The harvesting contribution (green) accounts for the small bump in $|\mathcal{M}|$ at small $\zB/\zA$.
    Meanwhile, the communication contribution (orange) accounts for the rapid increase of $|\mathcal{M}|$ at the onset of the branch-dominated phase.
    Deeper into the branch-dominated phase, $\mathcal{M}$ becomes a mixture of both $\mathcal{M}_C$ and $\mathcal{M}_H$, and is increasingly dominated by $\mathcal{M}_H$.
    }
    \label{fig:HCdecomp}
\end{figure}

As shown in figure \ref{fig:HCdecomp}, 
the magnitude of $|\mathcal{M}|$ in the saddle-dominated regime is predominantly accounted for by $|\mathcal{M}_H|$, the harvesting-driven correlations. Meanwhile, the rapid increase in $|\mathcal{M}|$ accompanying the immediate transition to the branch-cut dominated regime is predominantly accounted for by $|\mathcal{M}_C|$, the communication-driven correlations. This is strong evidence that the rapid increase in $|\mathcal{M}|$ corresponds to causal signalling.

Nevertheless, as figure \ref{fig:HCdecomp} makes clear, there is no natural identification between $\mathcal{M}_C$  and $\mathcal{M}_{\textrm{br}}$. Indeed, in the  case considered there, as $\zB$ becomes very large (\iee in the regime $10\lesssim z_\B/z_\A\lesssim20$, where $|\mathcal{M}|$ is reaching its peak), $|\mathcal{M}_{\textrm{br}}|$ is instead predominantly accounted for by $|\mathcal{M}_H|$.
To understand this intuitively, note that for Gaussian switching functions centered at time zero (the case considered throughout this paper),  one has $\mathcal{M}_H\in \mathbb{R}$ and $\mathcal{M}_C \in i\mathbb{R}$ when time-reflection symmetry is present \cite{Zambianco:2024dqn}. Meanwhile, there is no symmetry which constrains the phase of $\mathcal{M}_{\textrm{br}}$. Rather it can have substantial real and imaginary parts. Hence, $\mathcal{M}_{\textrm{br}}$ can in principle receive contributions from both $\mathcal{M}_C$ and $\mathcal{M}_H$.

We note that the case presented in figure \ref{fig:HCdecomp} has $\Delta=1$.
For values of $\Delta$ distinct from 1, 
similar features can be observed,
except that $\mathcal{M}_H$ does not always dominate the large-$\zB$ regime.
What appears universal is the mismatch between 
$\mathcal{M}_{\textrm{br}}$ and $\mathcal{M}_{C}$ (and hence between $\mathcal{M}_{s}$ and $\mathcal{M}_{H}$) after the initial moment of rapid increase in $|\mathcal{M}|$, which indicates the onset of signalling.

\section{Discussion}
\label{sec:discussion}

In this paper, we have used bulk reconstruction in AdS/CFT to give a boundary description of bulk Unruh--DeWitt detectors. Our main result is that the boundary dual of a local bulk detector is not an ordinary detector with finite spatial extent in the CFT. Rather, it is a detector with a nontrivial \textit{spacetime} smearing determined by an HKLL kernel. In this sense, HKLL reconstruction gives not only an operator dictionary for local bulk fields, but also an operational dictionary for detector protocols. This complements the study of 
ordinary pointlike detector protocols in CFTs, including holographic CFTs, in \cite{Wurtz:2026ofi}. 
Our work in sections \ref{sec:detector-setup} and \ref{sec:harvesting-geometries} uses the same entanglement harvesting framework as in \cite{Wurtz:2026ofi}, except that  we replace one or both pointlike boundary couplings with HKLL smearings that represent pointlike couplings in the bulk.

The original HKLL constructions represent a local bulk operator as a boundary operator smeared over an extended region which includes an entire Cauchy slice of the boundary manifold \cite{Hamilton:2005ju, Hamilton:2006az,Hamilton:2006fh}. 
In odd spacetime dimensions, this region is the entire boundary of the Poincar\'e patch.
For RQI protocols (\eg entanglement harvesting as studied here), it is more natural to use representatives supported on a finite causal diamond, so that the corresponding detectors are only smeared over a finite boundary region. 
As reviewed in section~\ref{sec:ads-rindler}, causal diamond reconstruction relies on rewriting the entanglement wedge of the boundary diamond in AdS-Rindler coordinates and using the corresponding Rindler smearing kernel.
We also noted that the diamond and  Poincar\'e descriptions are different boundary representatives of the same bulk operator in the semiclassical code subspace.
That is, when inserted in code-subspace correlators, the causal diamond and Poincar\'e representatives reproduce the same bulk-to-boundary and bulk-to-bulk Wightman functions \cite{Hamilton:2005ju,Hamilton:2006az}.

We emphasize that our HKLL detectors  are not within the class of typical detector models that one would use to probe the boundary CFT in an RQI protocol.
The detector degree of freedom is still an auxiliary two-level system, and this qubit still couples to a quantum field in a  local way in the bulk (or with some small smearing; see section \ref{sec:smeared-hkll}). 
However, when we rewrite this interaction as an HKLL detector in the boundary theory, the interaction between the qubit and the dual CFT operator takes an unconventional form. 
The HKLL smearing \eqref{eq:boundary-int-hkll} couples the monopole operator $\hat{\mu}(t)$ at detector time $t$ to CFT operators $\hat{\mathcal O}(t',x')$ at many boundary times $t'$. 
Thus, the boundary description involves a spacetime smearing distinct from the 
smearing of a standard spatially-extended detector \eqref{eq:smeared-hamiltonian}, where the qubit-operator coupling remains local in time. 
This nonlocal behaviour makes it unclear how to realize an HKLL detector in an actual laboratory experiment, and we consider this an interesting question for future work.

A second key difference is that the (pointlike) bulk detector is covariantly defined, with its interaction  supported on a single worldline and expressible  as a proper-time integral. Similarly, the free qubit Hamiltonian characterizes evolution in proper time along this worldline. 
When expressed in terms of boundary time, however, this proper-time generator is multiplied by \(d\tau/dt\), which need not be constant along a general bulk trajectory.
The simple bulk trajectories that we examined in section \ref{sec:detector-setup} remained at a fixed bulk depth (\iee $z$ remained constant), and hence the translation between the proper time along the worldline and the boundary time was easily accommodated by a multiplicative shift of the detector gap, \eg see eq.~\eqref{eq:H-int-0}. However, one can easily imagine situations where the bulk trajectory is not static, \eg  an observer falling towards the center of the AdS geometry or towards a black hole in the bulk. In such a setting, the redshift between the proper time along the worldline and the boundary time would itself be a function of time. 
Hence, from the boundary perspective, the free qubit Hamiltonian is time-dependent, with the gap varying in time to mimic the proper time evolution in the bulk. 
This again emphasizes the unconventional form that the HKLL detector takes in the boundary theory.

\subsection*{Time ordering and HKLL detectors}

As emphasized in section~\ref{sec:detector-setup}, the spacetime smearing of an HKLL kernel raises an important question concerning the ordering of the corresponding interaction \eqref{eq:boundary-int-hkll}.  For an ordinary detector with finite spatial extent, different spacetime foliations may order spacelike-separated portions of the interaction differently --- see discussion around eqs.~\eqref{dyson} and \eqref{eq:time-order-mess}. The resulting covariance violation is governed by the commutator of the interaction Hamiltonian density at the points whose ordering is reversed~\cite{Martin-Martinez:2020lul}.  
The discussion of time-ordering for HKLL detectors differs structurally from that of these conventional extended detectors. 

For instance, for conventional spatially smeared detectors,
the corresponding correction to the detector's reduced density matrix  for detectors initially diagonal in energy vanishes at order $\lambda^2$ \cite{Martin-Martinez:2020lul}. 
However, due to the nontrivial spacetime smearing,
this result does \textit{not} imply that one may chronologically reorder the individual CFT insertions within an HKLL detector according to their boundary times.
Doing so defines a different boundary protocol and may in principle modify the nonlocal coherence \(\mathcal M\) at order $\lambda^2$, even for energy-diagonal initial detector states.

To clarify the scope of this issue, we recall that
HKLL detectors are boundary representatives of pointlike (in our examples) bulk detectors, whose interactions are supported on timelike bulk worldlines and are naturally ordered by the bulk detector's proper time, or by a common bulk time function in a multi-detector protocol.
In the boundary representation, the detector at time $\tau$ couples to the composite operator
\begin{equation}
\hat{\mathcal O}_{A}(\tau)
\ \equiv \
\int_A \dd^d\mathsf{x}'\,
K_A\bigl(\sfX(\tau)|\mathsf{x}'\bigr)\,
\hat{\mathcal O}(\mathsf{x}')
\ =\
\hat{\Phi}\bigl(\sfX(\tau)\bigr)\,
\label{eq:composite-HKLL-operator}
\end{equation}
supported on the codimension-zero region $A$.
As specified in eq.~\eqref{eq:HKLL-evolution2}, the Dyson expansion inherited from the bulk protocol orders these boundary operators $\hat{\mathcal O}_{A}(\tau)$ according to the  bulk evolution parameter $\tau$. 
In other words, eq.~\eqref{eq:composite-HKLL-operator} gives a particular boundary operator dual to the bulk field $\hat{\Phi}\bigl(\sfX(\tau)\bigr)$ and interacts with the qubit system according to $\tau$. 
Eq.~\eqref{eq:composite-HKLL-operator} should not be regarded as giving a prescription in which the detector couples separately to each local CFT operator $\hat{\mathcal{O}}(\sfx')$ as the boundary time $(\mathsf{x}^0)'=t'$ passes. 
The upshot is that while reordering the local operators $\hat{\mathcal O}(\mathsf{x}')$ according to their boundary times can in principle modify the nonlocal coherence \(\mathcal M\) at order $\lambda^2$,
reordering the composite operators $\hat{\mathcal O}_A(\tau)$ according to alternative foliations of the bulk evolution cannot do so.
This result follows from applying the analysis of \cite{Martin-Martinez:2020lul} in the bulk,
and in particular is relevant in the case of spatially-smeared bulk detectors.

Again, while the evolution defined above is mathematically well-defined, it does not describe an ordinary boundary-local laboratory coupling.  Implementing it would require direct access to the nonlocal operator $\hat{\mathcal O}_{A}(\tau)$, 
or a novel protocol which in any case involves an unconventional level of quantum control. 
At higher orders in perturbation theory, expanding the time-ordered products of the $\hat{\mathcal O}_{A}(\tau)$ operators in terms of local CFT operators will also generate operator orderings that are not chronological with respect to the relevant boundary times.  Such terms might be expressed using general Wightman or out-of-time-ordered correlators.  These features reflect the unconventional and nonlocal form taken by a local bulk detector in the boundary theory.

\subsection*{Accelerated bulk detectors}

The issues raised above become especially concrete in the setting of uniformly accelerated bulk detectors and their HKLL representatives. As we now explain, this setup is instructive for understanding  how time ordering should be implemented.

Consider a uniformly accelerated bulk observer contained within a single AdS-Rindler wedge. 
We take the  AdS-Rindler wedge to be associated with a boundary causal diamond $D$ of radius $\mathcal{R}$,
centered at $(t_i,x_i)$.
In describing the bulk trajectory, one may use the same boundary causal diamond $D$ at all bulk times, rather than introducing a family of moving diamonds. Hence for all values of $\tau$, the corresponding operators $\hat{\mathcal O}_{D}(\tau)$ would involve a smearing over the same boundary region (\iee the causal diamond $D$) and they are all associated with the same boundary  algebra $\mathcal A(D)$.

More concretely, in Rindler coordinates,
let the detector follow a trajectory at fixed $(Z,X)$ while evolving in the Rindler time $T$. 
Passing from Rindler to Poincar\'e coordinates using the transformation~\eqref{eq:poincare-to-rindler}, one finds that the trajectory obeys
\begin{equation}
(z,t,x)\ 
\underset{T\to\pm\infty}{\longrightarrow}\ 
(0,t_i\pm \mathcal R,x_i)
\, .
\end{equation}
That is, the observers remain within the causal wedge $W[D]$ and asymptotically approach the future and past tips of $D$ as $T\to+\infty$ and $T\to-\infty$, respectively. From the metric~\eqref{eq:ads-rindler-metric}, the observer's proper time satisfies
\begin{equation}
\tau =
\ell\,\frac{\sqrt{1-Z^2}}{Z}\,T\equiv \frac{T}{\gamma(Z)}\,,
\label{eq:rindler-proper-time}
\end{equation}
where the integration constant has been chosen to align $\tau=0=T$. 

Choosing the same reconstruction diamond $D$ for all values of $\tau$, 
the boundary operator $\hat{\mathcal O}_{D}$ depends on $\tau$ only through the profile of the HKLL kernel on that diamond:
\begin{equation}
\hat{\mathcal O}_{D}(\tau)
\equiv
\int_D \dd^d\mathsf{x}'\,
K_D(Z,\,T\!=\!\gamma(Z)\,\tau,\,X|\mathsf{x}')\,
\hat{\mathcal O}(\mathsf{x}')
= \hat{\Phi}(Z,\gamma(Z)\tau,X)\,.
\label{eq:rindler-HKLL-operator}
\end{equation}
Translations in $T$ are generated on the boundary by the future-directed (global) conformal transformation that preserves $D$. 
For the CFT vacuum, this geometric flow agrees  with the modular flow on the diamond, up to a normalization \cite{Casini:2011kv}.
For the chosen fixed-\(Z\) orbit, let \(\hat H_D\) denote the generator of proper-time translations, normalized so that the vacuum modular Hamiltonian is \(2\pi\hat H_D/\gamma(Z)\).
Then, one has schematically
\begin{equation}
\hat{\mathcal O}_{D}(\tau)
=
e^{\ii \hat H_{D}\, \tau}\,
\hat{\mathcal O}_{D}(0)\,
e^{-\ii\hat H_D\, \tau}\,.
\label{eq:HKLL-modular-flow}
\end{equation}
Therefore, proper-time ordering for these accelerated bulk detectors can be understood as modular-time ordering of the $\hat{\mathcal O}_{D}(\tau)$ operators. 
We emphasize again that, for every value of the proper time $\tau$, the operator $\hat{\mathcal O}_{D}(\tau)$ in eq.~\eqref{eq:rindler-HKLL-operator} is smeared over the entire causal diamond $D$. Nevertheless, the time ordering is determined entirely by the label $\tau$ assigned to the boundary operators $\hat{\mathcal O}_{D}(\tau)$ and which determines the weighting of the smearing across $D$ through the corresponding HKLL kernel.

For example, the corresponding time-ordered two-point function is
\begin{align}
\left\langle
\mathcal T_\tau\left\{
\hat{\mathcal O}_{D}(\tau_1)
\hat{\mathcal O}_{D}(\tau_2)
\right\}
\right\rangle
&=\,
\theta(\tau_1-\tau_2)
\int_D\dd^d\mathsf{x}_1\,\dd^d\mathsf{x}_2\,
K_{\tau_1}(\mathsf{x}_1)\,K_{\tau_2}(\mathsf{x}_2)
\,G^+(\mathsf{x}_1,\mathsf{x}_2)
\nonumber\\
&\quad+\ 
\theta(\tau_2-\tau_1)
\int_D\dd^d\mathsf{x}_1\,\dd^d\mathsf{x}_2\,
K_{\tau_1}(\mathsf{x}_1)\,K_{\tau_2}(\mathsf{x}_2)
\,G^+(\mathsf{x}_2,\mathsf{x}_1)\,,
\label{eq:modular-time-ordered-HKLL}
\end{align}
where $K_{\tau_i}(\mathsf{x})
\equiv K_D(Z,T\!=\!\gamma(Z)\tau_i,X|\mathsf{x})$.
Thus the step functions depend on the proper-time labels $\tau_1$ and $\tau_2$ of the HKLL operators \eqref{eq:rindler-HKLL-operator}, rather than on the time coordinates of the boundary points integrated over within the kernels.

Moving away from time-ordering issues,
we also note that the connection to modular flow provides a natural perspective on the thermal response of accelerated detectors in AdS~\cite{Deser:1997ri,Jacobson:1997ux,Das:2001zc,Parikh:2012kg}. 
The CFT vacuum restricted to $D$ is a KMS state with respect to the modular flow generated by $2
\pi\hat H_D/\gamma(Z)$. With the normalization in eq.~\eqref{eq:HKLL-modular-flow}, the proper temperature with respect to $\tau$ is therefore\footnote{Note that at fixed dimensionless Rindler position $(Z,X)$, the proper acceleration and temperature are independent of the diamond radius $\mathcal R$. Changing $\mathcal R$ rescales the trajectory in Poincar\'e coordinates by an AdS isometry, so the resulting observers are distinct but related by the isometry. 
Nevertheless, the $\mathcal R$-dependence of the conformal transformation from boundary Rindler  to Minkowski coordinates causes
many boundary quantities expressed in terms of Minkowski time 
to retain an explicit dependence on $\mathcal R$, \eg see \cite{Casini:2011kv}.
}
\begin{equation}
T_{\rm proper}
=
\frac{\gamma(Z)}{2\pi}
=
\frac{Z}{2\pi\ell\sqrt{1-Z^2}}\,.
\label{eq:rindler-proper-temperature}
\end{equation}
This result agrees with the standard relation between acceleration and temperature in AdS\footnote{We note that this expression holds for AdS of any dimension.}
\begin{equation}
T_{\rm proper}
=\frac1{2\pi}
\sqrt{
a^2-\frac{1}{\ell^2}
}\,.
\label{eq:acceleration-temperature-general}
\end{equation}
In particular, substituting the proper acceleration of a fixed-$(Z,X)$ trajectory, $a =
1/\sqrt{\ell^2(1-Z^2)}$, 
we recover the result in eq.~\eqref{eq:rindler-proper-temperature}. It would be interesting to understand these results better from the boundary perspective with HKLL detectors.

Finally, the example of the uniformly accelerating observer also illustrates the unusual behaviour of non-static systems in the laboratory frame, 
\iee in terms of the Poincar\'e time $t$. 
For simplicity, we set $x=x_i$ (and hence $X=0$) and then we may use eqs.~\eqref{eq:dimensionless-diamond-coords} and \eqref{eq:poincare-to-rindler} to find
the relation between the proper time and the Poincar\'e time,
\begin{equation}
\tau(\Delta t) =
\frac{1}{\gamma(Z)}
\log\left[
\frac{\Delta t+\sqrt{(1-Z^2)\mR^2+Z^2(\Delta t)^2}}
{\sqrt{1-Z^2}\,(\mR-\Delta t)}
\right]\,,
\label{eq:tautot}
\end{equation}
where $\Delta t \equiv t-t_i$.
Hence, although the free detector Hamiltonian~\eqref{eq:freeH-detect} is designed to generate a simple evolution with respect to the proper time $\tau$, \eg see eq.~\eqref{eq:monopole-operator},
the corresponding evolution becomes considerably more complicated when expressed in terms of the laboratory time $t$ by substituting $\tau(\Delta t)$ from eq.~\eqref{eq:tautot}.
When the free evolution is parametrized by the Poincar\'e time \(t\), the
corresponding gap becomes time-dependent with
\begin{equation}
\Omega_t(t) \equiv \Omega\,\frac{\dd\tau}{\dd t} =
\frac{\Omega}{\gamma(Z)}\,
\frac{\mR\bigl[\mR+ \sqrt{(1-Z^2)\mR^2+Z^2(\Delta t)^2}\bigr]}{\bigl[\mR^2-(\Delta t)^2\bigr]\sqrt{(1-Z^2)\mR^2+Z^2(\Delta t)^2}}\,.
\label{eq:Omegat}
\end{equation}
This definition follows from $H^t_{\sA}=\frac{\dd\tau}{\dd t}\,H_{\sA}$,
where $H_{\sA}$ is given in eq.~\eqref{eq:freeH-detect}.
Note that near the future and past tips of the diamond, this boundary-time gap \eqref{eq:Omegat} diverges as
\begin{equation}
\Omega_t(t) \sim
\frac{\Omega}{\gamma(Z)\,(\mR-|\Delta t|)}\,,
\qquad {\rm with}\ \ \ \ \Delta t\to \pm (\mR-0^+)\,.
\label{eq:Omegalim}
\end{equation}
Thus, although the proper gap \(\Omega\) remains constant, the gap with
respect to Poincar\'e time is time-dependent, and diverges at the past and future tips of the causal diamond.
This divergence reflects the compression of an infinite detector proper-time
interval into a finite interval \(-\mR<\Delta t<\mR\).

\subsection*{Boundary harvesting from bulk protocols}

In our first harvesting example, examined in sections \ref{sec:exam12} and \ref{sec:bulk-boundary-harvesting}, we considered entanglement harvesting in the boundary theory with one HKLL detector, smeared over a causal diamond, and one pointlike detector. The coupling of the HKLL detector was chosen to scale as
\begin{equation}
  \lambda_\A = \lambda\, z_\A^{-\Delta}\, ,
\end{equation}
to cancel the near boundary falloff of the HKLL kernel,
\begin{equation}
  K_\Delta(z_\A,\sfx|\sfx') \longrightarrow z_\A^\Delta\, \delta^{(d)}(\sfx-\sfx')\, .
\end{equation}
This choice ensures that, as the bulk detector is pushed to the boundary, the protocol
reduces smoothly to the CFT harvesting setup studied in \cite{Wurtz:2026ofi}.
Correspondingly, the matrix elements in section~\ref{sec:bulk-boundary-harvesting} reproduce the ordinary boundary-to-boundary harvesting result in the limit \(z_\A\to0\). Thus example 1 provides a useful consistency check of our holographic dictionary for detectors. As the associated bulk point approaches the conformal boundary, the HKLL detector reduces to an ordinary pointlike detector in the boundary CFT.

Away from the boundary, the same calculation shows how harvesting is sensitive to
radial separation. 
For fixed boundary separation \(L\), the nonlocal coherence \(\cM\)
is approximately constant near \(z_\A=0\), but eventually decays as the depth becomes of order $L$ and the
bulk-to-boundary correlator is suppressed.
As measured by the negativity \eqref{eq:nega}, harvesting persists only while 
\begin{equation}
  |\cM| > \sqrt{\cL_{\A\A}\,\cL_{\B\B}}\, ,
\end{equation}
and shuts off once \(\cM\) falls below this threshold. In this sense, the harvested
entanglement provides an operational probe of the radial position of the HKLL
detector. 
The disappearance of harvested entanglement beyond a certain range of detector 
parameters is qualitatively reminiscent of the ``islands of separability" found for
pairs of detectors in AdS$_3$ \cite{Henderson:2018lcy}, where harvesting becomes impossible for particular
combinations of detector separation, energy gap, AdS curvature scale, and boundary
conditions.

\subsection*{Bulk harvesting from boundary protocols}

In the second harvesting example of sections \ref{sec:exam12} and \ref{sec:bulk-bulk-harvesting}, we illustrated a different protocol which is more
natural to the bulk. 
There, both boundary detectors are HKLL detectors, dual to
ordinary pointlike bulk detectors where the coupling, switching width, and energy gap are all fixed with respect to their proper times. The HKLL detectors are formulated in terms of the boundary coordinate time, and so in this representation, we require
\begin{equation}
  \mathcal{T}_i = \frac{z_i}{\ell}\mathcal{T}\, , \qquad
  \Omega_i = \frac{\Omega}{(z_i/\ell)}\, , \qquad
  \lambda_i = \frac{\lambda}{(z_i/\ell)}\, ,
  \label{eq:redshiftparm}
\end{equation}
for \(i=a,b\). From this boundary perspective, the $z_i$ simply appear as parameters in the diamond kernel smearing, while from the bulk perspective, they appear because of the physical  ``redshift'' in the AdS geometry \eqref{eq:metric} between Poincar\'e time and the proper time of the detectors. Clearly, this is not the same scaling as in example 1, and indeed the harvesting behaviour in example 2 is correspondingly different.
Note that the role of redshift in AdS harvesting protocols has also been examined in \cite{Ng:2018drz}, including for static detectors with unequal redshift factors. In that setting, the different relations between proper and coordinate time produce distinct coordinate frequencies and lead to nontrivial dependence of the harvested entanglement on the relative switching time.

Our findings were as follows.
In contrast to example 1, we found
the proper-time-matched detectors of example 2 have depth-independent local noise terms.
This is because stationary bulk worldlines at different radial positions are related by the dilatation isometry of the Poincar\'e patch once the detector parameters are measured in proper time. 
Hence, in example 2, unlike in example 1, it is the nonlocal coherence \(\cM\) which encodes all the sensitivity of the harvested entanglement to bulk geometry.

We then found that \(\cM\)  enjoys two qualitatively different regimes as detector B is pushed deeper into the bulk.
First note that according to this ``redshifted'' protocol, detector B interacts with the bulk field over larger  time-scales (as measured with respect to Poincar\'e time) as it gets pushed into the bulk.
The first regime of \(\cM\) takes place when $\zB$ is small, and hence the time smearing of detector B is small.
In this regime, 
\(\cM\)  takes on moderate, real values,
and its behaviour is well-approximated using saddle point methods which probe the analytic regions of the bulk two-point function.
The interpretation of this regime is that the two detectors remain mostly spacelike separated throughout their respective interactions with the vacuum, and so correlations are primarily extracted by harvesting. 
However, the amount of  entanglement harvested is relatively small.
As $z_\B$ grows, the time smearing for detector B becomes larger and a second regime takes over.
In the second regime,  \(\cM\)
becomes complex, 
and its absolute value undergoes  rapid initial growth (after which it peaks and  ultimately enters a polynomial decay phase --- see the discussion below eq.~\eqref{eq:M-branch-bb}).
The initial portion of the growth behaviour is well-approximated by the communication contribution  $\mathcal{M}_C$,
and hence corresponds to the onset of appreciable signalling between the two detectors.
For the $\Delta=1$ example shown, as detector B is pushed even deeper into the bulk, however, $\mathcal{M}_H$ dominates over $\mathcal{M}_C$, indicating that signalling is no longer the primary source of the correlations between the two detectors.
In future work, it would be interesting to analyze the causal structure between the detectors in more detail, 
and investigate whether there is a geometric explanation for the relative contributions of  
$\mathcal{M}_C$ and $\mathcal{M}_H$  as detector B is pushed deep into the bulk.

The contrast between examples 1 and 2 highlights an important lesson from our analysis. 
Both calculations use HKLL reconstruction, but correspond to different protocols and result in very different physics.
Example 1 defines a boundary-normalized protocol that reduces to ordinary CFT harvesting as \(z\to0\). Example 2 instead constructs the boundary dual of a standard detector protocol in the bulk, with parameters fixed in terms of the detector's proper time.
Their qualitatively different radial behaviour (decay in example 1 vs.~growth in example 2) shows that the boundary dual of a bulk detector is specified by more than the HKLL smearing kernel alone. 
One must also specify 
the detector clock, switching function, energy gap, and coupling,
and this freedom allows for multiple physically distinct scenarios.
In particular, the fixed 
proper-time switching width of the detector in example 2 is what causes the switching window to grow in terms of coordinate time, and consequently what generates causal signalling between the detector worldlines, which was negligible in example 1.

\subsection*{Causality from boundary cancellations}

The use of HKLL detectors raises an apparent tension with causality. A bulk
detector may be spacelike separated from another bulk or boundary detector, even while its
HKLL representative has boundary support containing points that are timelike related
to the other detector's boundary support. From the boundary perspective, the
commutator need not vanish pointwise under the smearing integrals. Instead,
microcausality is recovered only after the full HKLL smearing is performed:
\begin{equation}
  \int \dd^d \sy\; \dd^d \sy'\,
  K_\A(\sfX_\A|\sy) K_\B(\sfX_\B|\sy')\,
  [\mathcal{O}(\sy),\mathcal{O}(\sy')] = 0
\end{equation}
whenever the corresponding bulk points are spacelike separated. Thus bulk locality is
encoded in cancellations between various contributions to the integrals, rather than in a pointwise way.

\begin{figure}[ht]
  \centering
  \begin{subfigure}{.485\linewidth}
\includegraphics[width=\linewidth]{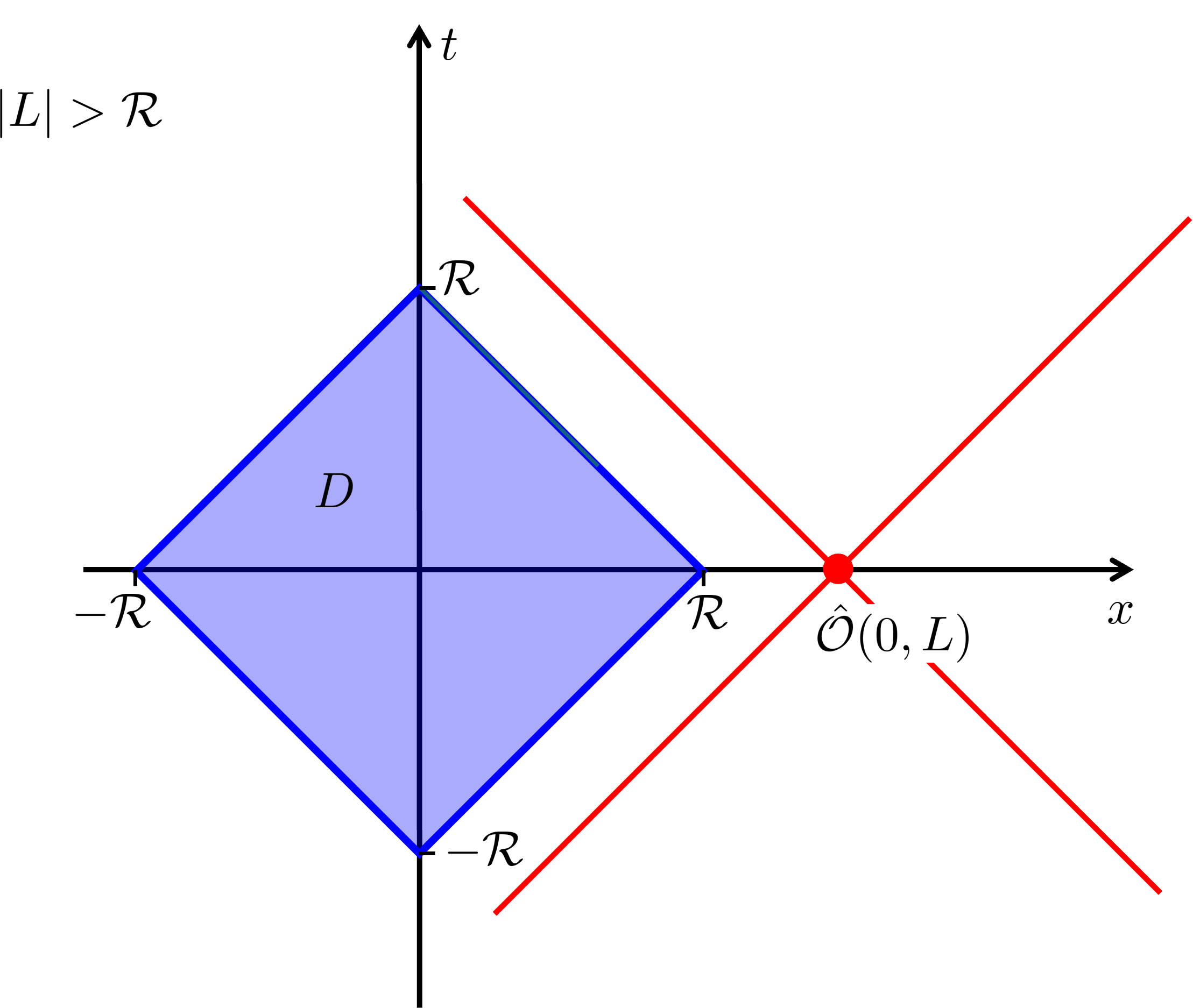}
    \caption{}
  \end{subfigure}
  \hfill
  \begin{subfigure}{.4\linewidth}
\includegraphics[width=\linewidth]{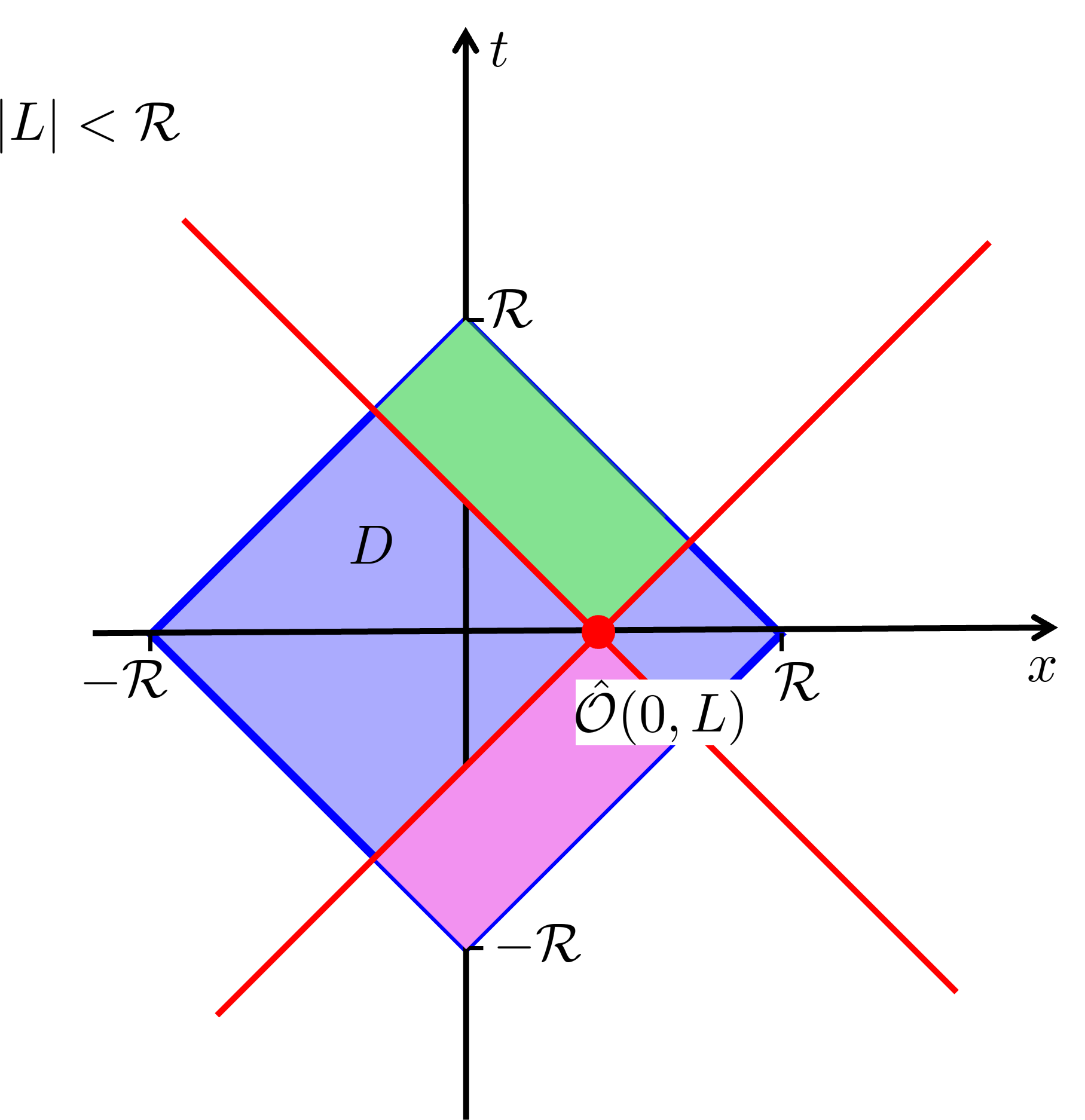}
    \caption{}
  \end{subfigure}
  \caption{(a) When $|L|>\mR$, $\hat{\mathcal O}(0,L)$ lies outside the diamond $D$. Therefore the two boundary operators in the commutator \eqref{eq:zero} are spacelike separated for all $\sfx'$ in the diamond. (b) When \(|L|<\mR\), $\hat{\mathcal O}(0,L)$ lies inside the boundary diamond, and the two operators in eq.~\eqref{eq:zero} are timelike separated when $\sfx'$ lies in the green or pink region. 
  For generic noninteger $\Delta$, the vanishing of the commutator then relies on a precise cancellation of the contributions coming from these two regions.}
  \label{fig:NewFig}
\end{figure}

A simple example illustrating this issue is as follows: consider a bulk operator
\(\hat\Phi(\zA,0,0)\), which we represent on the boundary using the diamond kernel supported on 
\begin{equation}
  D=\{(t,{x}):\ |t|+|{x}|\leq \mR\}\,.
  \label{eq:diamond00R}
\end{equation}
For simplicity, the diamond is centered at the origin \((t,x)=(0,0)\) and recall that \(0<\zA< \mR\). Now compare this operator with a boundary operator \(\hat{\mathcal O}(0,L)\). The bulk and boundary insertions are spacelike separated wherever \(\zA\) and \(L\) are nonvanishing, and so bulk microcausality requires
\begin{equation}
  [\hat\Phi(\zA,0,0),\hat{\mathcal O}(0,L)]=0\, . 
\end{equation}
Using the causal diamond representation, this becomes the boundary identity
\begin{equation}
  \int_D \dd^d \sfx'\,
  K_D(\zA,0,0\,|\,\sfx')\,
  [\hat{\mathcal O}(\sfx'),\hat{\mathcal O}(0,L)] =0\, . 
  \label{eq:zero}
\end{equation}
However, the vanishing of this expression is manifest only when $|L|>\mR$. In this case, $\hat{\mathcal O}(0,L)$ lies outside the diamond~\eqref{eq:diamond00R}, as shown in figure \ref{fig:NewFig}(a), so for all $\sfx'\in D$, the two boundary operators are spacelike separated and the commutator vanishes pointwise in the integral. In contrast, when \(|L|<\mR\), the pointlike operator lies inside the boundary diamond, and the two operators are timelike separated for the points \(\sfx'\in D\) which lie within the past and future lightcone of the point \((0,L)\), as illustrated with the green and pink regions in figure \ref{fig:NewFig}(b). 
Therefore, the boundary commutator does not vanish pointwise throughout the integration region.

The resolution is that the HKLL smearing produces cancellations between different contributions to the left-hand side of eq.~\eqref{eq:zero}. For the simple example considered above, it is straightforward to show the diamond kernel is invariant under the reflection $t'\to -t'$, \iee $K_D(\zA,0,0\,|\,-t',x')=K_D(\zA,0,0\,|\,t',x')$.
This follows directly from the construction of the kernel in section \ref{sec:ads-rindler}. From eq.~\eqref{eq:poincare-to-rindler}, we see the reflection in $t$ translates to 
Rindler time reflection, which is an isometry of the wedge. Similarly, the Weyl factor \eqref{eq:jacobian-omega} appearing in the pullback to the boundary is also reflection invariant (with $t_i=0$). On the other hand, the boundary commutator with \(\hat{\mathcal O}(0,L)\) is odd under this reflection. That is, the Wightman function \eqref{eq:Gbdybdy} obeys \(G^+(-t',x'-L)=[G^+(t',x'-L)]^*\), and hence the commutator is odd since it is $2i$ times the imaginary part of \(G^+\). Therefore, we have
\begin{equation}
  [\hat{\mathcal O}(t',x'),\hat{\mathcal O}(0, L)]
  \propto \operatorname{sgn}(t')\, .
\end{equation}
Thus the integrand of eq.~\eqref{eq:zero} is odd under \(t'\to -t'\), while the integration region, \iee the diamond \(D\), is invariant under this reflection. Therefore the integral vanishes, even though the boundary commutator is nonzero in parts of the integration. More precisely, the contributions from the portion of the diamond lying inside the future lightcone of \((0,L)\) (the green region) cancel exactly against those from the portion lying inside its past lightcone (the pink region) --- see figure~\ref{fig:NewFig}(b).

This simple example illustrates the general mechanism. Bulk locality is not
realized pointwise with boundary locality when using HKLL representatives. Rather,
it is recovered after integrating the boundary commutator against the full
HKLL kernel. 
A more general version of this argument, including arbitrary diamond representatives, is given in appendix~\ref{app:causality-overview}. The general lesson from HKLL reconstruction is that locality in the semiclassical bulk emerges from nonlocal and highly organized boundary expressions.

\subsection*{Distributional versus smooth detector representatives}

There is an important distinction between our HKLL detectors and the standard finite-size detectors encountered in RQI.
Namely, the causal diamond kernel \eqref{eq:K-diamond} is divergent at every boundary point!
This is nevertheless admissible in our calculations,
since this distributional kernel always appears under an integral and the resulting matrix elements become well defined.\footnote{More precisely, in the calculations of section \ref{sec:harvesting-geometries}, 
we use the distributional reconstruction as follows.
First, we use the HKLL identity \eqref{eq:double-smearing-mode-sum} to reconstruct the complete bulk correlator.
This identity is understood distributionally, as explained in appendix \ref{app:causality-overview}.
The resulting two-point function can then be restricted to the detector worldlines,
and integrated against smooth switching functions to yield the finite matrix elements of section \ref{sec:harvesting-geometries}.
}
However, a strictly operational finite-size boundary detector should be described by an ordinary smearing function. In this regard, the diamond kernel $K_D$ can be contrasted with other HKLL representations which do provide ordinary smearing functions. Here examples include the Poincar\'e kernel, which in odd spacetime dimensions smears over the full Minkowski boundary of the Poincar\'e patch, or the analytically continued diamond kernel, which has nontrivial support on complexified boundary loci \cite{Hamilton:2005ju,Hamilton:2006az,Hamilton:2006fh}. However, neither of these constructions is well-suited to defining the boundary detectors needed for RQI applications. Instead, the kernels $\kfd$ constructed in section~\ref{sec:smeared-hkll} by slightly smearing the bulk field provide such a description. They smooth out the distributional diamond representative and give a natural boundary dual of a (smeared) bulk detector with finite spacetime extent.

The structure of these smooth kernels is itself informative. As shown in figure \ref{fig:K_D_Hkll_full_numeric}, these smoothed diamond kernels develop oscillations whose wavelength depends on the radial position of the associated bulk point. 
In particular, in the regime that we studied, the HKLL kernels had a characteristic spatial wavelength proportional to $1/z$, where $z$ is the Poincar\'e coordinate of the bulk point.
Thus deeper bulk localization is encoded not only in the size of the boundary support, but also in increasingly intricate phase structure across the boundary diamond. 
This provides another way in which radial position is represented in boundary detector data.

\subsection*{Future directions}

Having established a bulk-boundary dictionary for detectors, we open the door to a range of further investigations of holography from the perspective of relativistic quantum information. Our analysis was restricted to describing bulk detectors within the Poincar\'e patch of the vacuum AdS spacetime.  It would be interesting to extend the detector dictionary to global AdS or to AdS black branes, where the response might be used to probe geometric bulk features such as horizons and singularities, or thermal correlations in the boundary theory.
In this vein, we note that the HKLL framework has also been applied to reconstruction beyond causal wedges. 
In particular, recent work \cite{Kaushal:2025wbn} constructed an explicit two-sided HKLL-like kernel for operators behind the black hole horizon.

Previous work has shown that the response of an Unruh--DeWitt detector in
global AdS$_2$ can depend nontrivially on the boundary conditions imposed at conformal infinity~\cite{Pitelli:2021oil}. In the present holographic setting, however, these boundary conditions are not freely adjustable. Instead, they are fixed by the AdS/CFT dictionary, \eg the asymptotic behaviour specified in eq.~\eqref{eq:extrapolate}. Nevertheless, it would be interesting to compare harvesting protocols for different admissible bulk quantizations, \eg \cite{Klebanov:1999tb,Witten:2001ua,Berkooz:2002ug}, understood holographically as corresponding to different choices of the dual boundary theory or its deformations.

Our framework also connects naturally to any efforts to formulate the experience of bulk observers directly in the boundary theory, \eg \cite{Papadodimas:2012aq,Jafferis:2020ora,deBoer:2022zps}. 
For instance, we speculate that our approach may provide a useful complementary perspective to the recent discussions of observers in gravitational path integrals, \eg \cite{Harlow:2025pvj,Abdalla:2025gzn,Akers:2025ahe}.

It would also be interesting to extend the present analysis beyond certain immediate limitations. For example, it would be interesting to include bulk interactions, which generate \(1/N\) corrections to the HKLL map and modify the bulk correlators that enter the detector response.
Including the known perturbative corrections to bulk reconstruction would be a natural next step. It would be interesting to understand how detector response functions diagnose the appearance of interactions or perhaps how they distinguish different gravitational dressings of the same bulk operator.
Similarly, 
in this paper, we have restricted our attention to the simple case of Unruh--DeWitt detectors coupled to a bulk scalar field;
however, 
a variety of extensions are possible here, such as coupling the detector to a gauge field or to the graviton in the bulk, which would correspond in the boundary theory to coupling to a conserved current or to the stress tensor, respectively. Of course, coupling to the stress tensor is especially interesting to produce detector protocols that probe gravitational physics in the bulk directly.

\acknowledgments

We thank Caroline Lima and especially Eduardo Mart\'in-Mart\'inez for useful discussions.
We also thank Eduardo Mart\'in-Mart\'inez for detailed comments on our draft.
Research at Perimeter Institute is supported in part by the Government of Canada through the Department of Innovation, Science and Economic Development Canada and by the Province of Ontario through the Ministry of Colleges and Universities. JC acknowledges the support from the Natural Sciences and Engineering Research Council of Canada (NSERC) through a Vanier Canada Graduate Scholarship [Funding Reference Number: CGV--192707].
RCM is also supported in part by an NSERC Discovery Grant, and by funding from the BMO Financial Group.

\appendix


\section{Review: Bulk Wightman function from HKLL}
\label{app:causality-overview}

In this appendix, we show that smearing both insertions of the
boundary vacuum two-point function $W_\partial^+(\sfy_a, \sfy_b)$ with HKLL kernels reproduces the expected bulk vacuum Wightman function $G^+(\sfX_a, \sfX_b)$. 
We work in Poincar\'e coordinates for AdS$_{d+1}$, 
\begin{equation}
    ds^2
    =
    \frac{\ell^2}{z^2}
    \left(
        dz^2-dt^2+d\vec{x}^{\,2}
    \right),
\end{equation}
and consider a free real scalar field with 
\begin{equation}
    \Delta \ = \ \frac{d}{2}+\sqrt{\frac{d^2}{4}+m^2\ell^2}\,,
\end{equation}
which is eq.~\eqref{eq:confdimen} in the main text.

\subsection{Mode argument}

In the bulk,
we choose a complete set of 
normalizable modes \(u_\lambda(\sfX)\) which are Klein--Gordon-orthonormal and positive-frequency with respect to Poincar\'e time-translations.\footnote{Note that a similar analysis would go through for the Rindler vacuum in an AdS-Rindler wedge, with the explicit mode functions presented in section \ref{sec:ads-rindler}.
Of course, the Poincar\'e vacuum and the Rindler vacuum are two different states, and so the actual form of the correlation functions $G^+(\sfX_a, \sfX_b)$ and $W_\partial^+(\sfy_a, \sfy_b)$ will differ between the two cases.}
The free bulk field then has the expansion
\begin{equation}
    \hat\Phi(\sfX)
    =
    \int d\mu(\lambda)\,
    \left[
    u_\lambda(\sfX)\,\hat a_\lambda
    +
    u_\lambda^*(\sfX)\,\hat a_\lambda^\dagger
    \right],
    \label{eq:bulk-field-mode-expansion}
\end{equation}
where  \(d\mu(\lambda)\) encodes an appropriate mode measure and
normalization, and the Poincar\'e vacuum is defined by 
\begin{equation}
    [\hat a_\lambda,\hat a_{\lambda'}^\dagger]
    =
    \delta(\lambda,\lambda'),
    \qquad
    \hat a_\lambda|0\rangle=0\,.
\end{equation}
It follows that the bulk vacuum Wightman function is 
\begin{equation}
    G^+(\sfX_a,\sfX_b)
    \equiv
    \langle0|
    \hat\Phi(\sfX_a)\hat\Phi(\sfX_b)
    |0\rangle
    =
    \int d\mu(\lambda)\,
    u_\lambda(\sfX_a)\,
    u_\lambda^*(\sfX_b)\,.
    \label{eq:bulk-wightman-mode-sum}
\end{equation}
The explicit form of $G^+(\sfX_a,\sfX_b)$ is given in eq.~\eqref{eq:Gbb} of the main text. The relation to explicit mode functions can be found in  \eg~\cite{Duetsch:2002hc}.

We normalize the boundary operator so that the extrapolate dictionary
takes the form
\begin{equation}
    \hat{\mathcal O}(\sy)
    =
    \lim_{z\to0}
    z^{-\Delta}\,\hat\Phi(z,\sy)\,.
    \label{eq:operator-extrapolate-dictionary}
\end{equation}
We then define the boundary imprint of each normalizable bulk mode by
\begin{equation}
    f_\lambda(\sy)
    \equiv
    \lim_{z\to0}
    z^{-\Delta}\,u_\lambda(z,\sy)\,.
    \label{eq:mode-boundary-imprint}
\end{equation}
Taking the boundary limit of the operator expansion
\eqref{eq:bulk-field-mode-expansion} then gives
\begin{equation}
    \hat{\mathcal O}(\sy)
    =
    \int d\mu(\lambda)\,
    \left[
    f_\lambda(\sy)\,\hat a_\lambda
    +
    f_\lambda^*(\sy)\,\hat a_\lambda^\dagger
    \right].
    \label{eq:boundary-operator-mode-expansion}
\end{equation}
In words, the creation and annihilation operators appearing in the bulk field expansion \eqref{eq:bulk-field-mode-expansion} are identified with the Fourier coefficients of the boundary field. (This  was also discussed around eq.~\eqref{eq:rindler-mode-coefficients} in the main text.)
Consequently, in the corresponding boundary vacuum state, we have
\begin{equation}
    W^+_{\partial}(\sy,\sy')
    \equiv
    \langle0|
    \hat{\mathcal O}(\sy)
    \hat{\mathcal O}(\sy')
    |0\rangle
    =
    \int d\mu(\lambda)\,
    f_\lambda(\sy)
    f_\lambda^*(\sy')\,.
    \label{eq:boundary-wightman-mode-sum}
\end{equation}
By inspection, this is the standard vacuum CFT two-point function, \iee
applying the extrapolate dictionary to eq.~\eqref{eq:Gbb} gives
eq.~\eqref{eq:Gbdybdy}.

By definition (see section \ref{sec:ads-rindler}), an HKLL kernel $K_A$ inverts the  boundary-limit map \eqref{eq:mode-boundary-imprint}:
\begin{equation}
    \int_A d^d\sy\,
    K_A(\sfX|\sy)\,
    f_\lambda(\sy)
    =
    u_\lambda(\sfX)\,,
    \qquad
    \sfX\in W_A,
    \label{eq:hkll-mode-reproducing}
\end{equation}
where $W_A$ denotes the bulk subregion reconstructible from $A$ using HKLL. Inserting the mode sum \eqref{eq:boundary-wightman-mode-sum} into the double smearing and applying eq.~\eqref{eq:hkll-mode-reproducing} gives the following relation:
\begin{align}
    \int_{A_a}d^d\sy
    \int_{A_b}d^d\sy'\,
    K_{A_a}(\sfX_a|\sy)
    K_{A_b}(\sfX_b|\sy')\,
    W^+_{\partial}(\sy,\sy')
    =
    \int d\mu(\lambda)\,
    u_\lambda(\sfX_a)
    u_\lambda^*(\sfX_b)\,,
    \label{eq:double-smearing-mode-sum}
\end{align}
where the right-hand side is precisely the bulk vacuum Wightman function \eqref{eq:bulk-wightman-mode-sum} in  $W_A$, as desired. 

All expressions above are understood as equalities of
operator-valued distributions. In particular, the exchange of the mode and boundary integrations used to obtain eq.~\eqref{eq:double-smearing-mode-sum} is justified after smearing the bulk
arguments with test functions.

\subsection{Corollary: Bulk microcausality from HKLL}
\label{app:bulk-microcausality-from-hkll}

We now use the result of the preceding subsection to generalize the microcausality argument of section~\ref{sec:discussion}.  

For $i=a,b$, let $A_i$ be a boundary region admitting an HKLL
reconstruction of the bulk field at
$\sfX_i\in W_{A_i}$:
\begin{equation}
    \hat\Phi_{A_i}(\sfX_i)
    \equiv
    \int_{A_i}d^d\sfy\,
    K_{A_i}(\sfX_i|\sfy)\,
    \hat{\mathcal O}(\sfy)
    =
    \hat\Phi(\sfX_i)\,.
    \label{eq:microcausality-region-reconstruction}
\end{equation}
 The commutator of the two boundary
representatives may be written as
\begin{align}
    \widehat{\mathcal C}_{ab}
    &\equiv
    \left[
        \hat\Phi_{A_a}(\sfX_a),
        \hat\Phi_{A_b}(\sfX_b)
    \right]
    \nonumber\\
    &=
    \int_{A_a}d^d\sfy
    \int_{A_b}d^d\sfy'\,
    K_{A_a}(\sfX_a|\sfy)
    K_{A_b}(\sfX_b|\sfy')
    \left[
        \hat{\mathcal O}(\sfy),
        \hat{\mathcal O}(\sfy')
    \right].
    \label{eq:general-hkll-smeared-commutator}
\end{align}
To evaluate eq.~\eqref{eq:general-hkll-smeared-commutator}, first take
its vacuum expectation value.  
Applying the double-smearing result
\eqref{eq:double-smearing-mode-sum} to the two operator orderings gives
\begin{align}
    \mathcal C_{ab}
    &\equiv
    \langle0|\widehat{\mathcal C}_{ab}|0\rangle
    \nonumber\\
    &=
    \int_{A_a}d^d\sfy
    \int_{A_b}d^d\sfy'\,
    K_{A_a}(\sfX_a|\sfy)
    K_{A_b}(\sfX_b|\sfy')
    \left[
        W^+_{\partial}(\sfy,\sfy')
        -W^-_{\partial}(\sfy,\sfy')
    \right]
    \nonumber\\
    &=
    G^+(\sfX_a,\sfX_b)
    -G^-(\sfX_a,\sfX_b).
    \label{eq:general-vacuum-commutator-from-hkll}
\end{align}
At the free-field limit under consideration, the bulk commutator is a $c$-number
distribution.  Consequently,
\begin{equation}
    \widehat{\mathcal C}_{ab}
    =
    \left[
        G^+(\sfX_a,\sfX_b)
        -G^-(\sfX_a,\sfX_b)
    \right]\mathbf 1\,.
    \label{eq:general-free-bulk-commutator}
\end{equation}

We can determine this difference from the
bulk Wightman function in
eq.~\eqref{eq:Gbb}.  Define
\begin{equation}
    \mathcal G_{d,\Delta}(\xi)
    \equiv
    2^{-\Delta}\xi^\Delta
    {}_2F_1\!\left(
        \frac{\Delta}{2},
        \frac{\Delta+1}{2};
        \Delta+1-\frac d2;
        \xi^2
    \right)
\end{equation}
and
\begin{equation}
    \xi_{\pm}
    \equiv
    \frac{2z_a z_b}{
        z_a^2+z_b^2
        +(\vec{x}_a-\vec{x}_b)^2
        -(t_a-t_b\mp\ie)^2}
    \label{eq:general-xi-two-wightman-prescriptions}
\end{equation}
for $\epsilon>0$.  The upper and lower signs correspond respectively
to the two operator orderings, so that
\begin{equation}
    G^\pm(\sfX_a,\sfX_b)
    =
    \lim_{\epsilon\to0^+}
    \mathcal G_{d,\Delta}\!\left(\xi_\pm\right).
    \label{eq:general-two-bulk-wightman-boundary-values}
\end{equation}

Writing $\sfX_i=(z_i,t_i,\vec{x}_i)$, the two bulk points are
spacelike separated in the Poincar\'e patch precisely when
\begin{equation}
    s_{ab}^2
    \equiv
    (z_a-z_b)^2
    +(\vec{x}_a-\vec{x}_b)^2
    -(t_a-t_b)^2
    >0\,.
    \label{eq:general-poincare-spacelike-condition}
\end{equation}
In this case,
\begin{equation}
\begin{split}
    z_a^2+z_b^2
    +(\vec{x}_a-\vec{x}_b)^2
    -(t_a-t_b)^2
    &=
    2z_a z_b+s_{ab}^2
    \\
    &>
    2z_a z_b\,.
\end{split}
\end{equation}
Since $z_a,z_b>0$, the invariant distance in
eq.~\eqref{eq:invardistance} satisfies
\begin{equation}
    0<\xi<1\,.
    \label{eq:general-spacelike-xi-range}
\end{equation}
For the standard quantization used here,
$\mathcal G_{d,\Delta}(\xi)$ is real and analytic throughout this
interval.  Reversing the sign of the $\ie$ prescription therefore does
not change its boundary value, and hence
\begin{equation}
    G^+(\sfX_a,\sfX_b)
    =
    G^-(\sfX_a,\sfX_b)
    \qquad
    \text{when $s_{ab}^2>0$}.
    \label{eq:general-spacelike-wightman-equality}
\end{equation}
Combining eqs.~\eqref{eq:general-free-bulk-commutator} and
\eqref{eq:general-spacelike-wightman-equality}, we obtain
\begin{equation}
    \left[
        \hat\Phi_{A_a}(\sfX_a),
        \hat\Phi_{A_b}(\sfX_b)
    \right]
    =0
    \qquad
    \text{for spacelike-separated $\sfX_a$ and $\sfX_b$}.
    \label{eq:general-hkll-microcausality}
\end{equation}
Equivalently, the corresponding double smearing of the boundary
commutator in eq.~\eqref{eq:general-hkll-smeared-commutator} vanishes.

Crucially, the cancellation need not occur pointwise on $A_a\times A_b$. Even
when the bulk points are spacelike separated, the boundary regions can intersect.
 The  contributions then cancel only after both HKLL smearings have been
performed. 
No use was made of the shape of either region or of an
explicit formula for either kernel.  Thus the causal-diamond example
of section~\ref{sec:discussion} is a special case of this result.

\section{Review: Asymptotics of the Rindler mode function}
\label{app:WKB}

In this appendix, we derive the large-$|k|$ behaviour of the Rindler mode function given in eqs.~\eqref{eq:V-rindler-asymptotic} 
and 
\eqref{eq:large-k-approx}.
The first step is to observe that the mode function $V(Z)\equiv V_{\omega k}^{R}(Z)$ (given in eq.~\eqref{eq:V-rindler}) obeys the following differential equation:
\begin{equation}
    (1-Z^2)V''
    - 
    \frac{1+Z^2}{Z}V'
    +
    \left[\frac{\omega^2}{1-Z^2}- 
    k^2
    -
    \frac{\Delta(\Delta-2)}{Z^2}\right]V
    =
    0\,.
    \label{eq:V-DE}
\end{equation}
To study the asymptotic behaviour of $V$, 
we introduce a new variable
\begin{equation}
    \sin\theta=Z
\end{equation}
and a new function
\begin{equation}
    \psi(\theta) = \frac{V(\theta)}{\sqrt{\tan\theta}}\,.
\end{equation}
Then, the differential equation \eqref{eq:V-DE} becomes the Schr\"odinger-type equation
\begin{equation}
    \psi''(\theta)
    = 
    \left[k^2 - \frac{\omega^2+\frac{1}{4}}{\cos^2 \theta}
    +
    \frac{(\Delta-1)^2-\frac{1}{4}}{\sin^2\theta}\right]\psi(\theta)\,.
    \label{eq:wkb-start}
\end{equation}

For a first taste of the large $|k|$ limit of $\psi$,
let us study eq.~\eqref{eq:wkb-start} near 
$\theta=0$, where it reads
\begin{equation}
   \psi''(\theta)
    = 
    \left[k^2 +
    \frac{(\Delta-1)^2-\frac{1}{4}}{\theta^2}\right]\psi(\theta)\,.
\end{equation}
The unique solution to the above equation which obeys the boundary condition discussed above eq.~\eqref{eq:V-rindler} reads
\begin{equation}
    \psi(\theta) = 
    2^{\Delta-1}\Gamma(\Delta)k^{1-\Delta}\sqrt{\theta}I_{\Delta-1}(k \theta)\,.
    \label{eq:small-theta}
\end{equation}
This function grows exponentially at large $|k|$, as $|k|^{\frac{1}{2}-\Delta}e^{\theta |k|}$.

Alternatively, we can take a direct large-$|k|$ limit in eq.~\eqref{eq:V-DE}, without first sending $\theta$ to zero.
To do this, we use WKB methods.
For simplicity, we take $k>0$, as the case $k<0$ works similarly.
Then, we employ the ansatz
\begin{equation}
    \psi(\theta) 
    \propto
    e^{k S_0(\theta) 
    +S_1(\theta) 
    + \frac{1}{k}S_2(\theta)+\ldots}\,,
\end{equation}
and solve eq.~\eqref{eq:wkb-start} order by order in $1/k$.
From the discussion above, the leading-order equation gives $S_0(\theta) = \theta$.
The subleading equation gives $S_1'(\theta)=0$, so $S_1$ just contributes an overall normalization constant.
The sub-sub-leading equation
reads
\begin{equation}
    S_2'(\theta) 
    =
    \frac{1}{2}
    \left[- \frac{\omega^2+\frac{1}{4}}{\cos^2 \theta}
    +
    \frac{(\Delta-1)^2-\frac{1}{4}}{\sin^2\theta}\right]
\end{equation}
and is solved by
\begin{equation}
    S_2(\theta) = -\frac{1}{2\tan\theta}\left((\Delta-1)^2-\frac{1}{4}\right)
    -
    \frac{1}{2}
    \left(\omega^2 + \frac{1}{4}\right)
    \tan\theta\,.
\end{equation}
Putting it all together, we have
\begin{equation}
    \psi(\theta) 
    \propto 
    e^{k \theta
    -
    \frac{1}{2k}\left(\frac{(\Delta-1)^2-\frac{1}{4}}{\tan\theta}
    +
    \left(\omega^2 + \frac{1}{4}\right)
    \tan\theta\right)
    +
    \mathcal{O}(1/k^2)}\,,
\end{equation}
where the overall prefactor is determined by the boundary conditions.
From the discussion of eq.~\eqref{eq:small-theta},
we know that the $k$ dependence of the prefactor reads $k^{\frac{1}{2}-\Delta}$.
Hence, restoring absolute values, we have 
\begin{equation}
    V(\theta) 
    \propto 
    |k|^{\frac{1}{2}-\Delta}\sqrt{\tan\theta}
    \;
    e^{|k| \theta
    -
    \frac{1}{2|k|}\left(\frac{(\Delta-1)^2-\frac{1}{4}}{\tan\theta}
    +
    \left(\omega^2 + \frac{1}{4}\right)
    \tan\theta\right)
    +
    \mathcal{O}(1/|k|^2)}\,.
\end{equation}
This is the general procedure used to derive eqs.~\eqref{eq:V-rindler-asymptotic} and 
\eqref{eq:large-k-approx}.


\bibliographystyle{JHEP}
\bibliography{biblio}

\end{document}